\documentclass[reprint,nofootinbib,prl]{revtex4-1}
\usepackage{amsmath,amssymb}
\usepackage{multirow} 
\usepackage{graphicx} 
\usepackage{epsfig}
\usepackage{subfigure}
\usepackage[english]{babel}
\usepackage{url}
\RequirePackage{color}
\usepackage[colorlinks]{hyperref}
\usepackage[T1]{fontenc}
\usepackage{ifpdf}
\begin{document}
\title{New results for electromagnetic quasinormal modes of black holes}
\author{Denitsa Staicova\thanks{Department of Theoretical Physics, Sofia University
"St. Kliment Ohridski",  5 James Bourchier Blvd., 1164 Sofia, Bulgaria} and Plamen Fiziev\thanks{Department of Theoretical Physics, Sofia University "St. Kliment Ohridski",  5 James Bourchier Blvd., 1164 Sofia, Bulgaria;JINR, Dubna}}
\affiliation{Department of Theoretical Physics, Sofia University "St. Kliment Ohridski",  5 James Bourchier Blvd., 1164 Sofia, Bulgaria and Laboratory of Theoretical Phisics, JINR, Dubna, 141980 Moscow Region, Russia }
\email{dstaicova@phys.uni-sofia.bg}
\email{fiziev@phys.uni-sofia.bg}

\begin{abstract}
The differential equations governing the late-time ring-down of the perturbations of the Kerr metric, the Teukolsky Angular Equation and the Teukolsky Radial Equation, can be solved analytically in terms of confluent Heun functions.  In this article, for the first time, we use those exact solutions to obtain the electromagnetic (EM) quasinormal spectra of the Kerr black hole . This is done by imposing the appropriate boundary conditions on the solutions and solving numerically the so-obtained two-dimensional transcendental system. 

The EM QNM spectra are compared with already published results, evaluated through the continued fraction method. The comparison shows that the modes with lower $n$ coincide for both methods, while those with higher $n$ may demonstrate significant deviations. To study those deviations, we employ the $\epsilon$-method which introduces small variations in the argument of the complex radial variable. Using the $\epsilon$-method one can move in the complex $r-$plane the branch cuts of the solutions of the radial equation and examine the dependence of the spectrum on their position. 
 
For different values of $\epsilon$, one can obtain both the frequencies evaluated through the well-established continued fraction method or somewhat different spectra calculated here for the first time. This result raises the question what spectrum should be compared with the observational data and why. This choice should come from better understanding of the physics of the problem and it may become particularly important  considering the recent interest in the spectra of the electromagnetic counterparts of events producing gravitational waves. 
\end{abstract}

\keywords{quasinormal modes,QNM, Schwarzschild metric, Regge-Wheeler equation, Teukolsky radial equation, Teukolsky angular equation, Heun functions, Kerr metric }
\maketitle
\section{Quasi-normal modes of black holes}
During the long history of the study of the quasi-normal modes (QNMs) of a black hole (BH) (\cite{RWE,ZRE,vish,Teukolsky0,Teukolsky_0, Teukolsky2, Teukolsky21, teukolsky_, chan0, QNM, chan1,det1,chan2,QNM2, QNM0, Leaver, Leaver0, Q_N_M, high, special1,extr1, Fiziev1, QNM1,extr2, special3, Fiziev2, Fiziev3, BHB1, QNM21, GW7}), the case of electromagnetic (EM) perturbations has been often ignored because of their perceived irrelevance to the problem of finding gravitational waves. The reasons behind this are that: 

1. The expected luminosity of the gravitational output in the most often studied process of BH binary merger is much bigger than the electromagnetic one (\cite{GW3}). 

2. The EM waves strongly interact with the surrounding medium. This may lead to essential deviation of the observed spectrum from the one expected from the no-hair theorem and makes the object's fingerprint harder to detect. 

3. Most importantly, because of this interaction, the electromagnetic perturbations are strongly absorbed by the interstellar medium, thus making the detection of the signal almost impossible at the predicted low frequencies for the electromagnetic QNMs. 

On the other hand, the gravitational waves (GW) interact very weakly with matter and thus they can be detected at big distances, without getting absorbed or scattered, i.e. without obscuring the signature of the body that emitted them. It is, therefore, reasonable to expect that the GW should be much better suited for studying the central engine of astrophysical events, such as gamma-ray bursts (GRBs), while the EM waves should be seriously influenced by its environment.

For now, however, there are no gravitational waves detected. Although both LIGO and VIRGO detectors already work at design sensitivity, both detectors still fail to ``see`` gravitational waves (\cite{LIGO1,LIGO2,LIGO3,LIGO4,LIGO6,LIGO7,LIGO8}). Particularly puzzling is the lack of GW detection from short GRBs (	\cite{LIGO5,LIGO9}) whose progenitors are expected to emit GWs in the range of sensitivity of the detectors. 

The simplest explanation of those negative results may be a new mechanism of generation  of short GRBs which in  good approximation preserves the spherical symmetry of the process in the central engine and admits only significant dipole radiation. As a result, no significant gravitational waves will be generated during the short GRBs, since the gravitational waves have a quadrupole character. A similar situation is observed in the long GRBs. Such a new hypothesis for short GRBs is supported by the strong observational indications that both types of GRBs may have the same nature and differ only in their time scales (\cite{GRB1,GRB2}). If so, we may expect that most of the energy release from GRBs is in the form of electromagnetic radiation.

A more traditional point of view is a physical process which yields both electromagnetic and gravitational radiation from GRBs. Actually, the ratio of the energy release in the form of electromagnetic waves and in the form of gravitational waves is still an open problem. Its solution strongly depends on the details of the hypothetical mechanism of GRBs which is still far from being well established. For example, the expected (but not yet observed) energy output in the case of GWs from a BH merger is $\sim10^{53}erg$, which coincides (up to a factor due to collimation) with the {\em observed} electromagnetic energy output of GRBs.

While hopes are laid on the Advanced LIGO and Advanced VIRGO which should start operating in the next years, this situation offers a good motivation for optimizing the GW search strategy and understanding better the physics of the GW sources. Particularly, this points to the advantages of studying the EM counterpart of the GW emission, which can help the localization of the source (improving on the big error box of the GW detector) and also it may give additional clues to the physics of the event (\cite{GW1,GW3}). Numerical simulations already explore the detectability of the EM counterpart in different cases (for the case of supermassive black holes mergers, for example, see \cite{GW4, GW5}, for neutron stars mergers \cite{GW1,GW_}). The first results of LIGO and VIRGO searches using such a multimessenger approach were also published \cite{LIGO10}. The idea behind those searches (for details, see \cite{GW8,LIGO10}) is to use the GRBs as the EM counterparts of the GW, since the suspected GRB progenitors (collapsars for long GRBs and binary system mergers of neutron stars and/or black holes for the short GRBs) should emit GWs as well as EM and there is already a well working mechanism for observing the extremely EM luminous GRBs. Although the theoretical results from the multimessenger approach are still being analyzed, the intensive activity in this field shows that the EM counterpart of the GW emission can both facilitate and improve the information obtained from the GW observations. 

The discrete spectrum of complex frequencies called QNMs describes only the linearized perturbations of the metric. Hence, they cannot describe completely the dynamics of the process during the early, highly intensive period of those events when the linearized theory is not applicable. On the other hand, it is known from full numerical simulations that it is the QNMs which dominate the late-time evolution of the object response to perturbations \cite{high}. Thus, from an observational point of view, the QNM are important, since we may observe only the tails from the corresponding events, being far from them. This conclusion is supported by the recent numerical observation of two lowest gravitational modes of QNMs in the spectrum of the signal obtained from the full 3d general relativistic head-on collision of non-spinning BH  (see \cite{headon}, and for further information \cite{headon0}). This result is not isolated -- there are number of works in which the QNMs approximate well the signal of full 3d general relativistic simulations of mergers (for example \cite{bin1,bin2,bin3,bin4,bin5} and also the pioneer works discussed in \cite{special3}). This clearly implies that studying QNMs can bring new insights to the physics in the processes which include strong-field regime. 

Those numerical results also point to another possible use of the QNMs in astrophysical observations. The QNMs correspond to particular boundary conditions characteristic for the object in question, and since in the case of BH the no-hair theorem states that they should depend only on the parameters of the metric (the mass $M$ and the rotation $a$ for the case of Kerr BH), measuring those frequencies observationally can be used to test the nature of the object -- a black hole or other compact massive objects like super-spinars (naked singularities), neutron stars, black hole mimickers  etc. \cite{special31,NB1,NB2,NB3,NB4,spectra}. It also can constrain additionally the no-hair theorem which was recently put into question in the case of black holes formed as a result from the collapse of rotating neutron stars \cite{GW6}. An interesting possibility is to find a way to use the damping times of the EM quasi-normal modes for comparison with observations. While the frequencies are subject to interaction with the surrounding matter which can significantly change the spectra, their damping times should be much less prone to deviation. A suggestion for such use can be found in simulations of jet propagation, which imply that the short-scale variability of the light-curve should be due to the central engine and not to the interaction of the jet with the surrounding medium (see \cite{time-scale} and reference therein). 

One more important application of the QNM spectrum can be found in the study of the central engines of the GRBs, whose extreme luminosity($\sim10^{51}-10^{53}erg/s$) and peculiar time-variability cannot yet be fully explained in the frames of current models.  Even though numerical simulations proved to be capable of describing {\em some} of the features of the GRBs light curves (for a recent review on GRBs, see \cite{GRB}), the biggest stumbling stone seems to be the lack of proper understanding of the central engine of the GRBs. 

Common ingredients of the existing GRB models include a compact massive object (black hole or a milli-second magnetar) and extreme magnetic field ($\sim10^{15}G$ ) which accelerate and collimate the matter via different processes. Although those processes are still an open question for both theory and numerical simulations, the very central engine can be studied approximately by the linearized EM (and also GW if data is available) perturbations of the Kerr metric. When finding the electromagnetic QNMs, one does not care for the origin of the perturbation, but only for its spin and the parameters of the compact massive object. In the idealized EM case, the perturbation is described by free EM waves in vacuum. While the astrophysical black holes are thought to be not charged, they are immersed into EM waves with different energy and origin. The black hole response to such EM perturbations in linear approximation will be then the QNM spectrum defined by the appropriate boundary conditions \footnote{Other   conditions more suitable for describing a primary jet were studied in \cite{spectra}.}. 

Studying the so obtained electromagnetic spectrum can give important insights into the key parameters of the physics occurring during  high-energy events as GRBs. In particular, the electromagnetic QNMs are subject to resonant amplification (the idea of the black hole bomb, \cite{BHB1,GW7,teukolsky_}) and additionally, it is known from previous evaluations of the spectrum, that they exhibit very low damping in the limit $a\to M$. For the moment, there are no observations of the rotations of the GRB progenitors, but the theoretical expectations are that they should be highly rotating in order to produce jets with such luminosity and collimation. Available observationally measured rotations of astrophysical compact massive objects show that there are many cases of near extremal values thus studying the extremal limit could be relevant to such objects. For example, recent evaluations of the spin parameter of astrophysical black holes give for the spin parameter $a=0.63,0.90$ and $a=0.89,0.99$ for $M=1,0.1M_{\odot}$ for Sw J1644+57 and Sw J2058+05 respectively (most probable values, see \cite{rot}), and also $a>0.98$ for GRS 1915+105 (\cite{rot1}) and $a=0.989$ for MCG-6-30-15 (\cite{rot2}).

Moreover, because of the relatively good coverage of the GRBs observations, there is a great amount of data, in a wide energy range (from optical to GeV energies) and from different epochs of the bursts which can be used to test the eventual applicability of the QNM spectrum in the late-time epoch of the burst. It may be hard to extract EM QNM spectra from the existing crude GRB spectra since the basic EM QNM frequencies are very low (from a small part of Hz --- for supermassive BH to several kHz -- for stellar mass BH) and the intensity of the higher EM QNM may be very low. 
To the best of our knowledge such attempts haven't been made. A new space mission, which will additionally help the EM observations in the radio range -- RadioAstron -- will offer unprecedented resolution (up to 1$\mu$arcsec) in a  wide range of high frequencies (from 0.3GHz to 18-25GHz) accompanied by continuum, polarized and spectral imaging (for details see RadioAstron website \footnote{\url{http://www.asc.rssi.ru/radioastron}}). One may hope to use this new mission for a more detailed study of the spectra of EM radiation from the compact objects but its sensitivity is also far from the area of the basic EM QNM.

Theoretical calculations of the QNMs, however, are not simple. The linear perturbations of the rotating BHs are described by two second-order linear differential equations: the Teukolsky radial equation (TRE) and  the Teukolsky angular equation (TAE) on which specific boundary conditions should be imposed (\cite{QNM,QNM0}). Until recently, solving those equations analytically was considered impossible in terms of known functions, so approximations with  simpler wave functions were used instead.  The resulting system of spectral equations -- a connected problem with two complex spectral parameters: the frequency $\omega$ and the separation constant $E$ -- has been solved using different methods (\cite{special3,special1,QNM1,QNM21}) with notably the most often used being the method of the continued fraction. This method was adapted by Leaver from the problem of the hydrogen molecule ion in quantum mechanics \cite{Leaver,Leaver0}. While being successful in obtaining the QNMs spectra, Leaver's method has the disadvantage of not being directly connected with the physics of the problem, thus making it harder to further explore the spectra -- for example studying its dependence on the choice of the branch cuts of the exact solutions of the radial equation. In addition, one has some specific numerical problems in calculation of particular modes, for example, in calculation of the  $9^{th}$ one in the gravitational case \cite{special3, special1,Q_N_M}.

The analytical solutions of the TRE and the TAE can be written in terms of the confluent Heun function (for $a\neq M$) as done for the first time in \cite{Fiziev1,Fiziev3, Fiziev4, Fiziev2}. Those functions are the unique local Frobenius solutions of the second-order linear ordinary differential equation of the Fuchsian type \cite{heun3_,heun,heun1_,heun2_} with 2 regular singularities ($z=0,1$) and one irregular ($z=\infty$) (for details see \cite{Fiziev3}) and in the \textsc{maple} notation, they are denoted as: 
$\text{HeunC}(\alpha,\beta,\gamma,\delta,\eta,z)$ (normalized to $\text{HeunC}(\alpha,\beta,\gamma,\delta,\eta,0)=1$). While the theory of the Heun functions is still far from being complete, they are implemented in the software package \textsc{maple} and despite the problems in that numerical realization (see the discussion in \cite{arxiv1}), the confluent Heun function was used successfully in our previous works \cite{Fiziev1,spectra,arxiv1,arxiv3}. The advantage of using the analytical solutions is that one can impose the boundary conditions on them {\em directly} (see \cite{Fiziev1,spectra}), and thus be able to control all the details of the physics of the problem. 

In a series of articles, we developed a method for solving numerically two-dimensional systems featuring the Heun functions (the two-dimensional generalization of the M\"uller method described in \cite{arxiv}) and we used it successfully in the case of gravitational perturbation $s=-2$ of the Schwarzschild metric \cite{arxiv1}. The so obtained frequencies repeat with high precision the results already published by other authors. Additionally, we used the epsilon-method (see below) to study the branch cuts of the solutions, which are particularly important in the case of the $9^{th}$ mode for $s=-2$. The latter, because of its very small real part, was often wrongly considered to represent the purely imaginary algebraically special mode. While the analysis of the potentials of the Regge-Wheeler equation (RWE) and the Zerilli equation (ZE) showed that there is a branch cut on the imaginary axis for this mode \cite{special2,AS} which leads to its interesting properties, this result is directly obtainable from the actual solutions of the RWE and ZE in terms of the confluent Heun functions.  Furthermore, the stability of the solutions with respect to movement of the position of those branch cuts was studied in the whole interval of applicability of the method. Such a study is impossible with the continued fraction method, where the radial variable does not explicitly enter the equations being solved and which cannot be used for purely imaginary frequencies (\cite{Leaver0}, p.8). If one looks at the equations used to derive this method in detail, it turns out that the angular equation \cite{Leaver} in the continued fraction method coincides with the the three-term recurrence defining the confluent Heun function, solution of the TAE, in the neighborhood of the two regular singular points, $u=-1,1$, where $u=\cos(\theta)$ (\cite{heun1_} Eq. (1.9-1.10). The radial equation in the continued fraction method, however, differs from the solution of the TRE in terms of the confluent Heun functions.  This is because in  Leaver's paper, the series from which the continued fraction are obtained are developed for the powers of $\frac{r-r_+}{r-r_-}$ (due to switching the places of the singular points, see \cite{Leaver0}, p.7), while the asymptotic three-term recurrence of the confluent Heun function at infinity is developed for $\frac{1}{r-r_-}$. Note that in \textsc{maple}, for $r>r_+$, the evaluation of the confluent Heun functions at infinity is obtained by numerical integration from the second singularity $r=r_+$.

In this article, we continue the exploration of the application of the confluent Heun functions by studying the QNMs of the Kerr BH. Our results show that using the confluent Heun function, one can obtain the QNMs for a wide range of modes and rotational parameters, and that there is very good agreement between our results and those obtained within other methods. Using the $\epsilon$-method made it possible, for the first time to study the dependence of the so obtained frequencies on small deviations in the phase condition and it is shown how this nontrivial dependence evolves with $n$ and $a$. In this case, some of the modes are independent of $\epsilon$ which should be expected since the frequencies should not depend on the radial variable. Other modes, however, depend critically on the value of $\epsilon$ and they can differ seriously from the already published results. Additionally, details how the modes change in the interval of validity of the steepest descent method are presented.

\section{The Teukolsky angular equation}
In Chandrasekhar's notation, the Teukolsky Master Equation (\cite{teukolsky_}), for $|s|=1$ is separable under the substitution $\Psi=e^{i(\omega t+m\phi)}S(\theta)R(r)$, where $m=0, \pm 1, \pm 2$ for integer spins and $\omega$ is the complex frequency. Because of the choice of this form of $\Psi$, the sign of $\omega$ differs from the one Teukolsky used, and the stability condition, guaranteeing that the perturbations will damp with time, reads $\Im(\omega)>0$. 

The TAE for EM perturbations ($s=-1$) has 16 classes of exact solutions $S(\theta)$ in terms of the confluent Heun functions (for full details see \cite{Fiziev3}). To fix the spectrum approximately, one requires an additional regularity condition for the angular part of the perturbation, which means that if we choose one solution, $S_1(\theta)$ regular around the one pole of the sphere ($\theta=0$) and another, $S_2(\theta)$, which is regular around the other pole ($\theta=\pi$), then in order to ensure a simultaneous regularity, the Wronskian of the two solutions should become equal to zero, $W[S_1(\theta),S_2(\theta)]=0$. This gives us one of the equations for the two-dimensional system that needs to be solved to obtain the QNMs of the Kerr BH. 

In \cite{Fiziev3}, there are four pairs of Wronskians, each pair being valid in a sector of the plane $\{s,m\}$. Ideally, using any of them should lead to the same spectrum. Numerically, the results obtained with the different Wronskians coincide within 11-13 digits of precision. The Wronskians used to obtain the spectrum are:
\begin{align}
 W[S_1,S_2]=\frac{\text{HeunC}'(\alpha_1,\beta_1,\gamma_1,\delta_1,\eta_1,\left( \cos \left( \pi/6  \right)  \right) ^{2})}{\text{HeunC}(\alpha_1,\beta_1,\gamma_1,\delta_1,\eta_1,\left( \cos \left( \pi/6  \right)  \right) ^{2})}+\notag\\
\frac{\text{HeunC}'(\alpha_2,\beta_2,\gamma_2,\delta_2,\eta_2,\left( \sin \left( \pi/6  \right)  \right) ^{2})}{\text{HeunC}(\alpha_2,\beta_2,\gamma_2,\delta_2,\eta_2,\left( \sin \left( \pi/6  \right)  \right) ^{2})}+ p=0
\label{Wr1}
\end{align}
\noindent where the derivatives are with respect to $z$ and the values of the parameters for the two confluent Heun functions for each $m$ are as follows:

For $m=0$:
$\alpha_1= 4\,a\omega,
\beta_1= 1,
\gamma_1=- 1,
\delta_1=4\,a\omega,
\eta_1=1/2-E-2\,a\omega-{a}^{2}{\omega}^{2}$ and

$\alpha_2=-4\,a\omega,
\beta_2=1,
\gamma_2=1,
\delta_2=-4\,a\omega,
\eta_2=,1/2-E+2\,a\omega-{a}^{2}{\omega}^{2}$, $p=\frac{1}{\left( \sin \left( \pi/6  \right)  \right) ^{2}}$

For $m=1$:
$\alpha_1=-4\,a\omega,
\beta_1=2,
\gamma_1=0,
\delta_1=4\,a\omega,
\eta_1=1-E-2\,a\omega-{a}^{2}{\omega}^{2}$
and

$\alpha_2= -4\,a\omega,
\beta_2=0,
\gamma_2=2,
\delta_2=-4\,a\omega,,
\eta_2=1-E+2\,a
\omega-{a}^{2}{\omega}^{2}$ and $p=-4\,a\omega$

For $m=2$:
$\alpha_1= -4\,a\omega,
\beta_1=3,
\gamma_1=-1,
\delta_1=4\,a\omega,
\eta_1=5/2-E-2\,a\omega-{a}^{2}{\omega}^{2}$
and

$\alpha_2= -4\,a\omega,
\beta_2=1,
\gamma_2=-3,
\delta_2=-4\,a\omega,
\eta_2=5/2-E+2\,a\omega-{a}^{2}{
\omega}^{2}$ and $p=8-4a\omega.$

\noindent where we use $\theta=\pi/3$ (the QNMs should be independent of the choice of $\theta$ in the spectral conditions). 

These Wronskians differ from those in \cite{Fiziev3}, most notably by the presence of the term $p$. The reason for this is that they were constructed using different two solutions $[S_1(\theta),S_2(\theta)]$ of the TAE (note that the sign convention in this paper differs from the one in \cite{Fiziev3}), each of which still being regular on {\em one} of the poles. The regularity condition in $\theta\in [0,\pi]$ is guaranteed by the Wronskian becoming 0 for some angle $\theta\in (0,\pi)$(\cite{Fiziev3}).  The different set of equations were used to improve the numerical convergence of the root-finding algorithm and to avoid \textsc{maple}'s problems with the evaluation of the confluent Heun function and its derivative for certain values of the parameters. 

\section{The Teukolsky radial equation}
The TRE differential equation is of the confluent Heun type, with $r=r_{\pm}$ regular singular points and $r=\infty$ -- irregular one. As it was noted in \cite{spectra}, the point $r=0, \theta=\pi/2$ is not a singularity for this equation and, therefore, it need not be considered when imposing the boundary conditions. The solutions of the TRE for $r>r_{+}$, are :
\begin{align}
&R(r)\!=\!C_1R_1(r)+C_2R_2(r), \text{for} \label{R2}\\
&R_1(r)=e^{\frac{\alpha\,z}{2}}(r\!-\!r_+)^{\frac{\beta\!+\!1}{2}}(r\!-\!r_-)^{\frac{\gamma\!+\!1}{2}}\text{HeunC}(\alpha,\beta,\gamma,\delta,\eta,z)\!\notag\\
&R_2(r)=e^{\frac{\alpha\,z}{2}}(r\!-\!r_+)^{\frac{\!-\!\beta\!+\!1}{2}}(r\!-\!r_-)^{\frac{\gamma\!+\!1}{2}}\text{HeunC}(\alpha,\!-\!\beta,\gamma,\delta,\eta,z),\notag
\end{align}
\noindent where $z=-\frac{r-r_+}{r_+-r_-}$ and the parameters are:

$\alpha =-2\,i \left( {\it r_{_+}}-{\it r_{_-}}
 \right) \omega$,  $\beta =-{\frac {2\,i(\omega\,({a}^{2}+{{\it
r_{_+}}}^{2})+am)}{{\it r_{_+}}-{\it r_{_-}}}}-1$,\\
$\gamma ={\frac {2\,i(\omega\,({a}^{2}+{{\it
r_{_-}}}^{2})+am)}{{\it r_{_+}}-{\it r_{_-}}}}-1$,\\
$\delta =-2i\!\left({\it r_{_+}}-{\it r_{_-}}
\right)\!\omega\!\left(1-i
 \left( {\it r_{_-}}+{\it r_{_+}} \right) \omega \right)$,\\
$\eta =\!\frac{1}{2}\frac{1}{{ \left({\it r_{_+}}\!-\!{\it r_{_-}}
\right) ^{2}}}\times\\
\Big[ 4{\omega}^{2}{{\it r_{_+}}}^{4}\!+ 4\left(i \omega
\!-\!2{\omega}^{2}{\it r_{_-}}\right) {{\it r_{_+}}}^{3}\!+\! \left(
1\!-\!4a\omega\,m\!-\!2{\omega}^{2}{
a}^{2}\!-\!2E \right) \times \\ \left( {{\it r_{_+}}}^{2}\!+\!{{\it r_{_-}}}^{2}
\right) \!+\! \\
 4\left(i\omega\,{\it r_{_-}} \!-\!2i\omega\,{\it r_{_+}}\!+\!E\!-\!{\omega}^
{2}{a}^{2}\!-\!\frac{1}{2} \right) {\it r_{_-}}\,{\it r_{_+}}\!-4{a}^{2} \left(
m\!+\!\omega\,a
 \right) ^{2} \Big].$

Here we have followed \textsc{maple}'s internal rules when constructing the general solution of the differential equation from the confluent Heun type. Accounting for the symmetries of the confluent Heun function, the solutions \eqref{Rbc} coincide with those in \cite{Fiziev3} (for $\omega$ replaced with $-\omega$). \footnote{It is important to emphasize that the so obtained solutions cannot be used for extremal KBH ($a=M$) since in this case the differential equation is of the double confluent type and its treatment differs, so it is outside the scope of this work.} 

The TRE has 3 singular points $r_-,r_+,\infty$ and in order to fix the spectrum, one needs to impose specific boundary conditions on two of those singularities (i.e. to solve the central two-point connection problem 
\cite{heun3_}). Different boundary conditions on different pairs of singular points will specify different physics of the problem. In our case, we impose the black hole boundary conditions (BHBC) -- waves going simultaneously into the event horizon ($r_+$) and into infinity -- following the same reasoning as in \cite{spectra} where additional details can be found. Then, the BHBC read:
\begin{enumerate}
 \item BHBC on the KBH event horizon $r_+$.

For $r\to r_+$, from $r(t) = r_+ +e^{\frac{-\Re(\omega)t+const}{\Im(n_{1, 2})}}\to r_{+}$, where $n_{1,2}$ are the powers of the factors $(r-r_+)^{n_{1,2}}$ in $R_{1,2}$, follows that for $m=0$, the only valid solution in the whole interval $(-\infty,\infty)$ is $R_2$, while for $m\neq0$, the solution $R_2$ is valid for frequencies for which $\Re(\omega) \not\in (-\frac{ma}{2Mr_+},0)$. This means that the rotation splits the area of validity of $R_2$ into two and if this condition is not fulfilled, then the spectrum corresponds to waves going out of the horizon -- a white hole. We won't pursue the spectrum in the case of a white hole, but it is important to keep in mind that the boundary conditions correspond to a BH, only in the ranges of validity of each solution. In our numerical work we use only $R_2$ since the confluent Heun function in $R_1$ is numerically unstable in \textsc{maple}. 

\item BHBC at infinity.

At $r\to \infty$, the solution is a linear combination of an ingoing ($R_{\leftarrow}$)
and an outgoing ($R_{\rightarrow}$) wave: $R=C_{\leftarrow}\,R_{\leftarrow}+C_{\rightarrow}\,R_{\rightarrow},$
where $C_{\leftarrow}$, $C_{\rightarrow}$ are unknown constants and $R_{\leftarrow}, R_{\rightarrow}$ are found using the asymptotics of the confluent Heun function as defined in \cite{heun3_,Fiziev3}. 

To ensure only outgoing waves at infinity, one needs to have $C_{\leftarrow}=0$. 

To achieve this, first one finds the direction of steepest descent in the complex plane $\mathbb{C}_r$ for which  $\lim\limits_{r\to\infty}\frac{R_{\rightarrow}}{R_{\leftarrow}}=r^{-4i\,\omega\,M+2} e^{-2i\omega\,r}= 0$ tends to zero most quickly: $\sin(\arg(\omega)\!+\!\arg(r))\!=\!-\!1$. This gives us a relation $r=\mid\! r\! \mid e^{3/2i\pi-i\,\arg(\omega)}$ (\cite{Fiziev1}) between $\omega$ and $r$ which is exact only if one uses the first term of the asymptotic series for the confluent Heun function (i.e. $\text{HeunC} \sim 1$). More details about this approximation can be found in the next section.

Then, it is enough to solve :
\begin{align}
C_{\leftarrow}\!=\!r^{2\!+\!i\,\omega\!+\!\frac{2i\,m\,a\!+\!i\,\omega}{r_+\!-\!r_-}}\text{HeunC}(\alpha,\!\!-\!\beta,\gamma,\delta,\eta,z)\!=\!0,\hspace{-6.5px}
\label{Rbc}
\end{align}
\noindent in order to completely specify the spectra $\{\omega_{n,m},E_{n,m}\}$, with $r=110\, e^{3/2i\pi-i\,\arg(\omega)}$ (we use $|r|=110$ as the actual numerical infinity and $M=1/2$).
\end{enumerate}

\section{The epsilon-method}
Equation \eqref{Rbc} relies on the direction of steepest descent defined by the phase condition $\arg(r)\!+\!\arg(\omega)\!=\!3/2\!\pi$. This approximate direction was chosen ignoring the higher terms in the asymptotic expansion of the solution around the infinity point, therefore, one can expect that the true path in the complex plane may not be a straight line but a curve. In principle, the spectrum should not depend on this curve, as long as $r$ stays in the sector of the complex plane where $\lim\limits_{r\to\infty}\frac{R_{\rightarrow}}{R_{\leftarrow}}= 0$, i.e. as long as $\pi<\arg(r)\!+\!\arg(\omega)\!<2\pi$, with only the convergence of the algorithm being affected. Numerical exploration of that limit evaluated with the first 3 terms in the asymptotic expansion of the appropriate confluent Heun functions for the modes $\omega_{n}, \, n=0..18$ when there is no BH rotation ($a=0$), and for some modes when there is rotation, confirms that indeed the limit remains approximately zero in the whole interval $\pi+0.1<\arg(r)\!+\!\arg(\omega)\!<2\pi-0.1$.

The spectrum obtained numerically in this interval, however, depends in a nontrivial way on this curve. The complications are partially due to the appearance of branch cuts in the numerical realization of the confluent Heun functions in \textsc{maple}. The branching points of the confluent Heun function in the complex z-plane are found at the singular points $z=1$ and $z=\infty$. In \textsc{maple}, the semi-infinite interval $(1,\infty)$ on the real axis is chosen as a branch cut. In the case of QNMs of nonrotating BHs \cite{arxiv1,arxiv3}, it was observed that when those branch cuts are found near a frequency (since the radial variable, $r$ depends on the frequency $\omega$, the branch cuts appear also in the complex $\omega$-plane), they have serious effect on it, leading to the disappearance or translation of certain modes. As seen from the numerical study, this effect is likely due to a transition to another sheet of the multivalued function. 

As a way to find the correct sheet and remain on it, we introduced the epsilon-method which consists in adding a small variation ($\mid\!\epsilon\!\mid<\!1$) in the phase condition:
\begin{equation}
\arg(r)\!+\!\arg(\omega)\!=\!\frac{3+\epsilon}{2}\!\pi.
\label{arg}
\end{equation}
 Using the $\epsilon$-method, one can change the location of the branch cut with respect to the eventual roots of the system and this way try to minimize the effect of the jump discontinuity of the radial function \footnote{Here, the radial function refers to the solutions of the radial equation and not to the differential equation itself.}. Using $\epsilon$, one can also explore the whole sector $\pi<\arg(r)\!+\!\arg(\omega)\!<2\pi$ , i.e. effectively moving $r=|r|e^{i\arg(r)}$ in the complex plane and this way test the numerical stability of the QNM spectrum, especially with respect to the position of the branch cuts of the radial function.  

Using the parameter $\epsilon$, the observed branch cuts in the realization of the confluent Heun function in \textsc{maple} are as follows: 
\begin{enumerate}
 \item For $r$--real, one encounters one of the branch cuts of the confluent Heun function. The equation of the line of this branch cut is: $\Im(\omega)/\Re(\omega)=\tan(3/2\pi+\epsilon\pi/2)=-\cot(\epsilon \pi/2)$. This line rotates when $\epsilon$ changes.
 \item If $\Im(\omega)=0$, then one encounters the branch cut of the argument-function. In this case the branch cut is defined for $\Re(\omega)=(-\infty,0)$. This branch cut, however, affects the solutions only very close to $a=M$ where the frequencies can become almost real.
 \item If $\Re(\omega)=0$ and $\Im(\omega)=2n$ , $n=1,2,3..$, then one can have $\Im(r)=0$ for certain values of $\epsilon$ and thus reach the branch cut of the confluent Heun function on the real axis. This condition can affect modes which are very near the imaginary axis (for example, similar condition holds around the algebraically special mode for a nonrotating BH). 
 \item Additional branch cuts may appear in the cases where $\Im(r^k)=0$, for k--noninteger or complex (where $z=1-r$). Those branch cuts depend on the numerical realization of the confluent Heun function in \textsc{maple} and the equations of their lines can be obtained numerically from the values of the function for each $\epsilon$. For example, one such branch cut was observed for  $\Im(\omega)/\Re(\omega)\approx\tan(1.419+\epsilon\pi/2)$.
\end{enumerate}

Although the study of the full effect from the movement of the branch cuts on the EM spectrum of KBH, as it was done in the case of gravitational perturbations of nonrotating black holes (\cite{arxiv1}), is outside the scope of this work, some preliminary results on the issue can be found in the Appendix.  

A review of the branch cuts of the retarded Green function in the Schwarzschild case, can be found in the recent article \cite{BC_new}. In it the authors use different asymptotic series expansions of the radial function suggested by Leaver in a novel way to study the properties of the so-called branch cut modes, which one needs to account for when evaluating the Green function when due to the existence of a branch cut, it cannot be determined solely by its QNM series. While Leaver's asymptotic series expansions are particular asymptotic expansions of the confluent Heun functions, we cannot draw a direct parallel between our results and theirs. For one, the asymptotics of the confluent Heun function in the whole complex plane consists of different series expansions, valid for different values of the parameters and having different properties. In the case of the KBH, the radial function is generally complex along the branch cuts and the analysis in terms of the radial potential besides in the limit cases $a=0, a=M$ is not straight forward. Thus, without an in depth study of the asymptotics of the radial function and its branch cuts, it is difficult to make analytical conclusions,  based on our results obtained with one particular numerical realization of the confluent Heun functions. Therefore, here we present the results of our numerical experiments, while a proper analytical study will be published elsewhere.

\section{Numerical algorithms}
The spectral equations we need to solve to find the spectrum $\omega_{n,m}(a)$ for $M=1/2$ are Eqs.\eqref{Wr1} and \eqref{Rbc}. This system represents a two-dimensional connected problem of two complex variables -- the frequency $\omega$ and the separation parameter $E$ -- and in both of its equations one encounters the confluent Heun function and in the case of the TAE -- their derivatives. 

A system like that cannot be easily solved by conventional methods like the Newton method and the Broyden method, as outlined in \cite{arxiv1} and \cite{arxiv}, since they do not work well with the confluent Heun function in \textsc{maple}. For this reason, our team developed a new method, namely the two-dimensional M\"uller algorithm which proved to be much better adapted to work with those functions. The details of the algorithm can be found in \cite{arxiv,arxiv1, arxiv3}, but for completeness, we will mention only that it relies on the M\"uller method which is a quadratic generalization of the secant method having better convergence than the latter. The new algorithm does not need the evaluation of derivatives, thus avoiding one of the biggest problems when using the confluent Heun function in \textsc{maple}. Clearly, in the system we solve the angular spectral equation (Eq. \eqref{Wr1}) includes derivatives, but in this case, they remain in the domain $|z|<1$, where they can be evaluated correctly (for most values of the parameters) and with precision comparable to that of the radial function. It is important to note that both $\omega,E$ are found directly from the spectral system (Eqs. [\eqref{Wr1} and \eqref{Rbc}]) and with equal precision. \footnote{The algorithm is realized in \textsc{maple} code and the numbers presented below are obtained using \textsc{maple} 13 on the computer cluster Physon. The software floating point number is set to 64 (unless stated otherwise), the precision of the algorithm -- to 15 digits. } \footnote{An important precaution when working with the confluent Heun function in \textsc{maple} is that its precision or over-all behavior may depend on different factors which are not always under user's control \cite{arxiv1}. From our observations, it seems that one can trust around 11-12 digits of the frequencies at the worst, usually around 13 digits. The points we present are the maximum we could get out of \textsc{maple}, but future improvements in \textsc{maple}'s code may significantly expand the area of application of the method and/or also its precision. }

\section{Numerical results for electromagnetic QNMs}
While the evaluation of QNMs is not new to physics, the actual numbers published for EM perturbations of KBH are scarce. Because of this, for comparison, we use the numbers published by Berti et al. \cite{special3,special31}, the numerical data can be found on \url{http://www.phy.olemiss.edu/~berti/qnms.html}. Those numbers were obtained using the continued fraction method which is still considered as the most accurate method for obtaining the QNMs from the KBH. The available control frequencies are $n=0..6$ for $l=1$ and $n=0..3$ for $l=2$. Using those ``control'' numbers denoted as $\omega_{n,m}^B,E_{n,m}^B$ one can easily check the precision of the method. 

The first 10 modes of the spectrum obtained using the new method in the interval $a=[0,M)$ can be found on \url{http://tcpa.uni-sofia.bg/conf/research}. In the Appendix, one can find some of the QNMs for specific values of $a$.

\subsection{Non-rotating BH}
It is already well known that when there is no rotation ($a=0$) the electromagnetic QNMs come in pairs symmetrical to the imaginary axis $\omega_{n,m}=\pm|\Re(\omega_{n,m})|+i\Im(\omega_{n,m})$ ($n=0,1..$ numbering the mode). In this case the system reduces to one equation -- the radial function \eqref{Rbc} (for $E=l(l+1),l=1,2..$) solved here using the one-dimensional M\"uller algorithm. 

\begin{figure}[!ht]
\vspace{-0cm}
\hspace{1.5cm}
\centering
\resizebox{150px}{!}{\subfigure[\, l=1]{\includegraphics{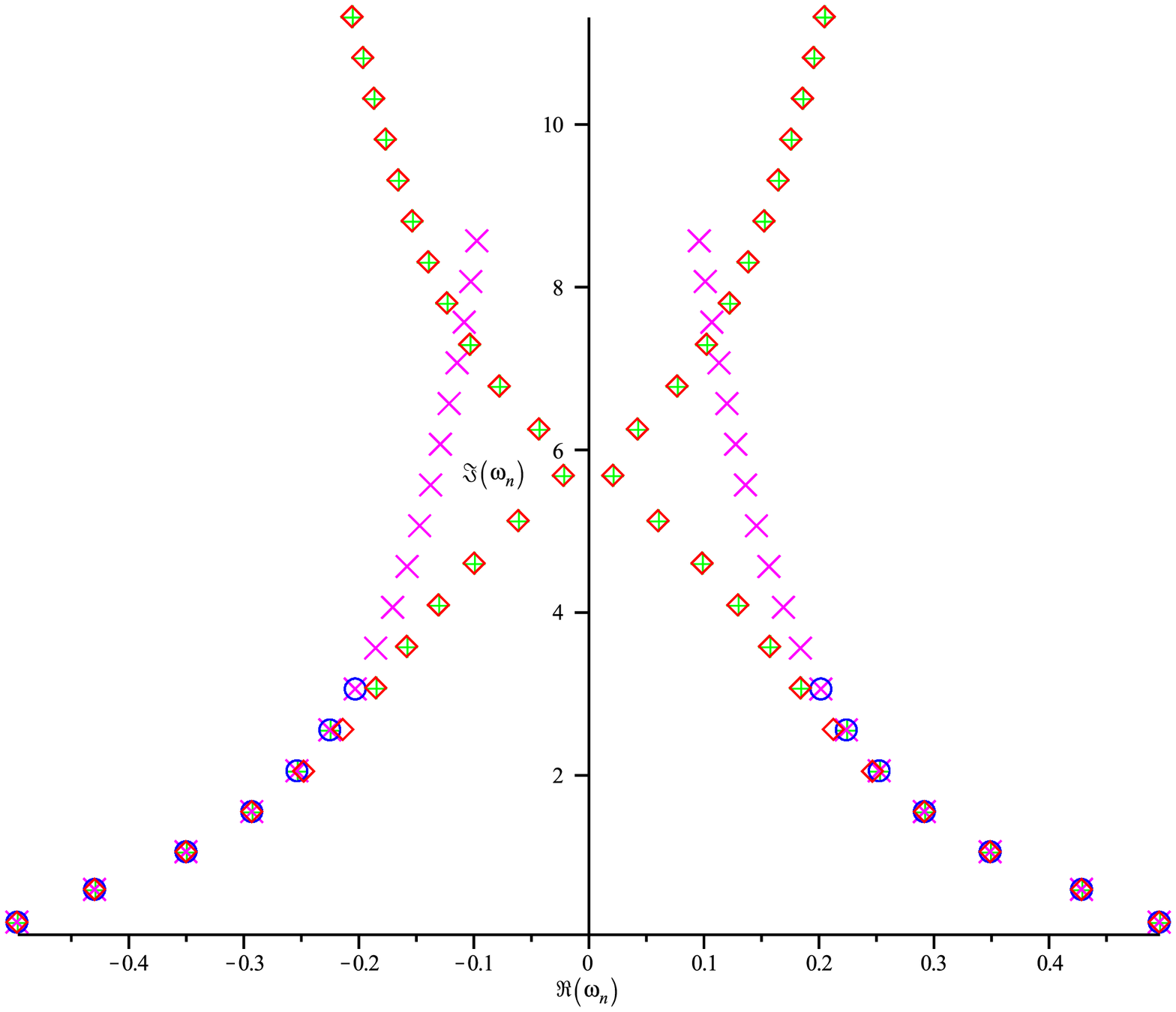}}}
\resizebox{150px}{!}{\subfigure[\, l=1,2]{\includegraphics{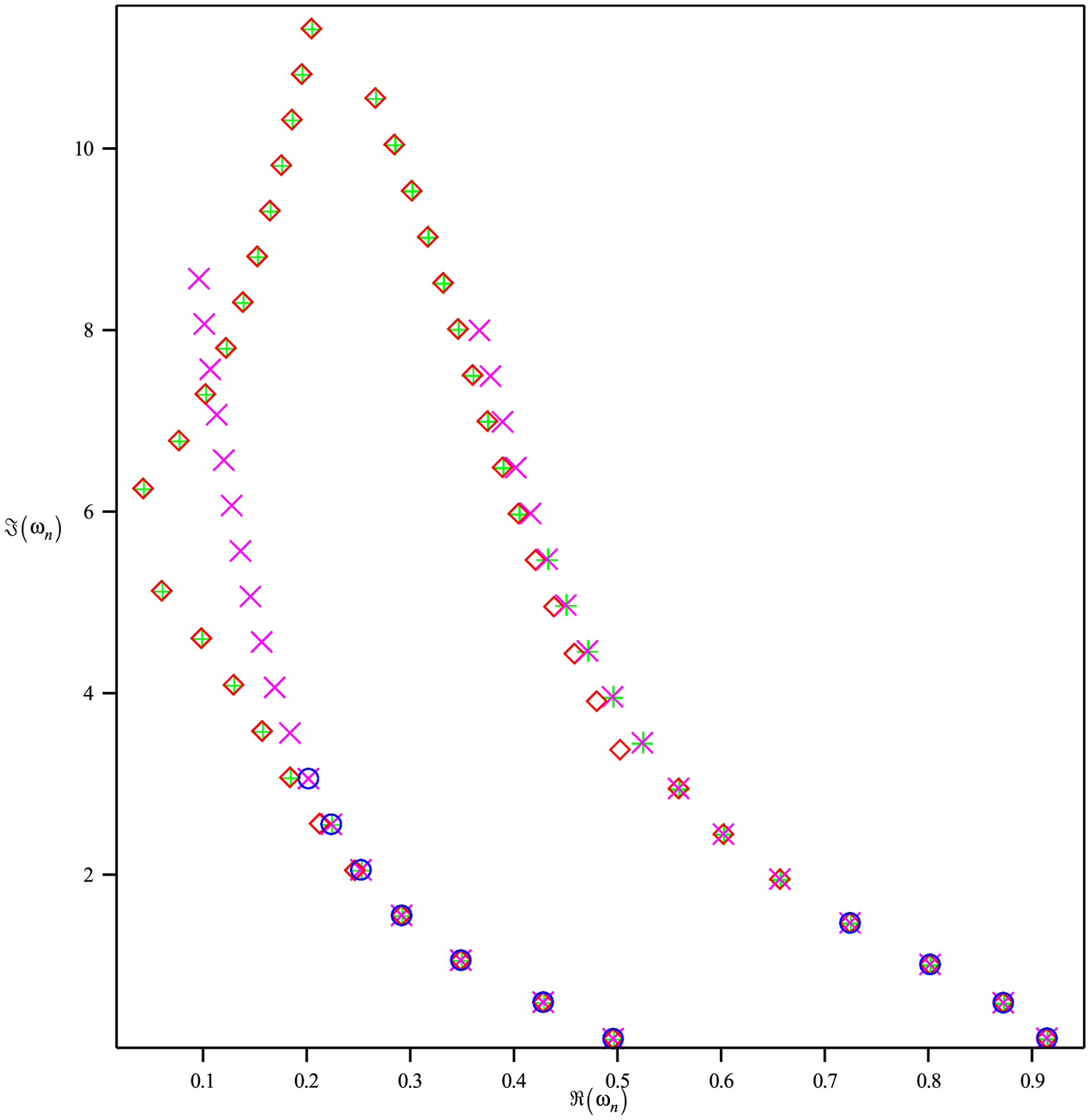}}}
\caption{QNMs for $a=0$, for $m=0,1,2$. The red diamonds are obtained for $\epsilon=0$, the green crosses -- for $\epsilon=0.05$, the magenta diagonal crosses -- for $\epsilon=0.15$. With blue circles are the control frequencies $\omega_{n,m}^B$. Some points cannot be differed because for them, the numbers for different $\epsilon$ coincide with precision higher than 10 digits. Clearly, for the higher modes, one obtains different spectra for different $\epsilon$}
\label{m0_0}
\end{figure}

The results can be seen on Fig. \ref{m0_0} a) and b), where we plotted the QNMs for $m=0,1,2$ and $l=1,l=2$. 
A numerical comparison of the frequencies obtained for $\epsilon=0$ with the frequencies obtained by Berti et al. shows that the average deviation is $|\omega_{n,m}^B-\omega_{n,m}| \approx 10^{-10}$ for the first 4 modes ($n=0..3$, $l=1,2$). For modes with $n>3$ (i.e. $n=4..6$ for $l=1$), however, there is an unexpected deviation which starts for $n=4$ from $|\omega_{4,m}^B-\omega_{4,m}| \approx 0.007$ and grows to $|\omega_{6,m}^B-\omega_{6,m}| \approx 0.022$ for the last available control mode. 

To study this systematic deviation, we employed the $\epsilon$-method to test the stability of those frequencies with respect to small deviations in the phase-condition. The results for $\epsilon$: $0,\pm0.05,\pm0.15$ are plotted on Fig. \ref{m0_0} a) and b). From there one can see that for $n>3$ the best coincidence with the control frequencies $\omega_{n,m}^B$ occurs for $\epsilon=\pm0.15$, while for $n<3$ the modes obtained for the different values of $\epsilon$ are equal and coincide with $\omega_{n,m}^B$. 

Similarly to the gravitational perturbations for nonrotating BH (\cite{arxiv1}), the dependence $\omega(\epsilon)$ in the electromagnetic case is not a trivial one. Here, because of the computational cost of the two-dimensional root-finding algorithm, only the case when there is no rotation ($a=0$) allows studying in detail this dependence. Exploring $a=0$, however, give us important intuition on the behavior of the QNMs under changes in $\epsilon$ for the case $a>0$. 
  
A more detailed study on both this case and the case with rotation can be found in the Appendix. In short, after examining the interval $\epsilon\in (-0.8,0.8)$ for roots of the radial function corresponding to certain mode (i.e. numerically close to certain control mode), one finds:
\begin{enumerate}
\item The symmetry with respect to the imaginary axis holds for all cases under consideration. In symmetrical parts of the interval $\epsilon\in (-0.8,0.8)$ (denoted here as $\,\Delta_{m,n}^ \epsilon=[\epsilon^{in}_n,\epsilon^{fin}_n]\subseteq (-0.8,0.8)$), one has frequencies with both signs of their real parts ($\omega_{m,n}=\pm|\Re(\omega_{m,n})|+i\Im(\omega_{m,n})$), which are roots of the radial function in $\,\Delta_{m,n}^ \epsilon$. 
\item For $n>0$, for each $\{m,n\}$, one has {\bf two} frequencies with {\em positive} real parts (and respectively two frequencies with negative real parts), i.e. $\omega_{n,m}^{1,2}=\pm|\Re(\omega_{n,m})|^{1,2}+i\Im(\omega_{n,m})^{1,2}$, both roots of the radial function in certain parts of the interval $\,\Delta_{m,n}^ \epsilon$.
\item The interval $\Delta_{m,n}^\epsilon$, where a mode $\omega_{m,n}$ with certain sign of its real part can be found, shortens with the increase of $n$, moving to the right (left) for frequencies with positive (negative) real parts.
\item The dependence $\omega_{m,n}(\epsilon)$  in the whole interval $\epsilon \in (-0.8,0.8)$ of applicability of the method, can be of three types -- 1) step-wise change of the values of the frequency for $\epsilon$ crossing one of the BCs.  2) an approximately smooth change of the value of the frequency in parts of the interval, not crossing a BC (with a change smaller than $10^{-9}$)  3) small fluctuations in the value of the mode, corresponding to the numerical error of the algorithm ($\sim10^{-11}$).
\end{enumerate}

It appears that finding each mode depends critically on the proximity of branch cuts in the complex $\omega$-plane. For a detailed description, please see the Appendix. In short, the BC defined by $\Im(r)=0$ limits the interval where a mode with certain sign of the real part can be found, while the BC defined by $\Im(r^k)=0$ seems to cause the difference between our frequencies,$\,\omega_{n,m}^{1,2}$, and the control results $\omega_{m,n}^B$.

\subsection{Rotating KBH}
In this case, we evaluated the QNMs for $a=[0..M)$ for 3 different values of $|\epsilon|=0,0.05,0.15$ (where the positive $\epsilon$ are used for frequencies with positive real part and the negative $\epsilon$ -- for the frequencies with negative real parts). 

\begin{figure}[!h]
\centering
\subfigure[$\Re(\omega_{0,3})(a)$]{\includegraphics[width=121px,height=110px]{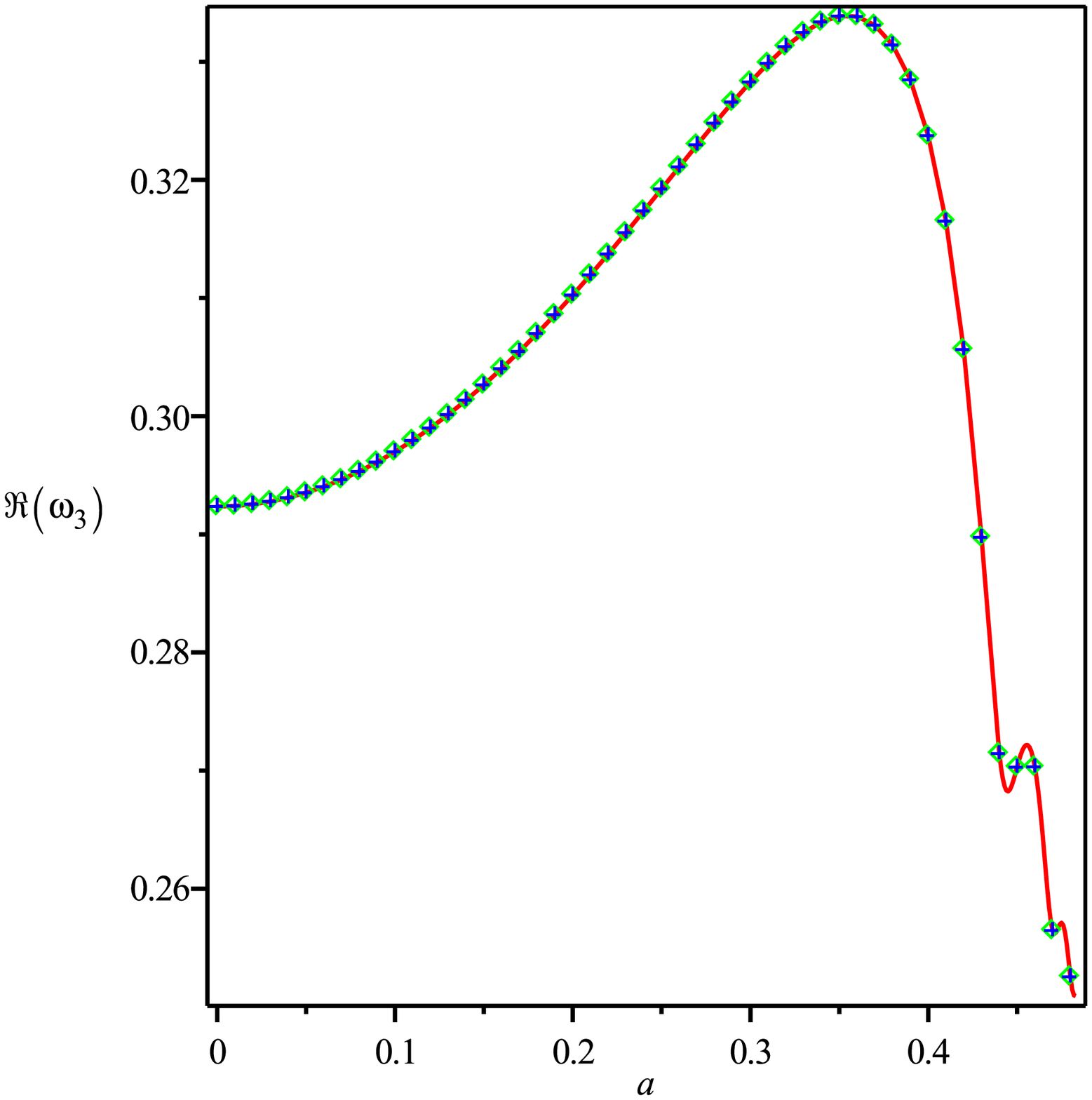}}
\subfigure[$\Im(\omega_{0,3})(a)$]{\includegraphics[width=121px,height=110px]{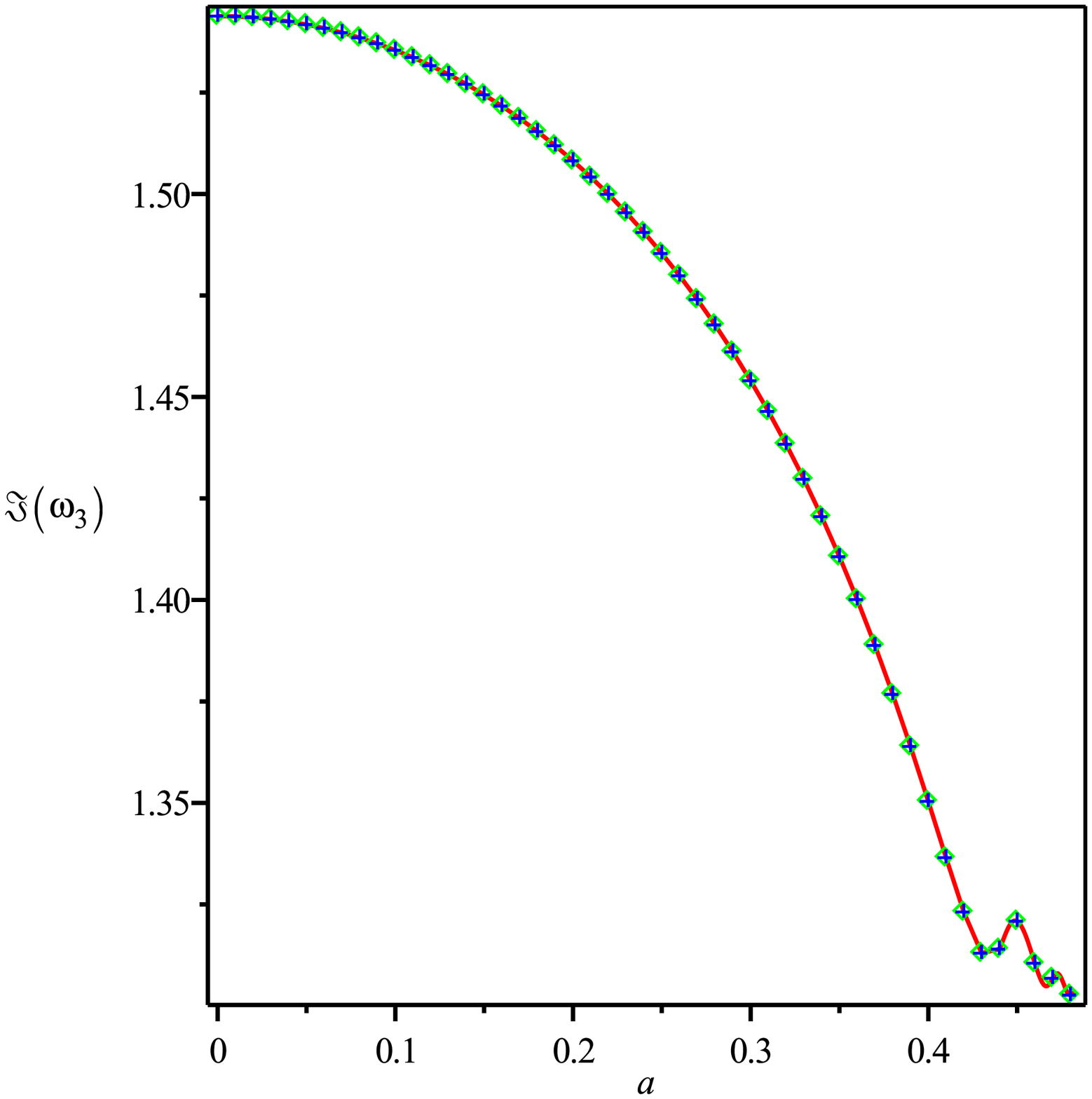}}
\subfigure[$\Re(E_{0,3})(a)$]{\includegraphics[width=121px,height=110px]{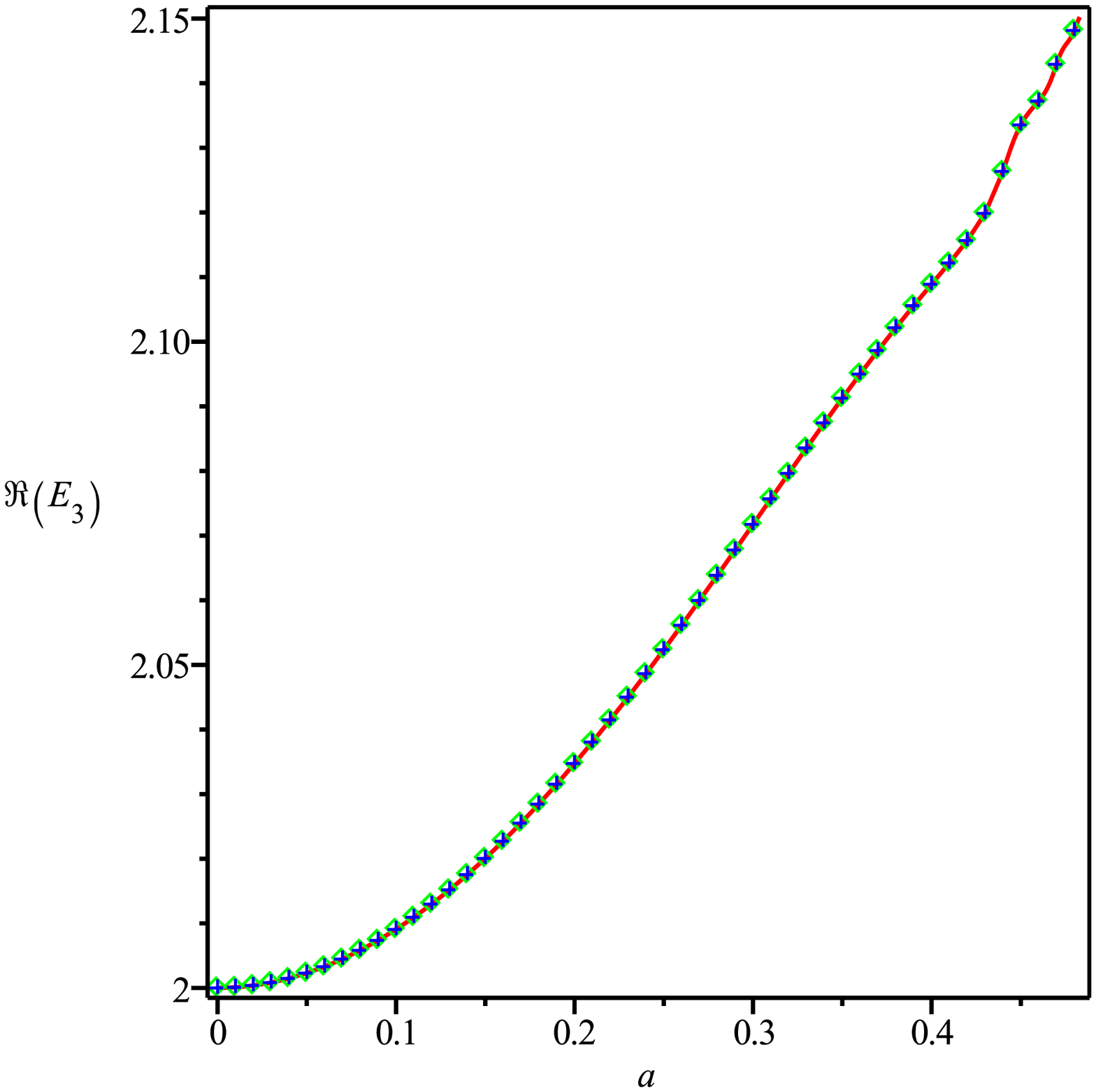}}
\subfigure[$\Im(E_{0,3})(a)$]{\includegraphics[width=121px,height=110px]{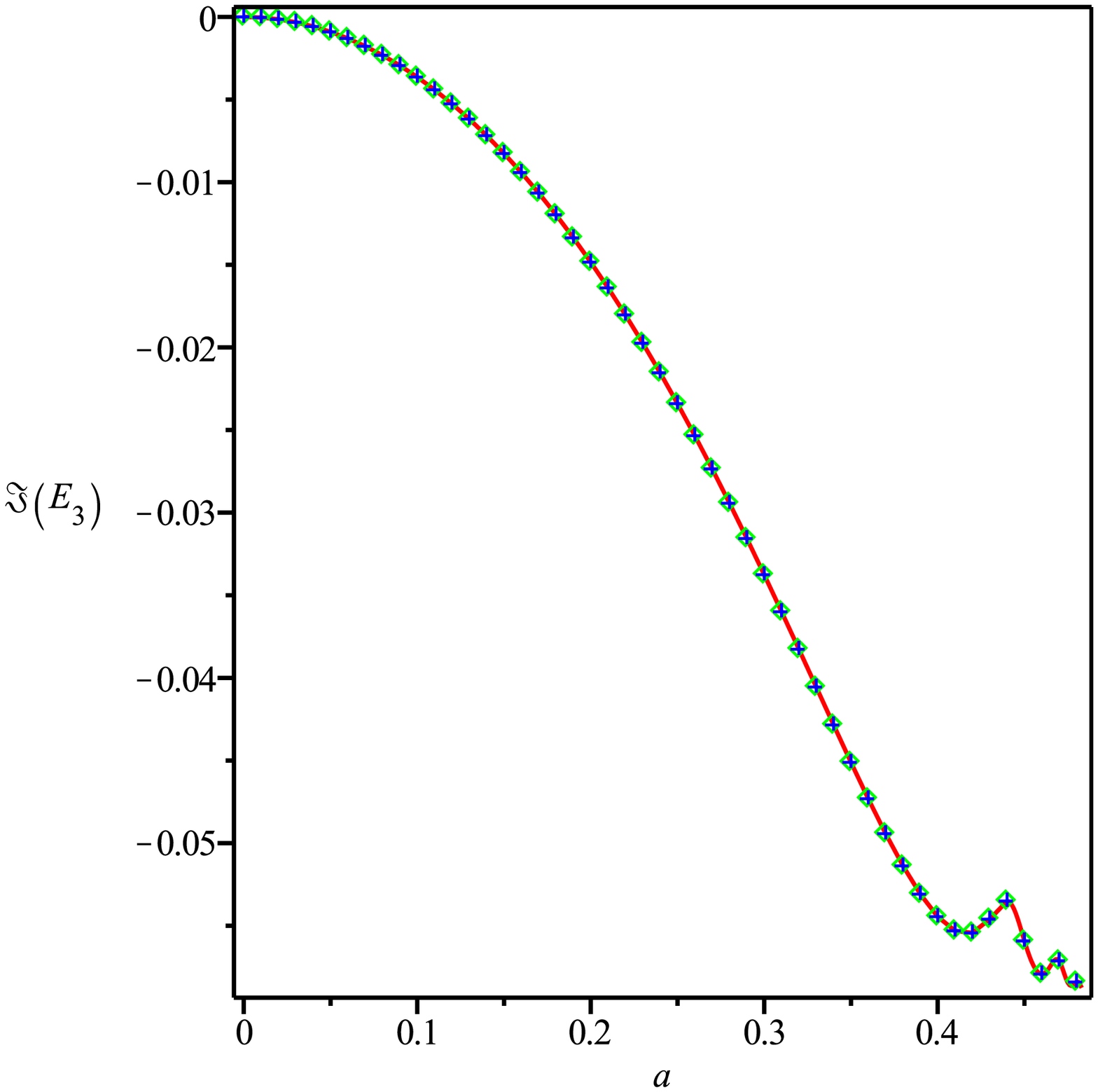}}
\caption{The plots depict the real and imaginary parts of $\omega_{0,3}(a)$ and  $E_{0,3}(a)$ for rotation changing in the interval $a=[0,M)$.  The red lines are the points obtained for $\epsilon=0$, the blue crosses -- those for $\epsilon=0.05$ and the green diamonds -- those for $\epsilon=0.15$. The the 3 types of points coincide}
\label{n3}
\end{figure}

\begin{figure}[!h]
\centering
\subfigure[$\Re(\omega_{0,4})(a)$]{\includegraphics[width=121px,height=110px]{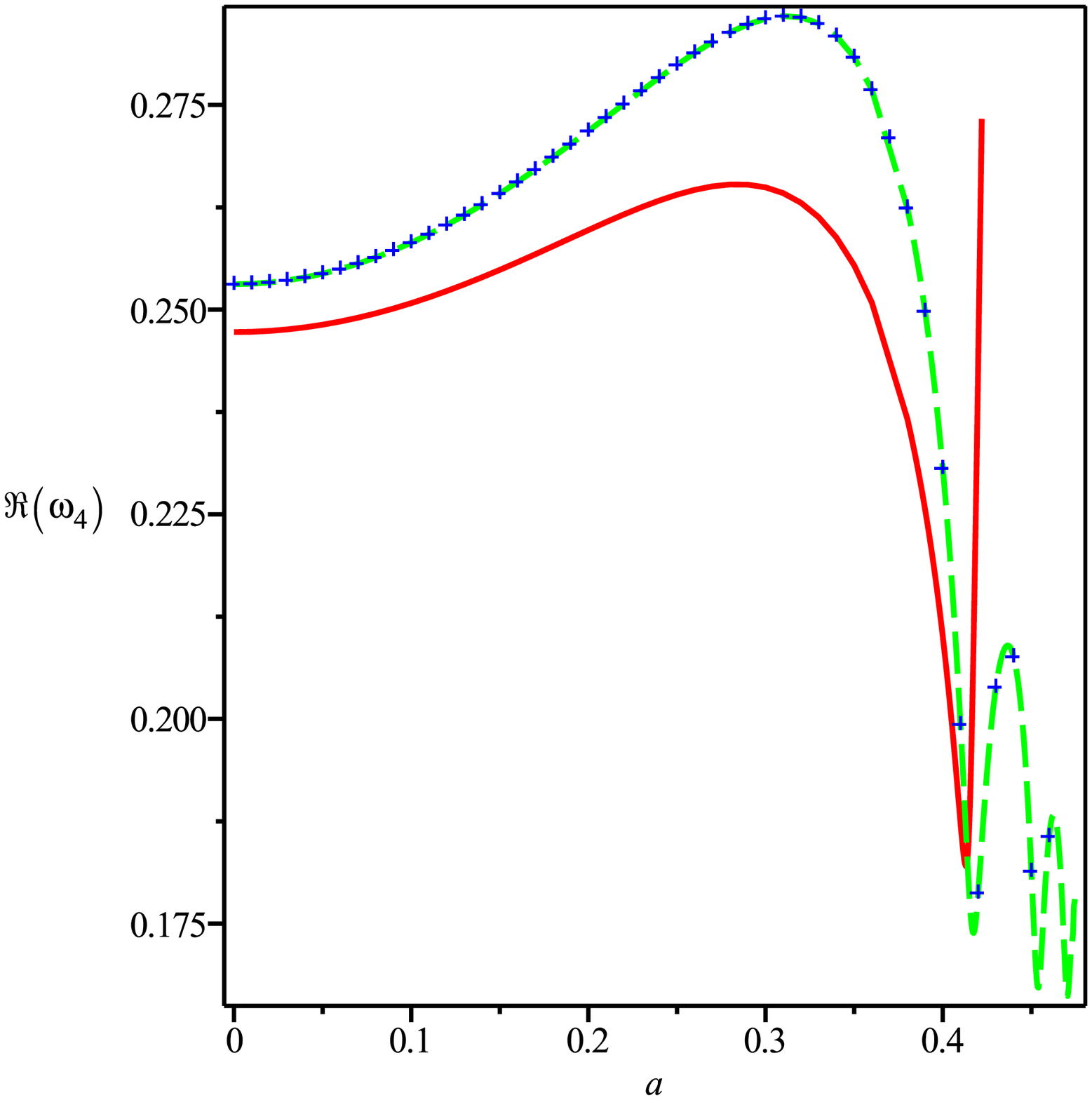}}
\subfigure[$\Im(\omega_{0,4})(a)$]{\includegraphics[width=121px,height=110px]{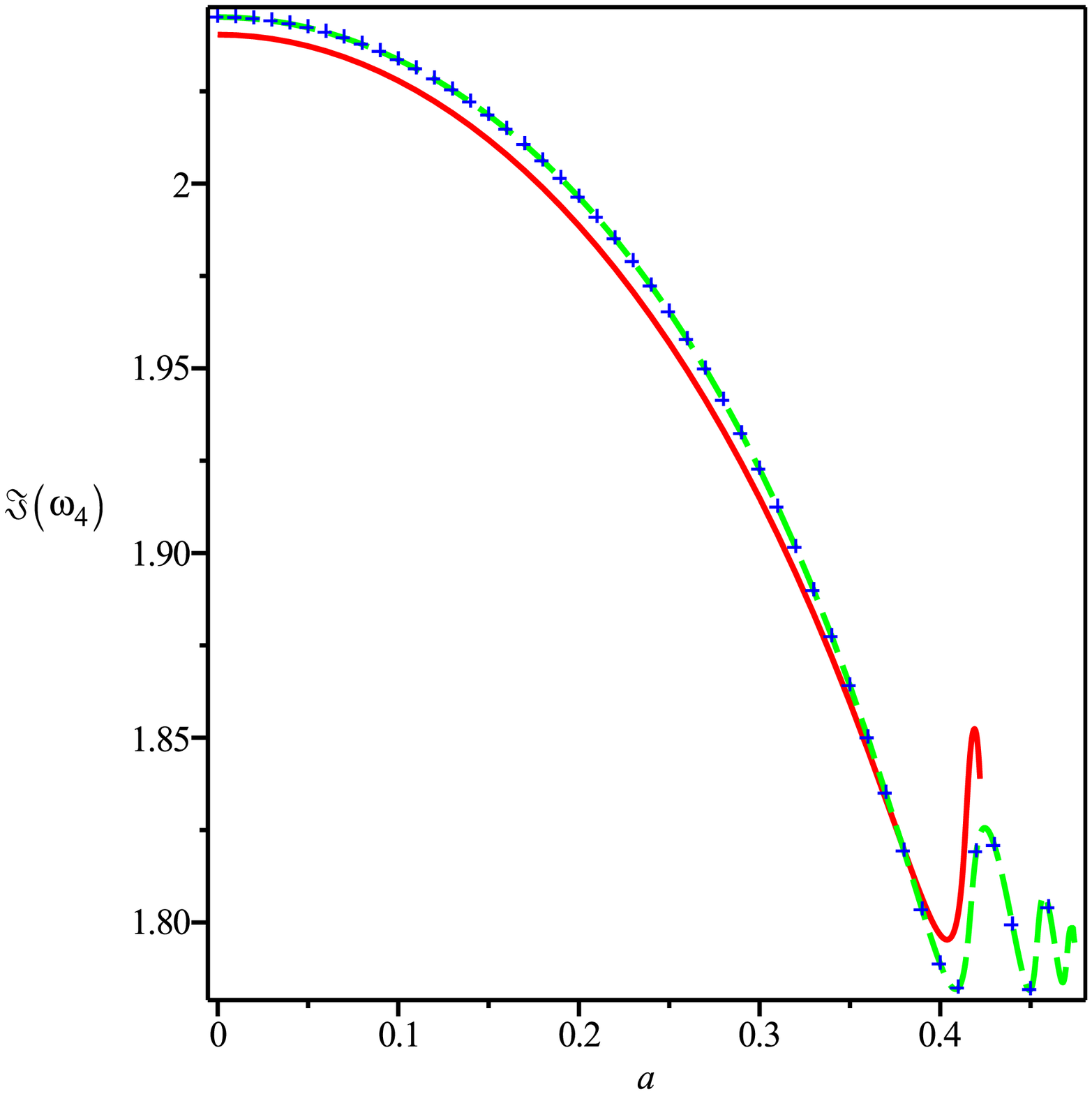}}
\subfigure[$\Re(E_{0,4})(a)$]{\includegraphics[width=121px,height=110px]{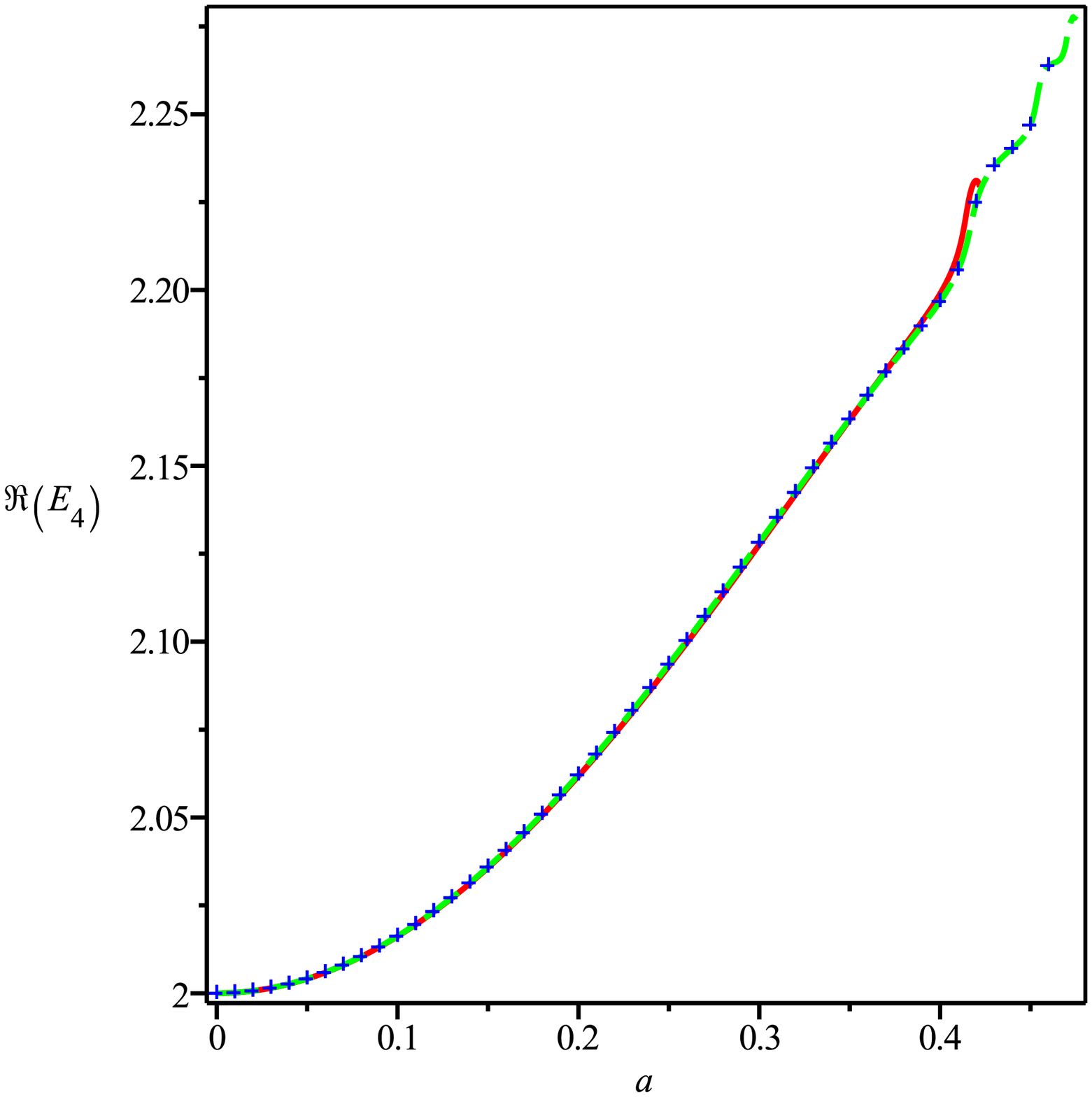}}
\subfigure[$\Im(E_{0,4})(a)$]{\includegraphics[width=121px,height=110px]{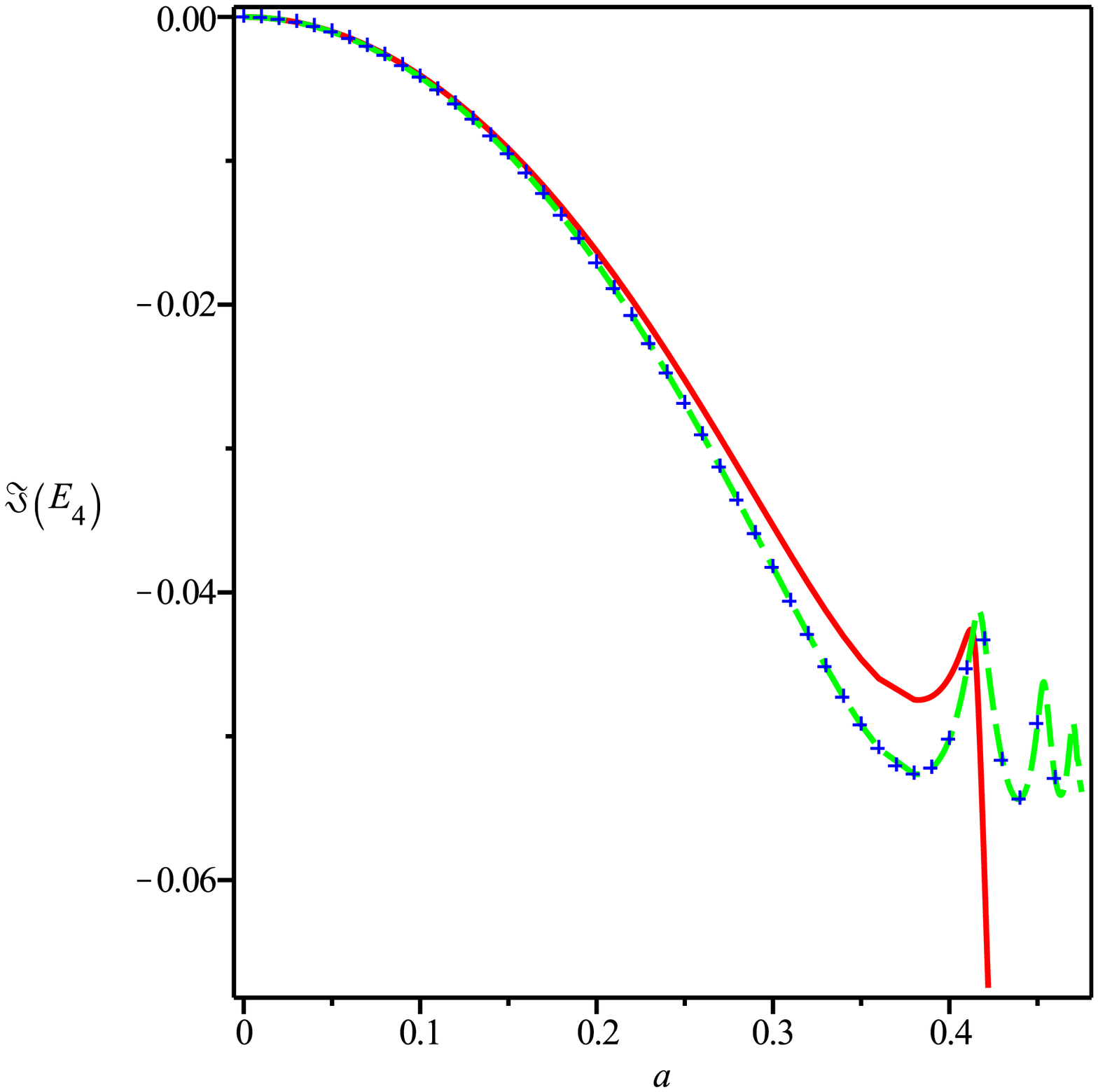}}
\caption{The plots depict the real and imaginary parts of$\omega_{0,4}(a)$ and  $E_{0,4}(a)$. With red lines are denoted the points obtained for $\epsilon=0$, with blue crosses -- those for $\epsilon=0.05$ and with green dashed lines -- those for $\epsilon=0.15$. The points for the different values of $\epsilon$ differ }
\label{n4}
\end{figure}
The results can be seen on the figures $2-11$. 

When $a\neq0$, the symmetry with respect to the imaginary axis $\omega_{m,n}(0)=\pm|\Re(\omega_{m,n})|+i\Im(\omega_{m,n})$ turns into:
 $$\{\Re(\omega_{m,n}^\pm),\Im(E_{m,n}^\pm),m\}\to \{-\Re(\omega_{m,n}^\pm),-\Im(E_{m,n}^\pm),-m\},$$ 
where $\pm\!$ corresponds to the sign in front of the real part of $\omega_{m,n}(0)$. Therefore, to study the complete behavior of the modes for $a\in[0,M)$, it is enough to trace both symmetric frequencies in the pair corresponding to each $\{m,n\}$ for $a=0$, for only $m>0$ (the index $l$ here is omitted to simplify the notation, but everywhere in the text, if not explicitly stated otherwise, we compare only frequencies with the same $l$.)

The parameter $\epsilon$ for rotating BH has an even more significant role than the nonrotating case, since the two distinct frequencies, corresponding to the modes $\,\omega_{n,m}^{1,2}(a=0)$, evolve differently with the increasing of the rotation. Some details can be found in the Appendix. Once again, those modes are roots of the radial function in different parts of the interval over $\epsilon$, depending on the position of the two branch cuts for each mode. The second mode appears for $n>N$, where N depends on ${m,n}$ and usually $N=2..4$ and it is a function of the gap between the two branch cuts (defined by the equations discussed above). For a case study of one such mode, see the Appendix. \footnote{where $\omega_{m,n}(\epsilon)= \omega_{m,n}^{\epsilon}$ to avoid confusion with $\omega_{m,n}(a)$}  (Fig. \ref{n3}) $\epsilon$ differ (Fig. \ref{n4}). Note that while here we discuss mostly the frequencies, the separation parameters $E_{m,n}$ also depend on $\epsilon$ as the figures show.

\begin{figure}[tb]
\centering
\subfigure{\includegraphics[width=121px,height=110px]{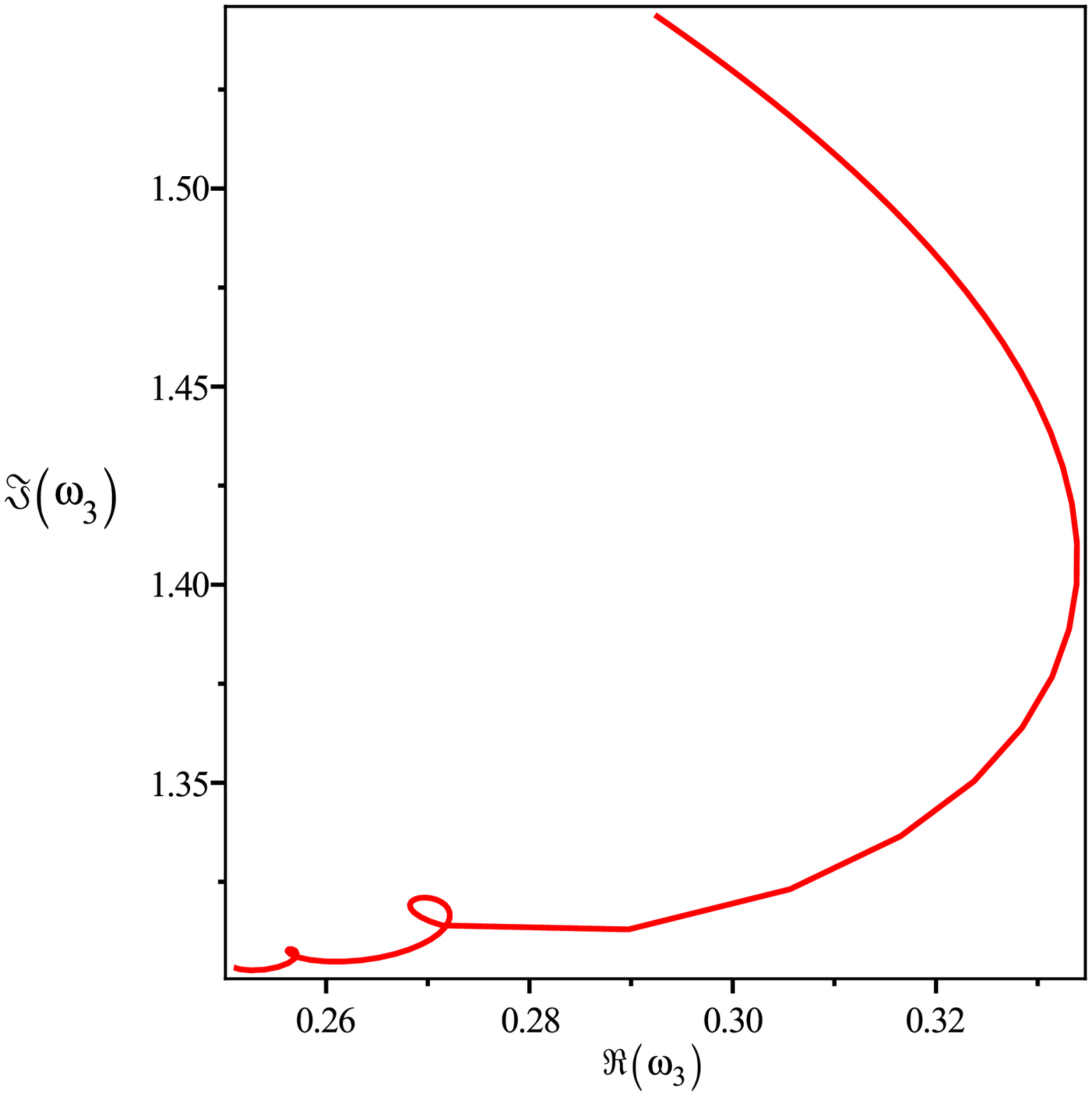}}
\subfigure{\includegraphics[width=121px,height=110px]{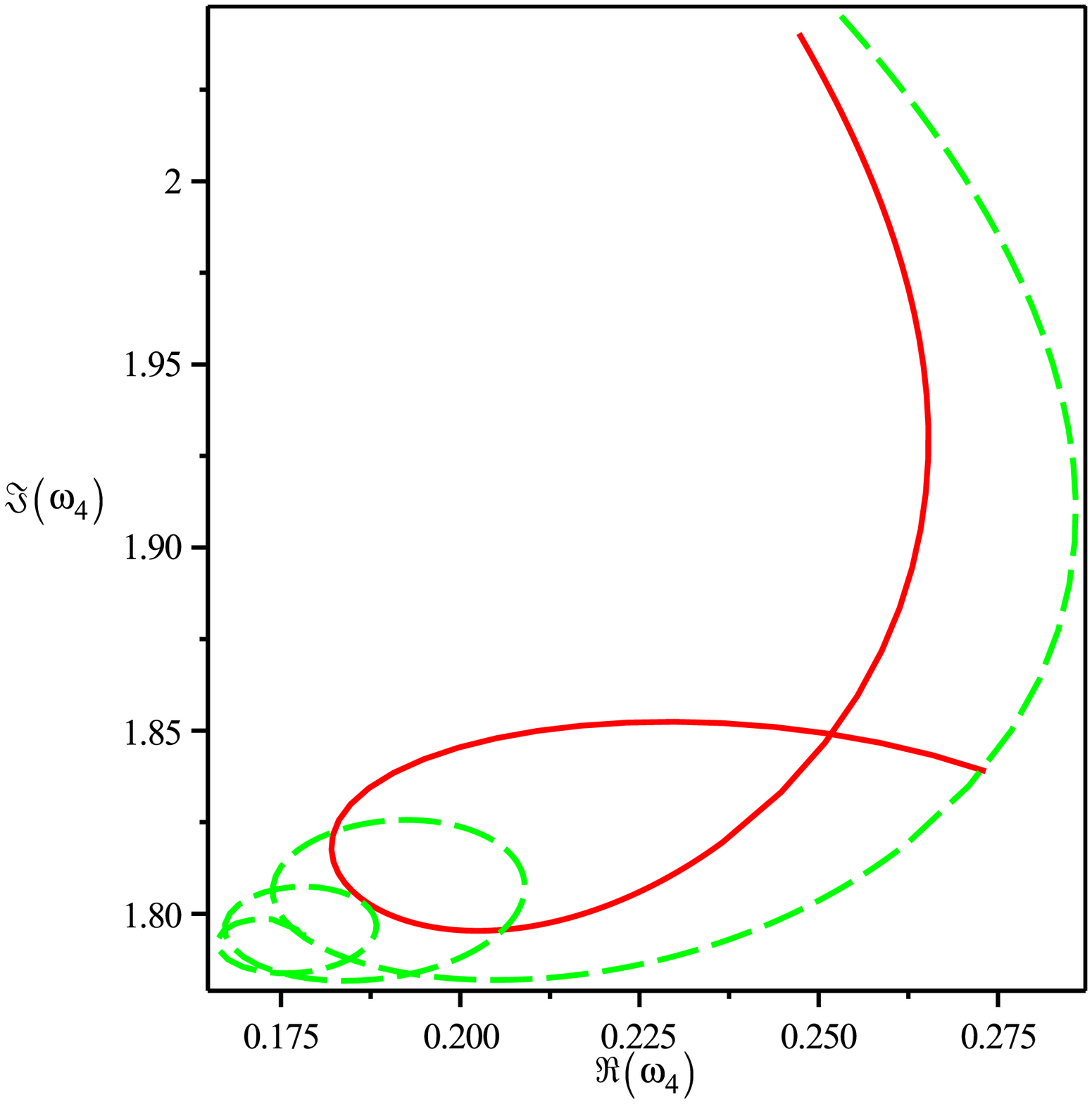}}
\vspace{-0.9cm}
\caption{Example of the loops observed for $m=0$. The figure shows the complex plots of $\omega_{0,3}(a)$ and $\omega_{0,4}(a)$, where the red lines are the points corresponding to $\epsilon=0$, the green dashed line -- those to $\epsilon=0.15$. For $n=3$, the results for $\epsilon=0,0.05,0.15$ coincide and thus only the points for $\epsilon=0$ are plotted}
\label{n3_loops}
\end{figure}

\begin{figure}[!htb]
\vspace{-0cm}
\centering
\subfigure{\includegraphics[width=150px,height=140px]{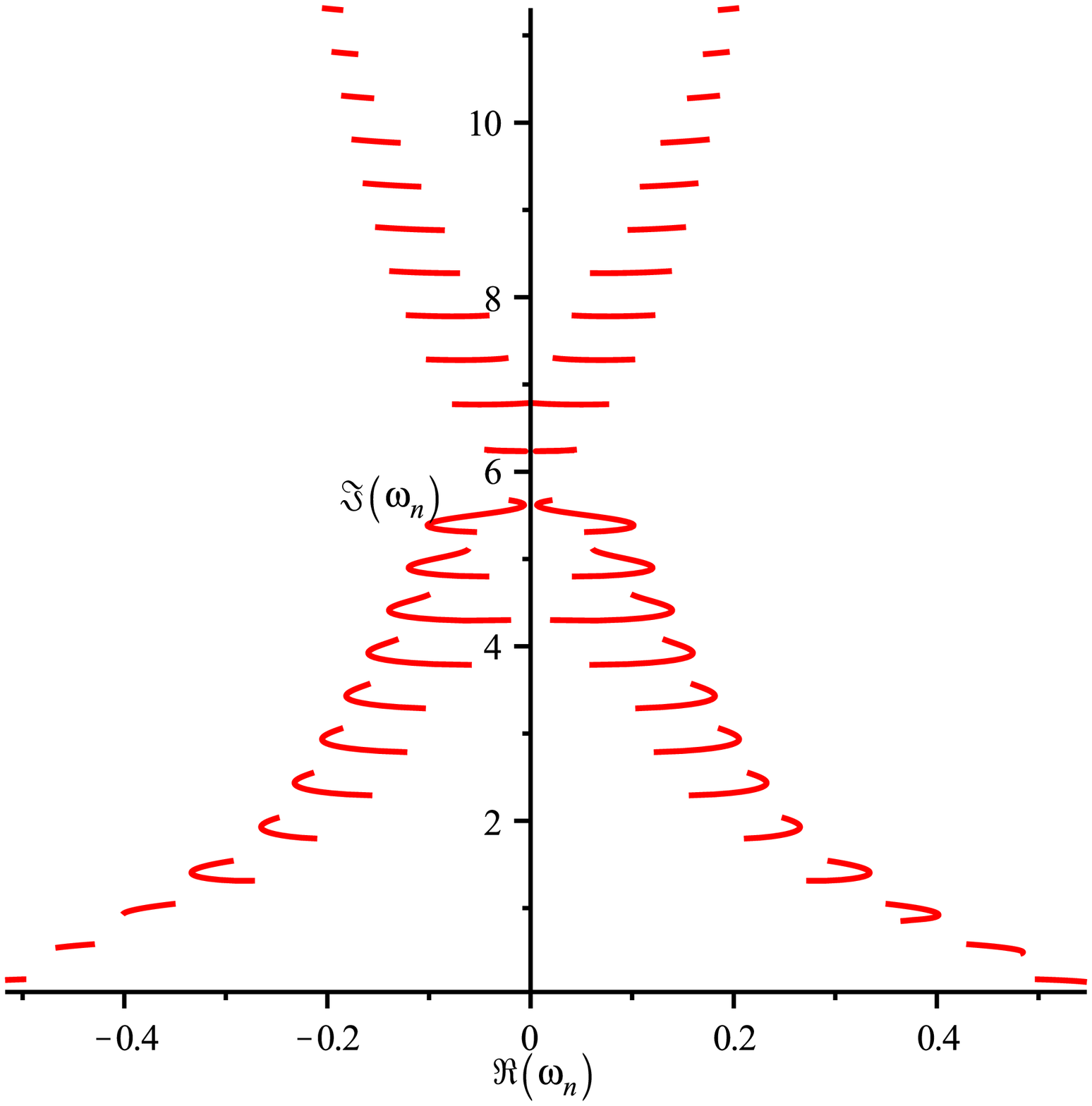}}
\subfigure{\includegraphics[width=150px,height=140px]{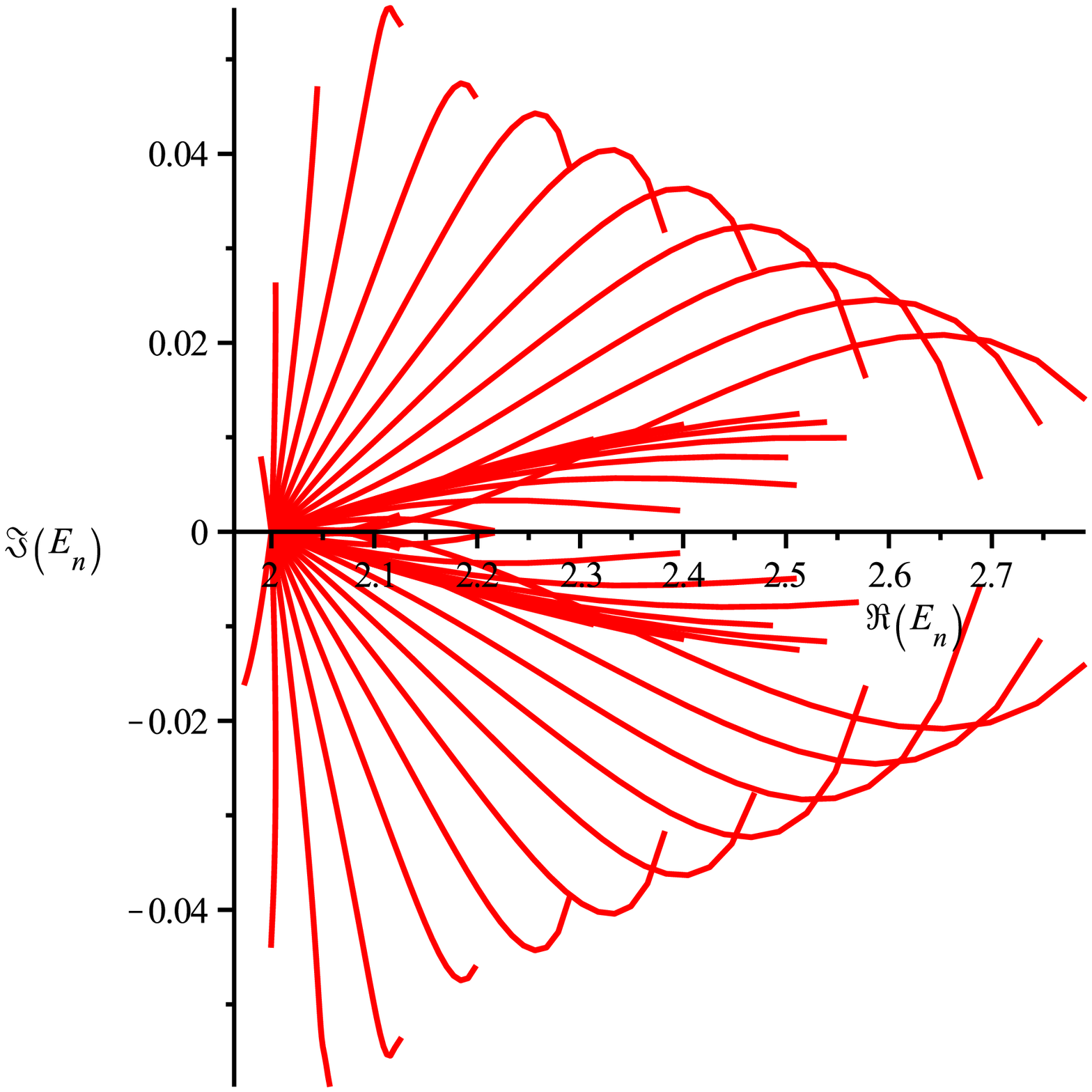}}
\caption{Complex plots of $\omega_{0,n}(a)$ and $E_{0,n}(a)$ for $a=[0,M)$ for the first 22 modes with both positive and negative real parts}
\label{m0_all_o}
\end{figure}

\begin{figure}[!htb]
\vspace{-0cm}
\centering
\subfigure{\includegraphics[width=150px,height=140px]{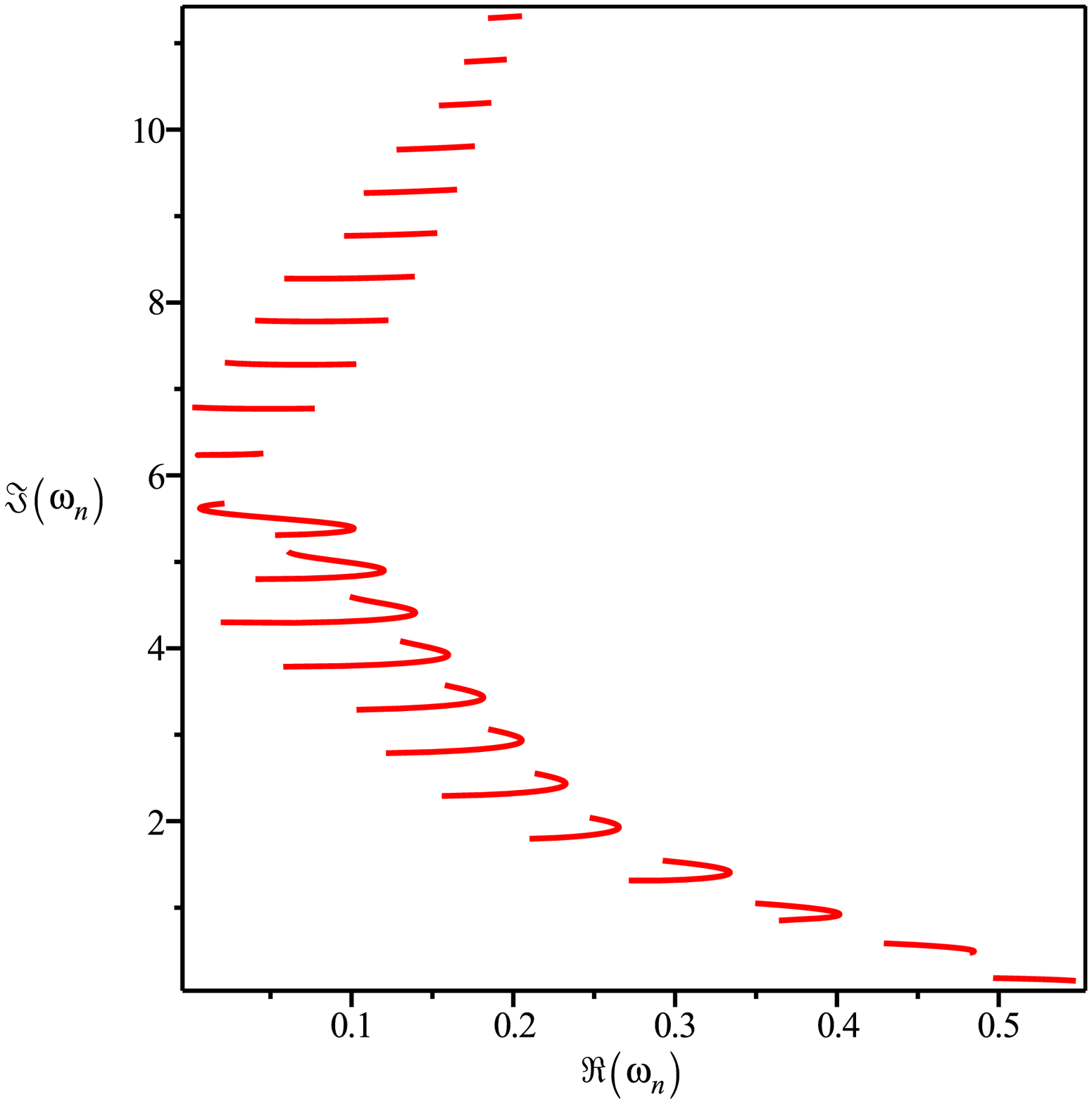}}
\subfigure{\includegraphics[width=150px,height=140px]{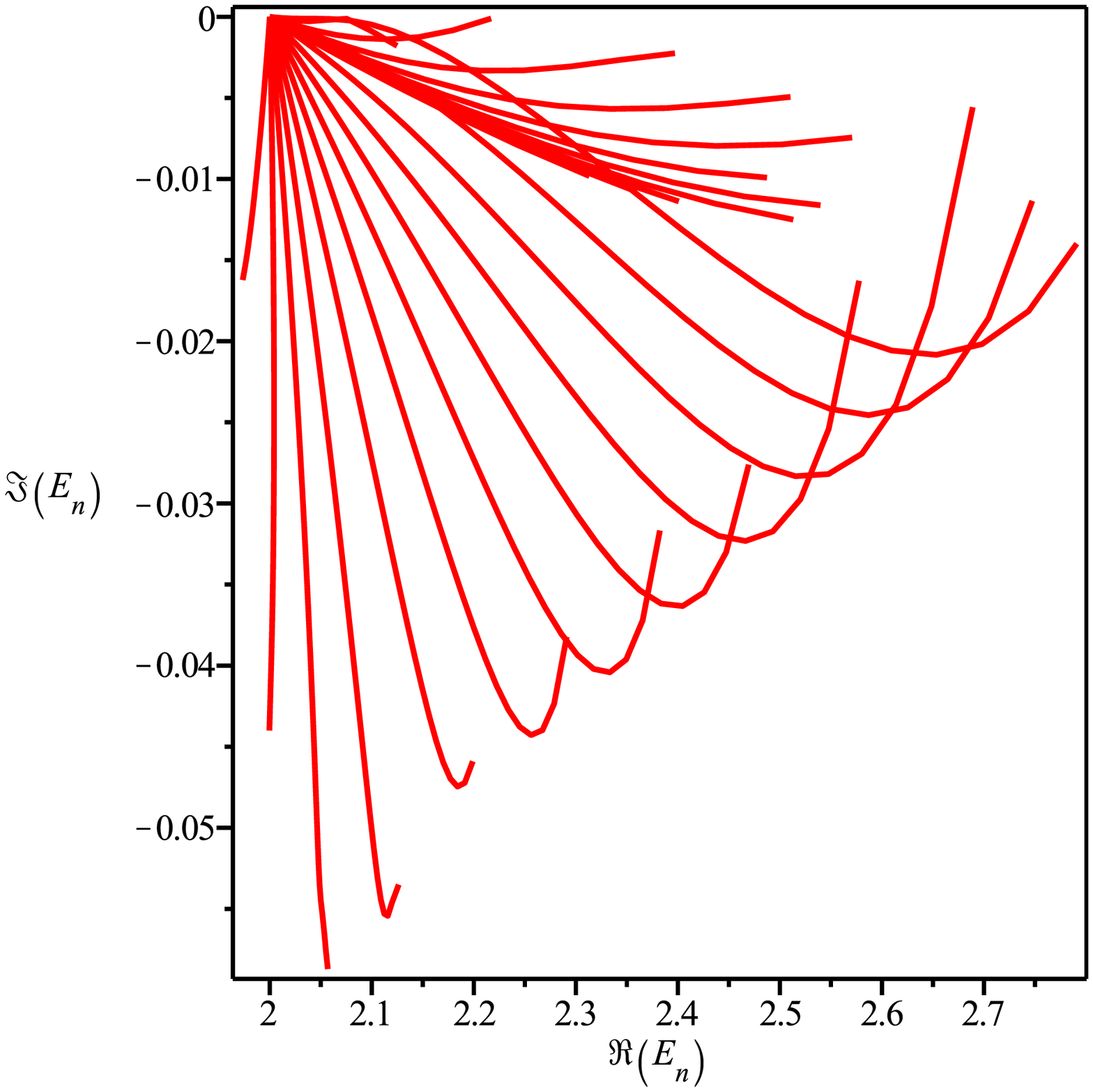}}
\caption{Complex plots of $\omega_{0,n}(a)$ and $E_{0,n}(a)$ for $a=[0,M)$, $l=2$, $\epsilon=0$}
\label{m0l12}
\end{figure}

\begin{figure}[!htb]
\centering
\subfigure{\includegraphics[width=150px,height=140px]{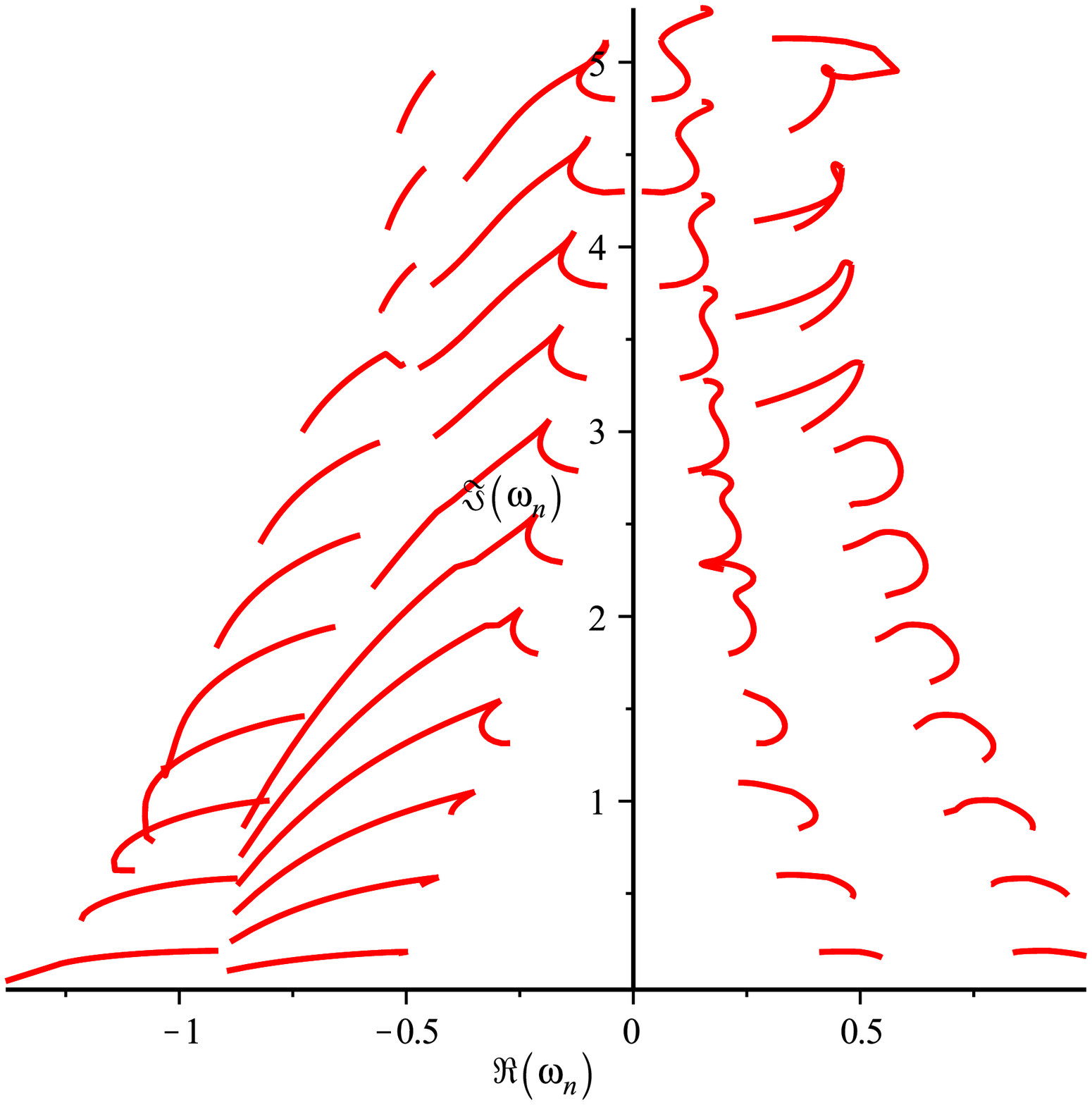}}
\subfigure{\includegraphics[width=150px,height=140px]{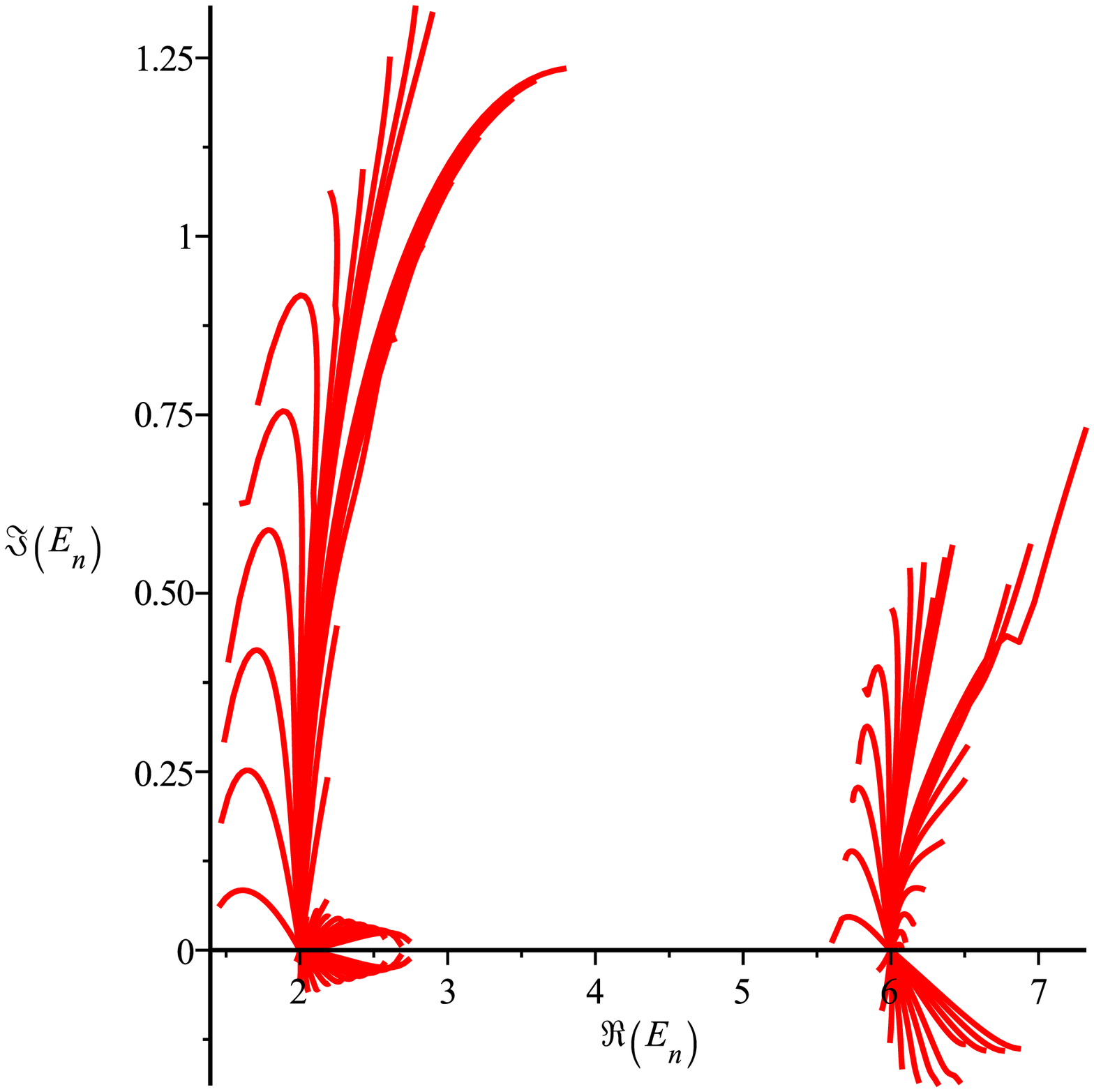}}
\caption{A complex plot of all the $\omega_{m,n}(a)$ and $E_{m,n}(a)$ obtained for $a=[0,M)$ for $m=0,1,l=1,2$ $n=0..10$}
\label{m01_all}
\end{figure}

It is important to note that for $m=0$, in the modes $n\ge 3$, one observes loops. An example can be seen on Fig. \ref{n3_loops}. Those loops appear in all the higher modes, and their position depends on $n$. Because those loops require a finer structure of the plot (i.e. smaller step), on the plots Fig. \ref{m0_all_o}, Fig. \ref{m0l12} and Fig. \ref{m01_all}, we will plot only the points before the first loop observed in each curve. On Fig. \ref{m01_all} one can see all the results plotted together. 

\begin{figure}[!htb]
\centering
\subfigure{\includegraphics[width=121px,height=120px]{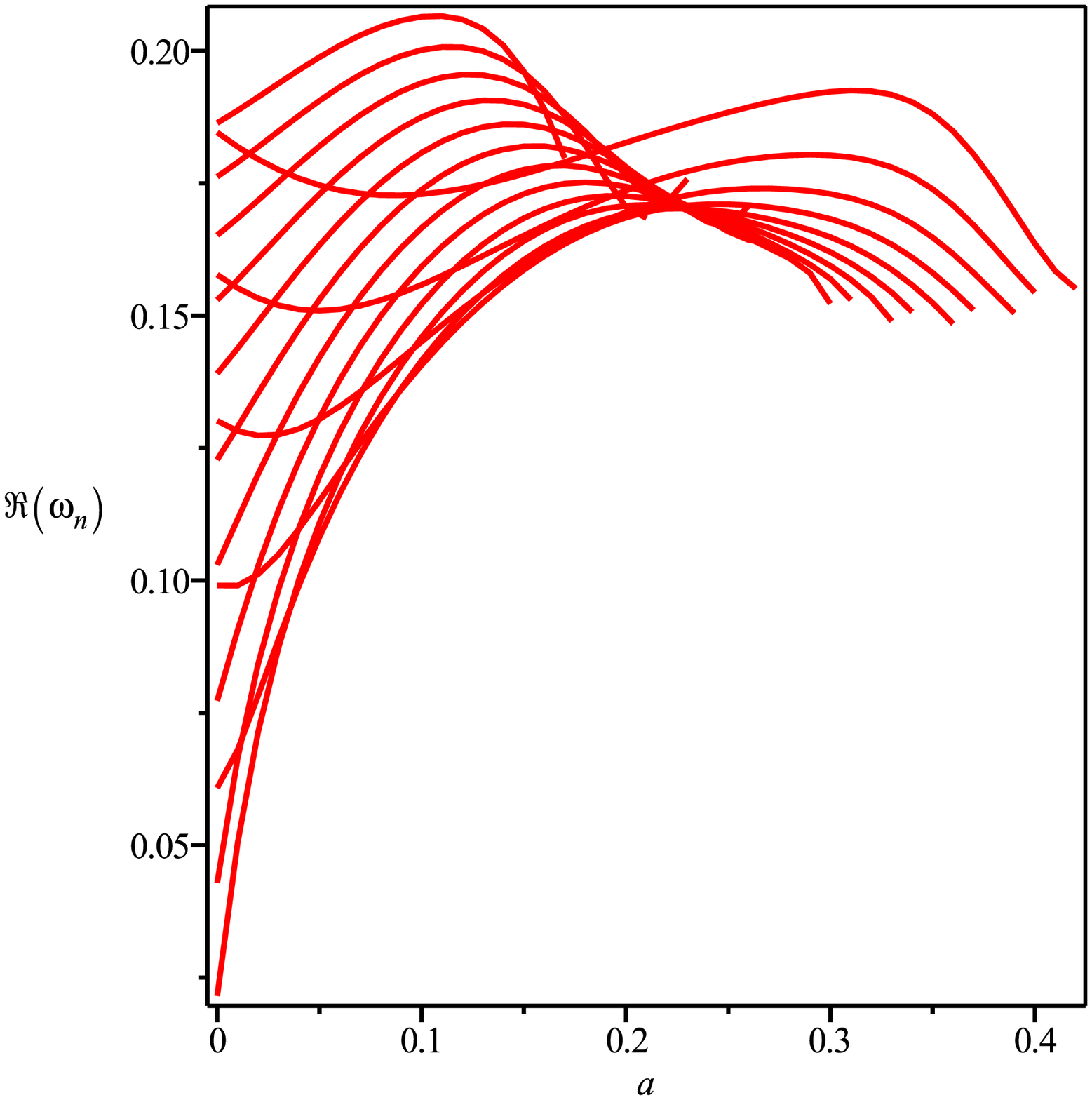}}
\subfigure{\includegraphics[width=121px,height=120px]{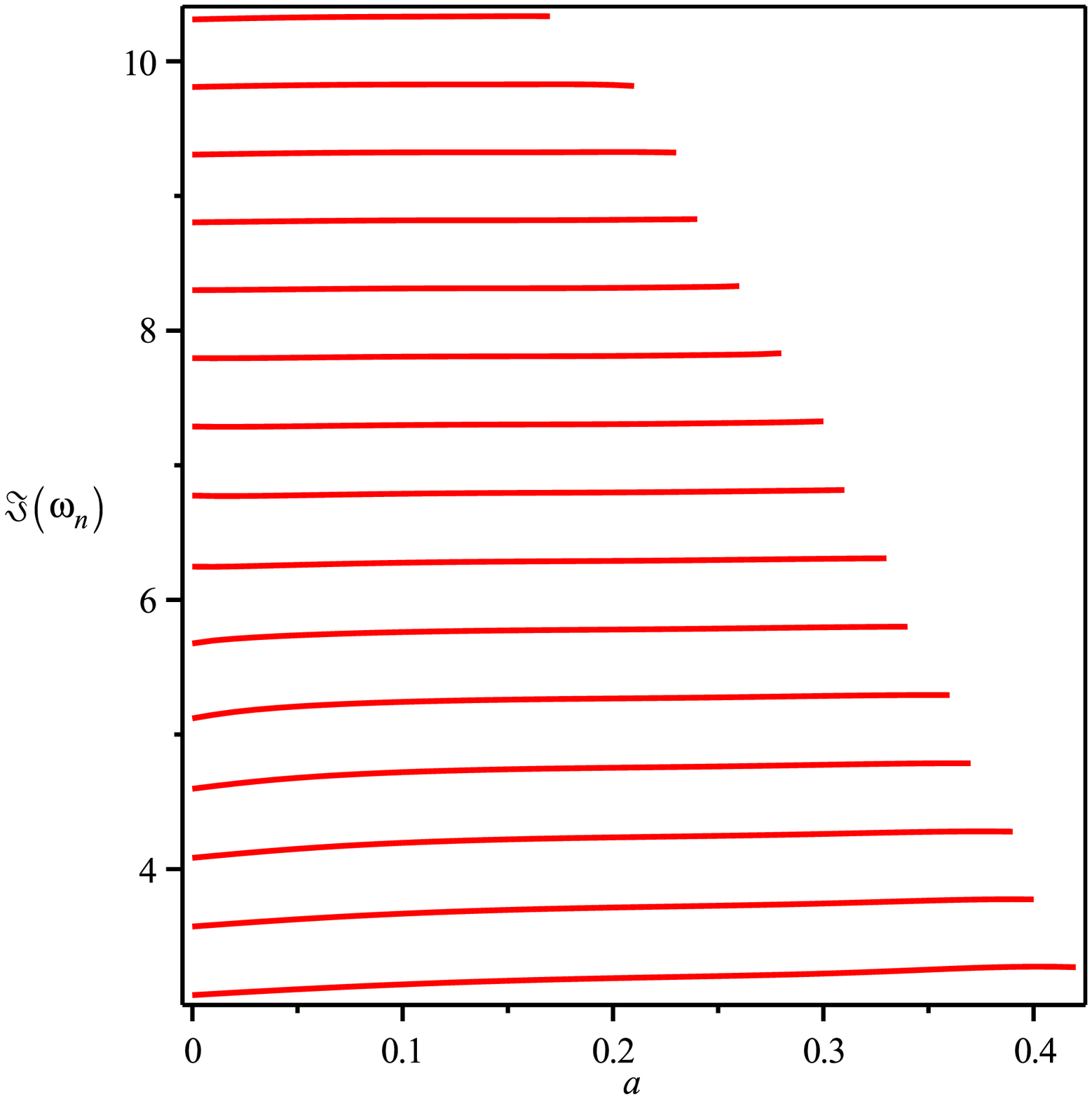}}
\subfigure{\includegraphics[width=121px,height=120px]{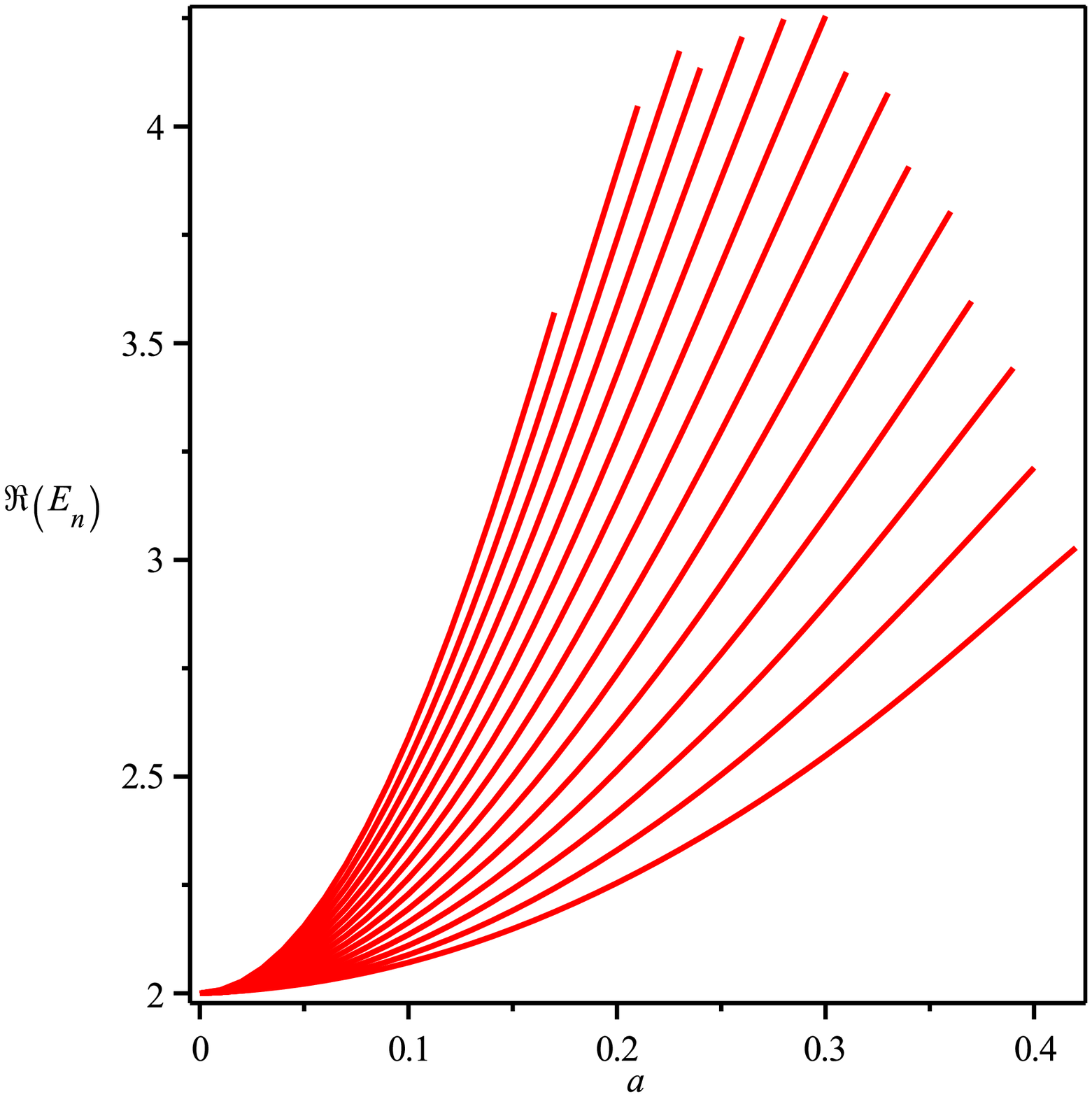}}
\subfigure{\includegraphics[width=121px,height=120px]{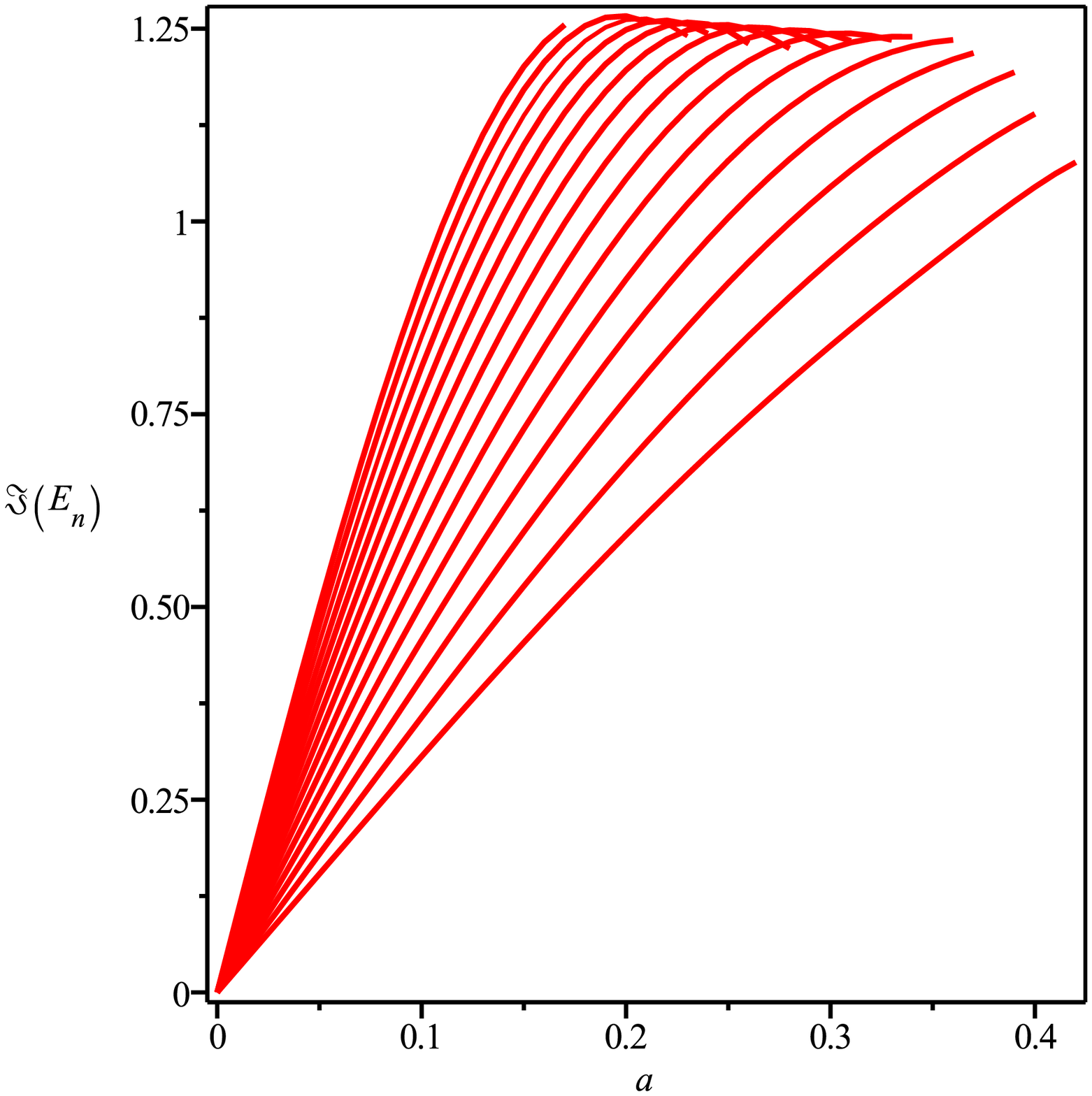}}
\caption{The plots show the real and the imaginary parts of $\omega_{1,n}(a)$ and $E_{1,n}(a)$ for $a=[0,M)$ for the modes $n=6..20$. }
\label{m1n7_21}
\end{figure}

\begin{figure}[!htb]
\centering
\subfigure{\includegraphics[width=121px,height=120px]{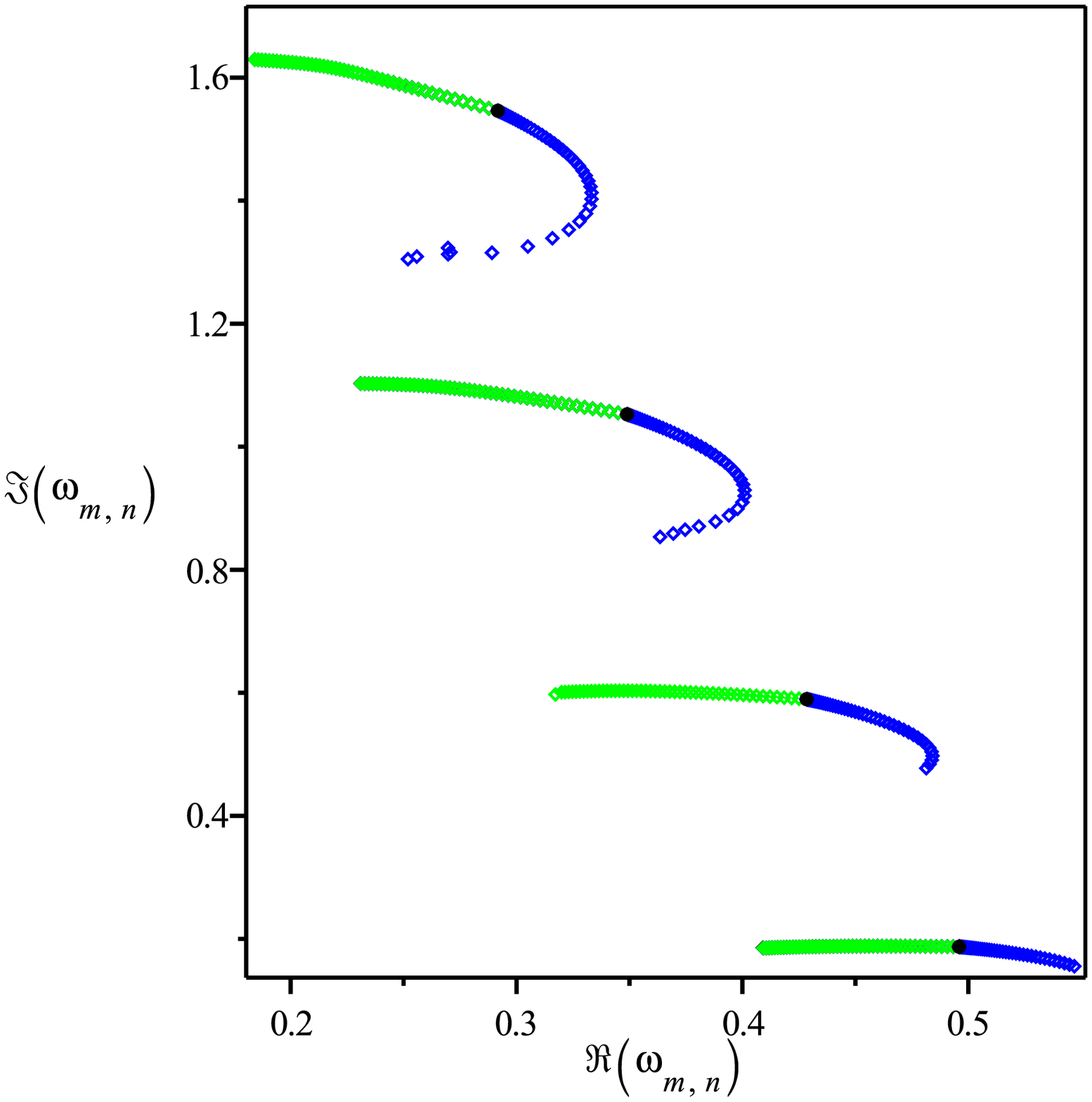}}
\subfigure{\includegraphics[width=121px,height=120px]{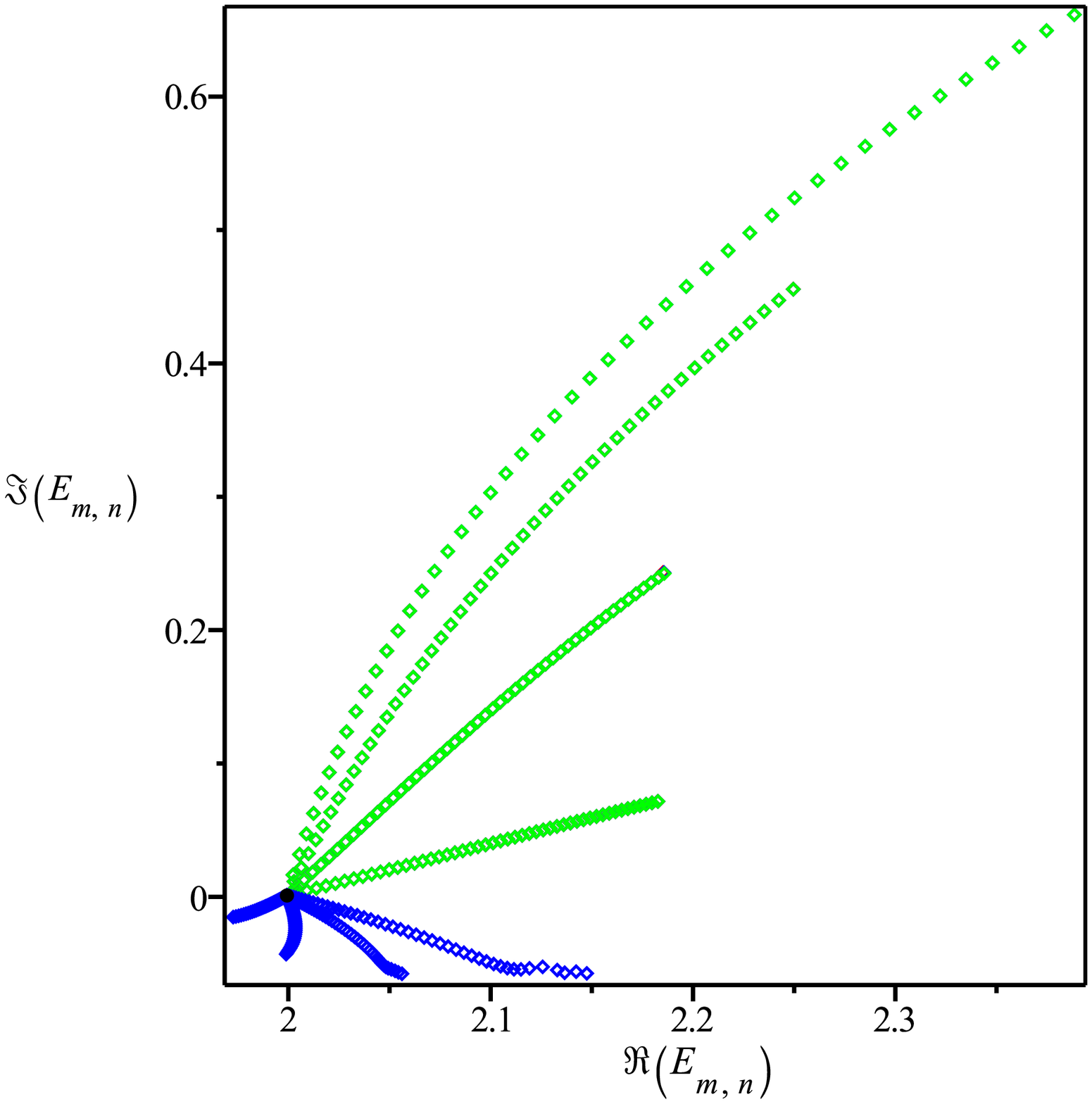}}
\vspace{-0.5cm}
\caption{Complex plots of $\omega_{m,n}(a)$ and $E_{m,n}(a)$ for $a=[0,M)$, $n=0..3$. With blue is $m=0$, with green $m=1$. The black solid circle denotes $a=0$}
\label{m01}
\end{figure}

\begin{figure}
\centering
\subfigure{\includegraphics[width=120px,height=120px]{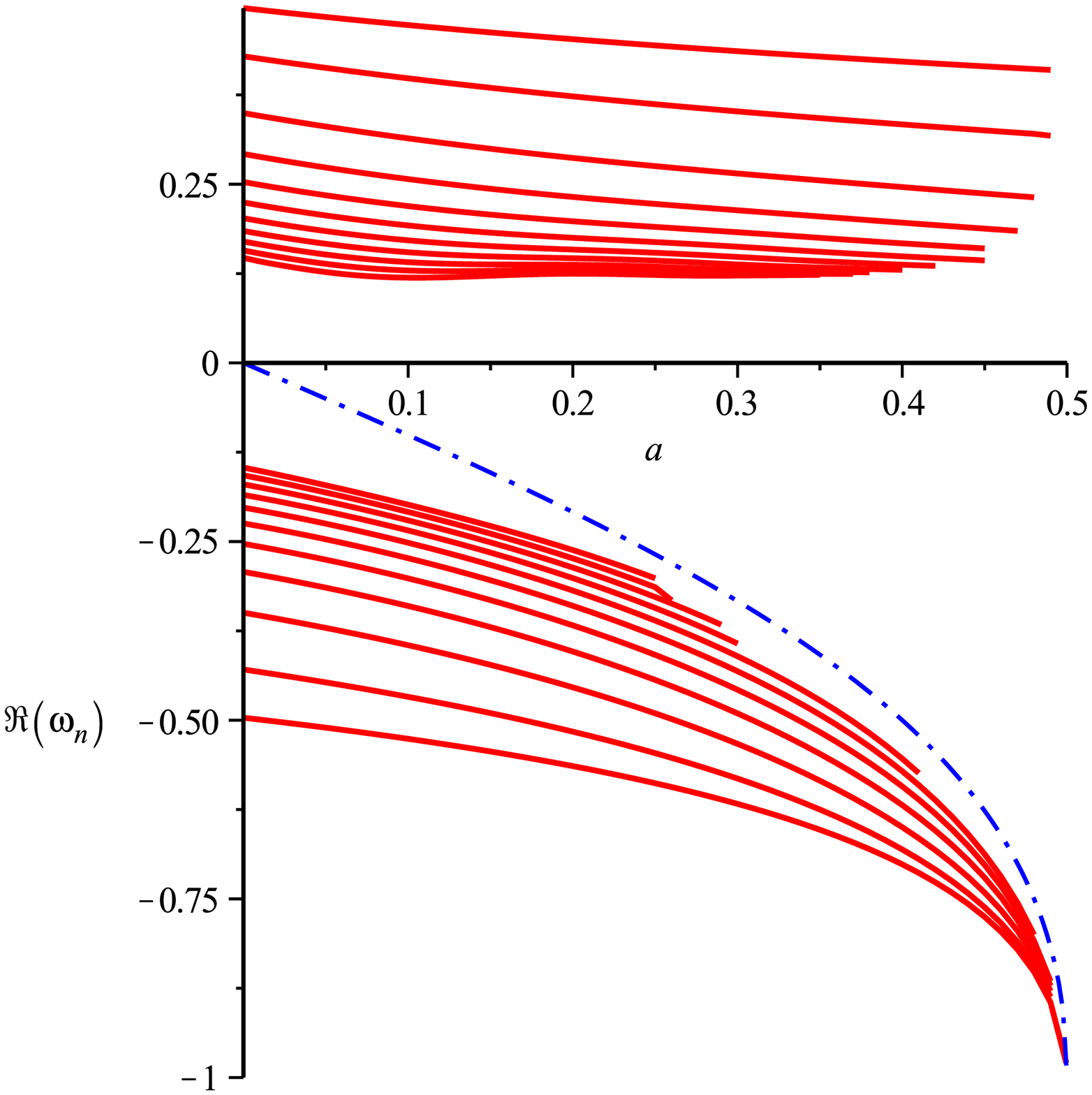}}
\subfigure{\includegraphics[width=120px,height=120px]{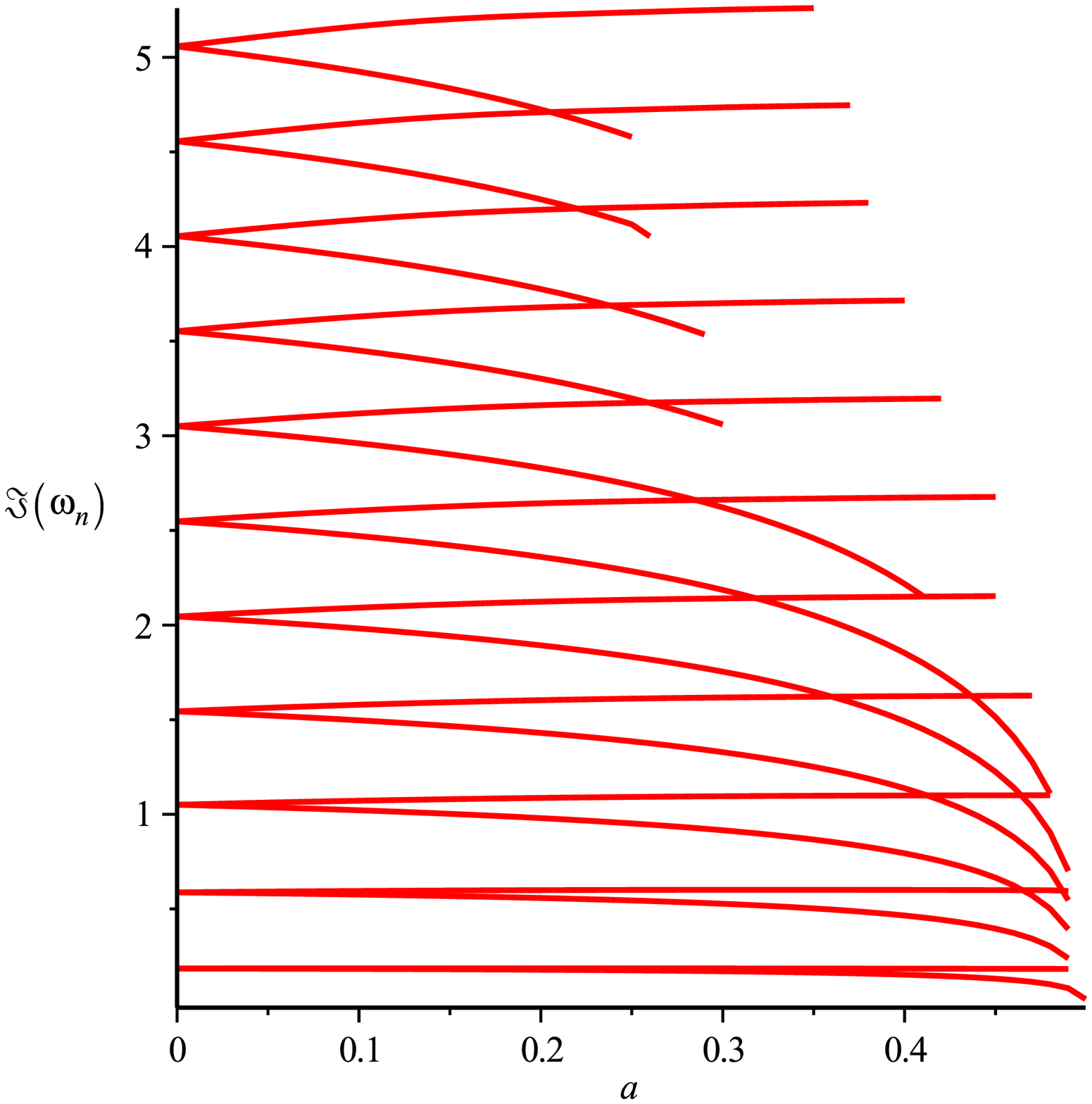}}
\vspace{-0.5cm}
\caption{On the plots: with red lines -- $\Re(\omega_{1,n})(a)$ and $\Im(\omega_{1,n})(a)$ for $a=[0,M)$, $n=0..10$ and with blue dashed line $-m\frac{a}{2Mr_+}$ for $m=1$}
\label{m1_cr}
\end{figure}

From the radial boundary conditions it follows that only frequencies for which $\Re(\omega)\not\in (0,-m\frac{a}{2Mr_+})$ correspond to black hole boundary conditions. Figure \ref{m1_cr} a) shows that the so obtained spectrum obeys this condition. A deviation from this condition was observed in \cite{spectra}, where some of the frequencies describing primary jets crossed the line defined by $-m\frac{a}{2Mr_+}$ , thus corresponding to a white hole solution. For the QNM spectrum, however, this is not the case and the spectrum corresponds to perturbation of  a black hole. 
  
From the same figure one can see in the negative sector of the plot that the real parts of the QNMs for increasing $n$ seem to tend to the line $-m\frac{a}{2Mr_+}$, which requires further investigation for $n>10$. For the positive sector (i.e. the frequencies with positive real parts) we were not able to trace the frequencies with high $n$ near $a\to M$, thus we cannot confirm the relation $\Re(\omega)=m$ for $a\to M$ observed in \cite{high}.

\begin{figure}
\vspace{-0cm}
\centering
\subfigure{\includegraphics[width=121px,height=120px]{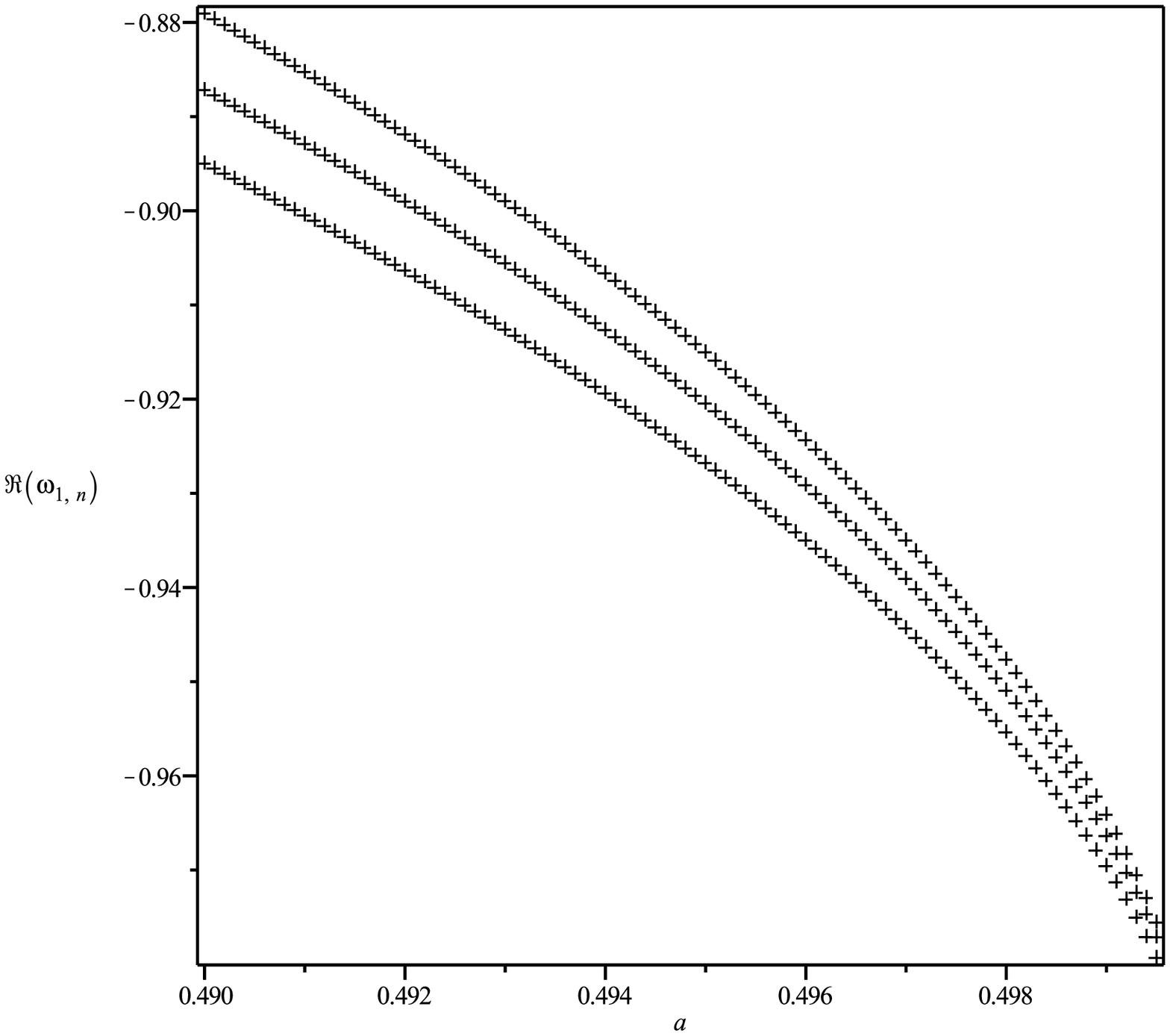}}
\subfigure{\includegraphics[width=121px,height=120px]{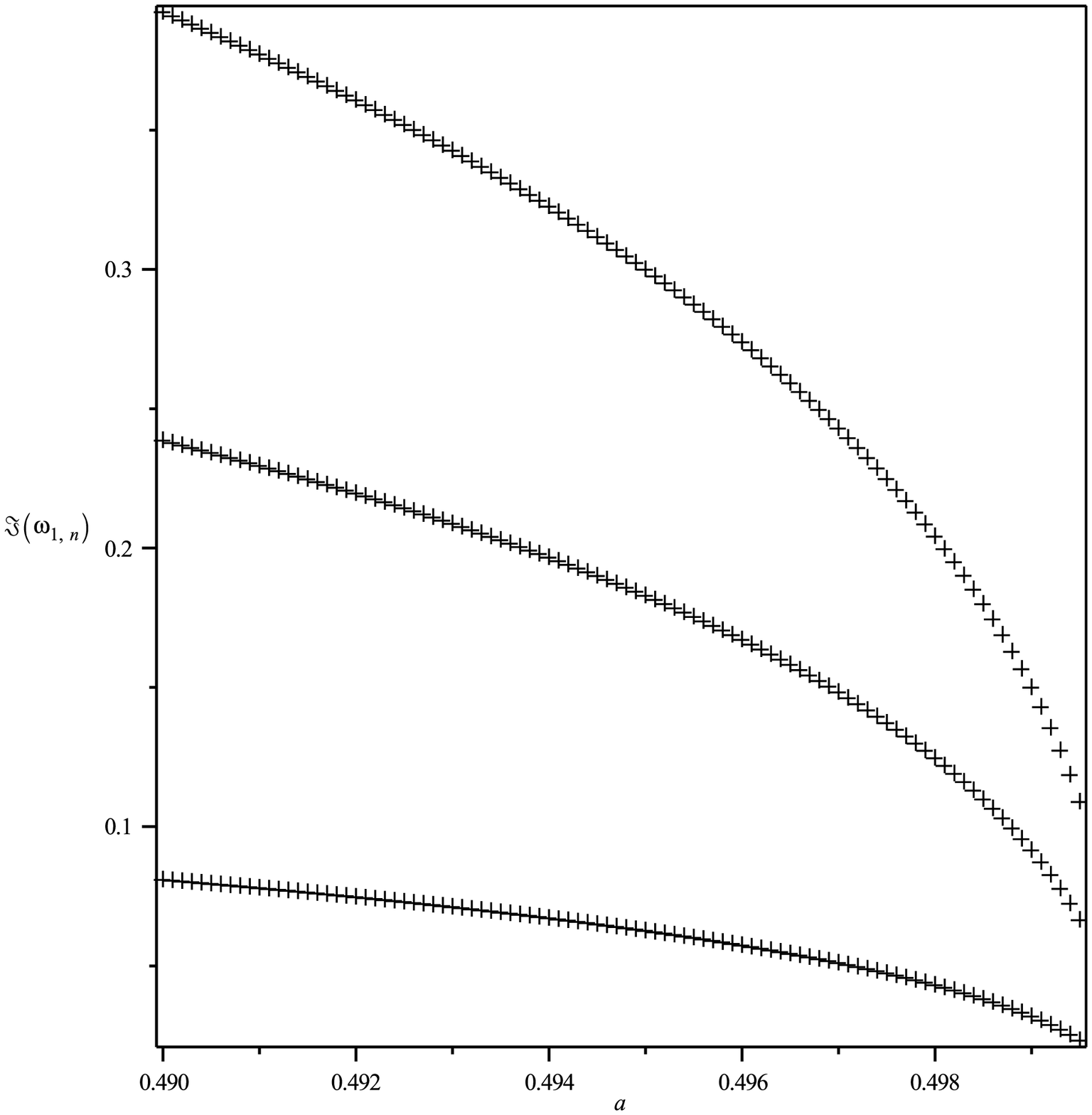}}
\caption{On the plot $\Re(\omega_{1,n})(a)$ and $\Im(\omega_{1,n})(a)$ for $a=[0.49,0.4995]$ for the modes $n=0,1,2$, $m=1$}
\label{aM}
\end{figure}

Finally, obtaining the modes in the limit $a\approx M$ could be of serious interest, if one is to compare the EM QNMs with the spectra obtained from astrophysical objects, but it is also technically challenging. This happens because for $a=M$ the TRE changes its type and near this limit, the confluent Heun function becomes numerically unstable since these functions are transforming to the double confluent Heun ones. Because of this, the examination of the limit $a\to M$  for modes with high $n$ is impossible with current numerical realization of that function in \textsc{maple}. For the lowest modes, however, the function is stable enough in the interval $a\in[0.49,0.4995]$ and the results of the numerical experiment for $m=1$ are plotted on Fig. \ref{aM}. As expected, for $n=0$, for $a>0.91M$ the imaginary part of the frequency quickly tends to zero, thus proving that for extremal objects, the perturbations damp very slowly. For the other two modes, it also seems to tend to zero, although somewhat slower than $n=0$. In physical units, the difference between the 3 modes for $a=0.4995$ is only 6Hz ($\omega_{1,1}\approx 1.582kHz$), but the damping times of the first mode is approximately 4.86 times bigger than that of the third and is $t^{damp}_{1,1} \approx 4.2ms$ for KBH with mass $M=10 M_\odot$. The frequencies in physical units, for some other values of the rotational parameter, can be found in Table \ref{table_n0_om} in the Appendix. 

While the analytical study of the extremal case is outside the scope of this work, one can find such analytical treatment of the issue in \cite{extr1,extr2}. In both articles, one deals with approximations of the exact solutions of the radial equation, obtained under certain assumptions -- be it through the continued fraction method in the limit $a\to M$ (\cite{extr1}) or through a particular case in which the solution of the radial equation for $a\to M$ can be written in terms of confluent hypergeometric functions (\cite{extr2}). Both methods seem to describe well the numerical results, but further investigation through the exact solutions written in terms of double confluent Heun functions and their properties, could further expand our knowledge of this limit case, especially with respect to effect of the branch cuts in those cases. 

For example, a recent study by Yang et al. (\cite{BC_New2}) showed that the nearly extremal Kerr black holes have two distinct sets of QNMs for the same $n$, coexisting for certain $\{m,l\}$. The authors derive those results using the matched expansions method, the continued fraction method and a WKB analysis. Although in our case we also observe different coexisting sets of QNMs, the important difference, is that while in our case, we see the different sets of QNMs starting from $a=0$ and persisting with the increasing of the rotation, in the Yang study, the branching appears when $a\to M$ and the modes merge as the rotation decreases. Understanding whether our results can be applied to this regime and this behavior corresponds to new physics or it is just a numerical artefact, requires an in depth study of the asymptotic behavior of the confluent Heun function near its branch cuts. 

\subsection{Algebraically special modes and branch cuts}
The algebraically special (AS) modes are obtained from the condition that the Starobinsky constant vanishes (\cite{QNM0}) and they correspond to the so called total transmission modes (TTM) -- modes moving only in one direction: to the right or to the left. In the case of gravitational perturbations ($s=-2$) from nonrotating BH, since the $9^{th}$ QNM coincides approximately with the theoretically expected purely imaginary AS mode, there were speculations that the two modes coincide (see \cite{high} for a review, and also \cite{special2,AS}). A study of this mode in the case of gravitational perturbations of KBH showed numerical peculiarities as the ``doublet'' emerging from  the ``AS mode'' for $m>0$ (see \cite{high}) and also unexplained ``spurious'' modes, blamed to numerical inaccuracies.  

For electromagnetic perturbations, the algebraically special modes have not been considered, since in the limit $a\to 0$, the Starobinsky constant does not vanish for purely imaginary modes (in fact, for $a=0$, the Starobinsky constant does not depend on $\omega$ at all, see Eq. (60) \cite{QNM0} p.392) and there appears to be no correlation between TTM and QNM modes \cite{AS1}. There is, however, one important parallel between the electromagnetic and the gravitational case. For the nonrotating gravitational case, Maassen van den Brink \cite{special2,AS} found that the peculiarities of the $9^{th}$ mode are due to the branch cut in the asymptotics of Regge-Wheeler potential, which the method of the continued fraction is not adapted to handle. This result was confirmed by the use of the $\epsilon$-method in our previous work \cite{arxiv3} where the AS character of the $9^{th}$ mode was disproved. Using the $\epsilon$-method, one sees that this result is not limited to the gravitational case and the branch cuts play an important role for the electromagnetic QNMs as well. 

Using $\epsilon$ as a parameter controlling the location of the branch cut with respect to certain QNM \footnote{Recall that $r\sim \exp(i \arg(\omega))$ and thus by changing $\omega_{m,n}$ and $\epsilon$, one changes the position in the $r$-complex plane}, from the equations of the branch cuts discussed in section ``The epsilon-method'', one can relate the appearance of peculiarities in the numerical results, with the proximity of one or more branch cuts in the complex $\omega$-plane. In the gravitational case \cite{arxiv3}, the supposed AS mode is the one with the smallest real part, for which the value of $\epsilon$ for which one observes the jump discontinuity is also very small.  Therefore, one can expect that for this mode, very small variations in the phase-condition can change the leaf of the multivalued function and thus to lead to a different $\omega$ from the expected.  

In the EM case, one can also find a mode with a very small real part -- $n=11$ with $\Re(\omega_{0,11})=.0215$ (evaluated for  $\epsilon=0.15$), for which one encounters the jump discontinuity very close to the imaginary axis at $\epsilon=0.0024$. In this case, one observes particularly interesting dependence on $\epsilon$ -- as showed on Fig. \ref{m0_0} a) -- for $\epsilon\le0.05$ the mode $n=11$ separates the lower QNM branch from the upper branch similarly to the way the so-called AS mode separates the QNM branches in the gravitational case \cite{special1}, but if one uses $\epsilon=0.15$ there is no such separation. Thus, for $\epsilon=0$ one finds a similarity between the EM and the gravitational cases. This similarity, however, is due to the appearance of branch cuts in the radial function in both cases and not to some special properties of the mode in question (i.e. $n=11$ for $s=-1$ and $n=8$ for $s=-2$). This is because in the EM case, $n=11$ is not an AS mode. In fact when the real part of that mode is very small ($\epsilon\le0.05$), its imaginary part deviates from the value of $6i$ and vice versa -- when the real part is not so small ($\epsilon=0.15$), the imaginary part tends to $6i$, see Fig. \ref{m0_0}. Therefore, for all values of $\epsilon$, this mode deviates from a purely imaginary, integer number. 

If we are to continue the analogy with the gravitational case, studying how the mode $n=11$ evolves with the increase of the rotation shows that it does not differ from the other modes. Although we didn't search specifically for doublets like the ones mentioned in \cite{high}, most of the modes with $n>0$ can be considered as  doublets, since we obtain distinct curves in the complex $\omega-$plane for different $\epsilon$ and in certain ranges of $\epsilon$ those two distinct frequencies, corresponding to the same $n$ (i.e. having similar imaginary parts), coexist. 

Focusing on the other peculiarity observed in \cite{high} -- the so called ``spurious'' modes the study of the frequencies in the interval $\epsilon=-0.8..0.8$ for $a=0$ showed that indeed when varying $\epsilon$ one may encounter a number of frequencies around a certain mode. Those ``transition'' frequencies appear to smoothly connect the two frequencies $\omega_{m,n}^{1,2}$, sometimes with precision of 15 stable digits.  Those frequencies, however, are isolated points in the ``transition-interval'' of $\epsilon$ and thus we consider them as a result of instability in the evaluation of the confluent Heun function and thus not trust-worthy. Such unstable transitions usually occur for $n\le4$, 
while for $n>4$, the transition appears to be step-wise in the scale of variation of $\epsilon$ we studied. 

We see that even though we use different method than the one in \cite{high}, on different physical problem, we obtain very similar qualitatively results . This implies that the real reason for the observed peculiarities in the behavior of the spectrum of QNMs in both cases, may be the complex character of the used analytical functions (the confluent Heun functions) in the vicinity of the irregular singular point $r=\infty$ in the complex $r$-plane.  

Considering all the numerical peculiarities demonstrated above, the use of the $\epsilon$-method poses a very serious question in front of the astrophysical application of that spectra -- if one is to compare the numerical results with some observational frequencies, which $\epsilon$ should be trusted? In our numerical experiments, we were able to obtain both the frequencies obtained with well-established methods with a precision bigger than 7 digits, and also other, significantly deviating from them frequencies, which have qualitatively quite different behavior with respect to changes in the rotation of the KBH. Both results are stable in different ranges for $\epsilon$, thus requiring new criteria for sifting out the physical modes based on better understanding of the behavior of the radial function in the complex plane of the radial variable. Such a study is outside the scope of the current work which aims to demonstrate the dependence of the method for obtaining the frequencies with respect to changes in the phase-condition and thus to provoke work in this area. 

\section{Conclusion}
From the recent developments in the field of gravitational waves detection it is clear that finding the EM counterpart to those events can prove to be very useful. In this case, it is needed to better understand the fundamental physics of quasi-normal ringing. In this paper, our team offered a new approach to finding the QNMs for the KBH based on directly solving the system obtained by the analytical solutions of the TRE and TAE in terms of the confluent Heun function. This approach has the advantage of being more traditional  (i.e. imposing directly the corresponding boundary conditions on the exact analytical solutions of the problem) and hence it should allow better understanding of the peculiar properties of the EM QNMs and the physics they imply. 

It was shown that using this approach, one can reproduce the frequencies already obtained by other authors, but without relying on approximate methods. Particularly important is the ability to impose the boundary condition {\em directly} on the solutions of the differential equations. We require the standard regularity condition on the TAE and we explore in detail the radial boundary condition (the BHBC). Critical in it is the use of the direction of steepest descent, which secures the purely outgoing wave at infinity. By using small deviations from this direction (and the phase-condition it defines), we were able to move around the branch cut in the solutions of the radial equation and thus to study its effect on the so obtained spectra. While this movement had no significant effect for the lower modes $n<3$, for the higher modes it led to significant deviations from the already published results. This behavior is persistent for the modes with $m=0,1,2$ and $l=1,2$. This observation raises the important question: What are the electromagnetic QNMs for which one has to look in astrophysical data.  Also interesting is that while the $\epsilon$-method leads to significant changes of the  frequencies $\omega_{m,n}$, it affects much less the separation parameter $E_{m,n}$ which here for the first time was obtained directly as a solution of the two-dimensional system without any prior approximations for it.  
 
Another general result is that the confluent Heun function proved to be an effective tool for physical problems. Even though its \textsc{maple} realization still has many flaws, its precision proved to be good enough to repeat  the already published results, and also studying the solutions, we were able to reveal new properties of the numerical stability of the EM QNMs with respect to changes in the phase-condition.

An interesting question is the results obtained using this method for $a>M$ or the so called naked singularity regime. Preliminary results show that the method is applicable in this case as well and the results will be published elsewhere. 

\section{Acknowledgements}
The authors would like to thank Emanuele Berti for discussion of the numerical values of the EM QNM frequencies  obtained through Leaver's method which was important for the comparison of our method with this already well-established one. 

P.F. is deeply indebted to Irina Aref'ieva, Alexei Starobinsky, Bruno Coppi and Michail Sazhin for interesting comments and suggestions about the relations of EM QNM to  BH physics and the possibilities
to discover such modes in astrophysical observations.

The authors are also thankful to Luciano Rezzolla for drawing our attention to the references \cite{headon0} and \cite{GW_} and to Shahar Hod for drawing our attention to the references \cite{extr1} and \cite{extr2}.    

This article was supported by the Foundation "Theoretical and
Computational Physics and Astrophysics", by the Bulgarian National Scientific Fund
under contracts DO-1-872, DO-1-895, DO-02-136, and Sofia University Scientific Fund, contract 185/26.04.2010.

\section{Author Contributions}
P.F. posed the problem of the evaluation of the EM QNMs of rotating BHs as a continuation of previous studies of the applications of the confluent Heun functions in astrophysics. He proposed the epsilon method and supervised the project.

D.S. is responsible for the numerical results, their analysis and the plots and tables presented here.

Both authors discussed the results at all stages. The manuscript was prepared by D.S. and edited by P.F..

\appendix{}
\section{Appendix: Tables of the obtained EM QNMs}
In the table \ref{table_n0_om} are presented some of the values obtained for the EM QNM, converted to physical units using the relations: 
$$\omega^{phys}=\Re(\omega)\frac{c^3}{2\pi\,G\,M}$$ 
$$\tau^{phys}=\frac{1}{\Im(\omega)}\frac{GM}{c^3}.$$ 

Note that in those formulas a factor of $2$ is missing because the EM 	QNMs were obtained for $M_{KBH}=1/2$ and not for $M_{KBH}=1$. Then if $M$ is the mass of the object in physical units, $M_{\odot}$ -- the mass of the Sun ($M_{\odot}=1.98892\,10^{30}[kg]$) and  $G=6.673\,10^{-11}[\frac{m^3}{kg\,s^2}], c=2.99792458\,10^8[m/s]$, one obtains $$\omega^{phys}\approx \frac{32310}{M/M_{\odot}}\Re(\omega)[Hz],$$
$$\tau^{phys}\approx \frac{0.4925\,10^{-5}\,M/M_{\odot}}{\Im(\omega)}[s].$$

The frequencies and the damping times in the table are calculated for $M=10M_{\odot}$. 

\begin{table*}[!htb]	
\footnotesize
\begin{tabular}{|l | l  | l | l | l  | l | l | l|l|l|}
\hline  \multicolumn{10}{|c|}{$n=0$}\\
\hline 
& \multicolumn{3}{|c|}{$m=0$} & \multicolumn{3}{|c|}{$m=-1$} & \multicolumn{3}{|c|}{$m=1$}\\
\hline \multirow{1}{*}{$a/M$}  & $\omega^{phys}_{m=0}$[Hz] & $\tau^{phys}_{m=0}$[ms] &$\delta(\omega^\epsilon)$&  $\omega^{phys}_{m=-1}$[Hz] & $\tau^{phys}_{m=-1}$[ms] &$\delta(\omega^\epsilon)$&  $\omega^{phys}_{m=1}$[Hz] & $\tau^{phys}_{m=1}$[ms]& $\delta(\omega^\epsilon)$\\ 
\hline
0&    802.1512449166& 0.5325890917&$10^{-10}$& 802.1512449166& 0.5325890917&$10^{-10}$&  802.1512449167& 0.5325890917&$10^{-10}$\\
0.2& 804.9393652797& 0.5343356142&$10^{-10}$& 849.8315682698& 0.5388677452&$10^{-10}$ &   763.6902591869& 0.5299212818&$10^{-10}$\\
0.6&  829.2637578502& 0.5526525743&$10^{-10}$&   996.9258848852& 0.5772488810&$10^{-10}$& 704.6451920585& 0.5313970112&$10^{-9}$\\
0.98& 884.6086875757& 0.6427140687&$10^{-9}$& 1445.8841670353& 1.2178343064&$10^{-11}$& 661.8628389523& 0.5373275077&$10^{-4}$\\
\hline
\multicolumn{10}{|c|}{$n=3$}\\
\hline 
& \multicolumn{3}{|c|}{$m=0$} & \multicolumn{3}{|c|}{$m=-1$} & \multicolumn{3}{|c|}{$m=1$}\\
\hline \multirow{1}{*}{$a/M$} & $\omega^{phys}_{m=0}$[Hz] & $\tau^{phys}_{m=0}$[ms] &$\delta(\omega^\epsilon)$&  $\omega^{phys}_{m=-1}$[Hz] & $\tau^{phys}_{m=-1}$[ms] & $\delta(\omega^\epsilon)$&  $\omega^{phys}_{m=1}$[Hz] & $\tau^{phys}_{m=1}$[ms]&$\delta(\omega^\epsilon)$\\ 
\hline
0&    472.3043572607& 0.0638131626&$10^{-10}$& 472.3043572599& 0.0638131626&0.3& 472.3043572609& 0.0638131626&$10^{-11}$ \\ 
0.2& 479.7880705182& 0.0641624606&$10^{-10}$& 549.5382706431& 0.0658306413&0.3& 415.5213067645& 0.0624052214&$10^{-11}$ \\ 
0.6& 530.3546588895& 0.0677548111&$10^{-11}$& 792.4033820584& 0.0741706889&--& -- & --&-- \\
0.98&408.1590806664& 0.0756272782&--& -- & -- &--& -- & --&--\\
\hline
\multicolumn{10}{|c|}{$n=7$}\\
\hline 
& \multicolumn{3}{|c|}{$m=0$} & \multicolumn{3}{|c|}{$m=-1$} & \multicolumn{3}{|c|}{$m=1$}\\
\hline \multirow{1}{*}{$a/M$}  & $\omega^{phys}_{m=0}$[Hz] & $\tau^{phys}_{m=0}$[ms] &$\delta(\omega^\epsilon)$&  $\omega^{phys}_{m=-1}$[Hz] & $\tau^{phys}_{m=-1}$[ms] &$\delta(\omega^\epsilon)$&  $\omega^{phys}_{m=1}$[Hz] & $\tau^{phys}_{m=1}$[ms]&$\delta(\omega^\epsilon)$\\ 
\hline
0&    254.8509577191& 0.0275665662&0.03& 254.8509577190& 0.0275665662&0.34& 254.8509577194& 0.0275665662&0.03\\ 
0.2& 264.6648749590& 0.0277638854&0.31& 334.8615860768& 0.0285332670&0.44& 251.7604458626& 0.0268474465&0.04\\ 
0.6& 268.3651895175& 0.0294003836&0.02& 610.5848397689& 0.0318459819&--& 291.2755338407& 0.0263079796&0.06\\ 
0.98& --& --&--& --& --&--& --& --&--\\
\hline
\end{tabular}
\caption{Table of the frequencies, $\omega^{phys}$, in Hz , the damping times, $\tau^{phys}$, in milliseconds   for some of the modes $n=0,4,7$ and for some chosen values of the rotational parameter in the case $l=1$. Presented is also $\delta(\omega^\epsilon)$, the maximal difference between the modes obtained for the 3 values of $\epsilon=0,0.05,0.15$ for each $a,n,m$. The numbers presented here correspond to 10 $M_\odot$.}
\label{table_n0_om}
\end{table*}

\begin{table*}[!htb]	
\footnotesize
\begin{tabular}{|l | l  | l | l | l |l |l |}
\hline  \multicolumn{7}{|c|}{$n=0$}\\
\hline 
& \multicolumn{2}{|c|}{$m=0$} & \multicolumn{2}{|c|}{$m=-1$} & \multicolumn{2}{|c|}{$m=1$}\\
\hline \multirow{1}{*}{$a/M$}  & $E_{m=0}$ &$\delta(E^{\epsilon})$&  $E_{m=-1}$ &$\delta(E^{\epsilon})$&  $E_{m=1}$&$\delta(E^{\epsilon})$\\ 
\hline
0& 2.0000000000 + $9.4 10^{-63}i$& $10^{-64}$&  2.0000000000 + $4.82 10^{-63}i$& $10^{-64}$&                                                               2.0000000000 + $4.83 10^{-60}i$& $10^{-64}$\\
0.2& $1.9991429248 - 0.0007350608i$& $10^{-12}$& $1.9460508578 + 0.0193509134i$& $10^{-11}$& 
      $2.0462372214 + 0.0176327659i$& $10^{-11}$\\
0.6& $1.9916552351 - 0.0066017264i$& $10^{-11}$& 
      $1.7970919424 + 0.0620215234i$& $10^{-10}$& 
      $2.1232126190 + 0.0478089014i$& $10^{-10}$\\
0.98&$1.9734060595 - 0.0162271167i$& $10^{-10}$& 
      $1.4500680177 + 0.0605778034i$& $10^{-11}$& 
      $2.1833506585 + 0.0705481152i$& $10^{-5}$\\
\hline
\multicolumn{7}{|c|}{$n=3$}\\
\hline 
& \multicolumn{2}{|c|}{$m=0$} & \multicolumn{2}{|c|}{$m=-1$} & \multicolumn{2}{|c|}{$m=1$}\\
\hline \multirow{1}{*}{$a/M$}  & $E_{m=0}$ &$\delta(E^{\epsilon})$&  $E_{m=-1}$ &$\delta(E^{\epsilon})$&  $E_{m=1}$&$\delta(E^{\epsilon})$\\ 
\hline
0& 2.0000000000+$9.44 10^{-31}$i& $10^{-64}$& 2.0000000000+4.82 $10^{-30}$i& $10^{-64}$& 2.0000000000+4.83 $10^{-30}i$& $10^{-64}$\\ 
0.2&$2.0090686631-0.0036398098i$& $10^{-12}$& $1.9778301381+.1550332409i$& 0.003& $2.0389149509+.1531306042i$& $10^{-11}$\\ 
0.6&$2.0717320628-0.0337575846i$& $10^{-12}$& $1.9332832523+.4598448835i$& $10^{-11}$&--&-- \\
0.98&$2.1483751533-0.0584019594$&--&$1.6286222388+.6109228338i$&--&--&--\\
\hline
\multicolumn{7}{|c|}{$n=7$}\\
\hline 
& \multicolumn{2}{|c|}{$m=0$} & \multicolumn{2}{|c|}{$m=-1$} & \multicolumn{2}{|c|}{$m=1$}\\
\hline \multirow{1}{*}{$a/M$}  & $E_{m=0}$ &$\delta(E^{\epsilon})$&  $E_{m=-1}$ &$\delta(E^{\epsilon})$&  $E_{m=1}$&$\delta(E^{\epsilon})$\\ 
\hline
0& 2.0000000000+$9.44 10^{-31}$i& $10^{-64}$& 2.0000000000+4.82 $10^{-30}$i& $10^{-64}$& 2.0000000000+4.83 $10^{-30}i$& $10^{-64}$\\ 
0.2& $2.0499475534-0.0045930025i$& $10^{-3}$ &$2.0450200861+.3500363139i$& 0.04&$2.0888497837+.3569822795i$&0.004\\ 
0.6& $2.3836324485-0.0361782348i$&0.004& $2.3691252941+.9806955183i$&--& $2.7122160605+.9492827382i$&0.03\\ 
0.98& --& --&--&--&--& --\\
\hline
\end{tabular}
\caption{Table of the separation parameter $E$ for some of the modes $n=0,4,7$ and for some chosen values of the rotational parameter in the case $l=1$. Also $\delta(E^\epsilon)$ is presented, the maximal difference between the $E$ obtained for the 3 values of $\epsilon=0,0.05,0.15$ for each $a,n,m$. }
\label{table_n0_E}
\end{table*}

\section{Appendix: The $\epsilon$-method for $a=0$}

Let us denote the dependence $\omega(\epsilon)$ as $\omega_n^{\epsilon}$, so that it differs from $\omega_n(a)$.

The comparison of the frequencies obtained for $\epsilon=0,0.05,0.15$ for different $m$ shows with precision of at least $10^{-10}$ one has:

 for $m=0$: $\omega_n^0=\omega_n^{0.05}$ for all $n$, but $\omega_n^0=\omega_n^{0.15}$ only for modes with $n<4$, and $\omega_n^{0.05}=\omega_n^{0.15}$ for $n<6$. 

 for the case $m=1$: $|\omega_n^0=\omega_n^{0.05}$ and $\omega_n^0=\omega_n^{0.15}$ for $n=0..3$ and $n=6..11$,  and $\omega_n^{0.05}=\omega_n^{0.15}$ for $n<12$. 

 for the case $m=2$: $\omega_n^{0}=\omega_n^{0.05}$ for $n=0..3$ and $n=6..14$,  
$\omega_n^{0}=\omega_n^{0.15}$  for $n<4$, and $\omega_n^{0.05}=\omega_n^{0.15}$ for $n<6$. 

In all the other cases, the modes evaluated at different $\epsilon$ deviate significantly from one another. 	

Such dependency on $m$ is unexpected, since in equation \eqref{R2}, $m$ is always coupled with $a$, so for $a=0$, those frequencies should coincide. This indeed happens with precision  $10^{-12}$ for frequencies evaluated for the same $\epsilon$. For $\epsilon\neq 0$, however, deviation may occur due to instabilities in the numerical algorithm evaluating the confluent Heun functions in proximity to a BC. 

Similarly to the case of gravitational perturbations ($s=-2$) of nonrotating BH, here we can also trace the origin of the deviations to the BCs in the {\em radial} function. By studying the values of each mode in the interval $\epsilon\in(-0.8..0.8)$, it was found that the points where one observes a step-wise change in the frequency, always coincide with passing trough one of the two BCs corresponding to this mode. 

\begin{figure}[!ht]
\vspace{-0cm}
\hspace{1.5cm}
\centering
\subfigure[]{\includegraphics[width=240px,height=150px]{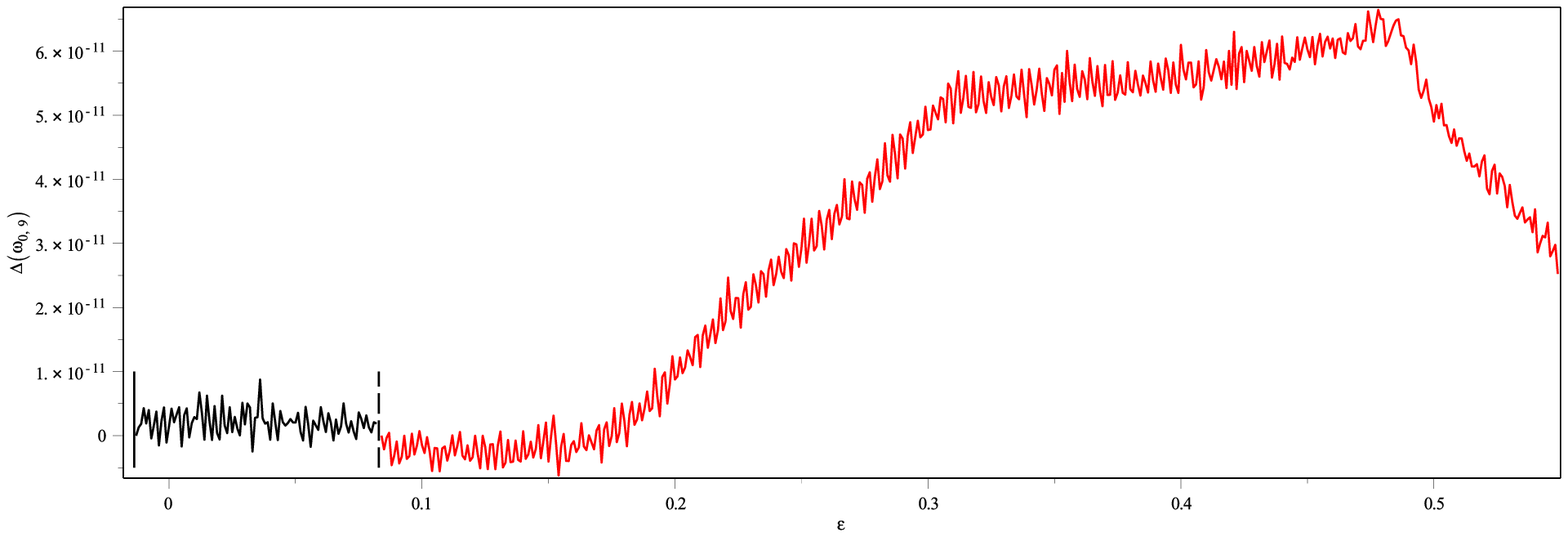}}
\subfigure[]{\includegraphics[width=121px,height=120px]{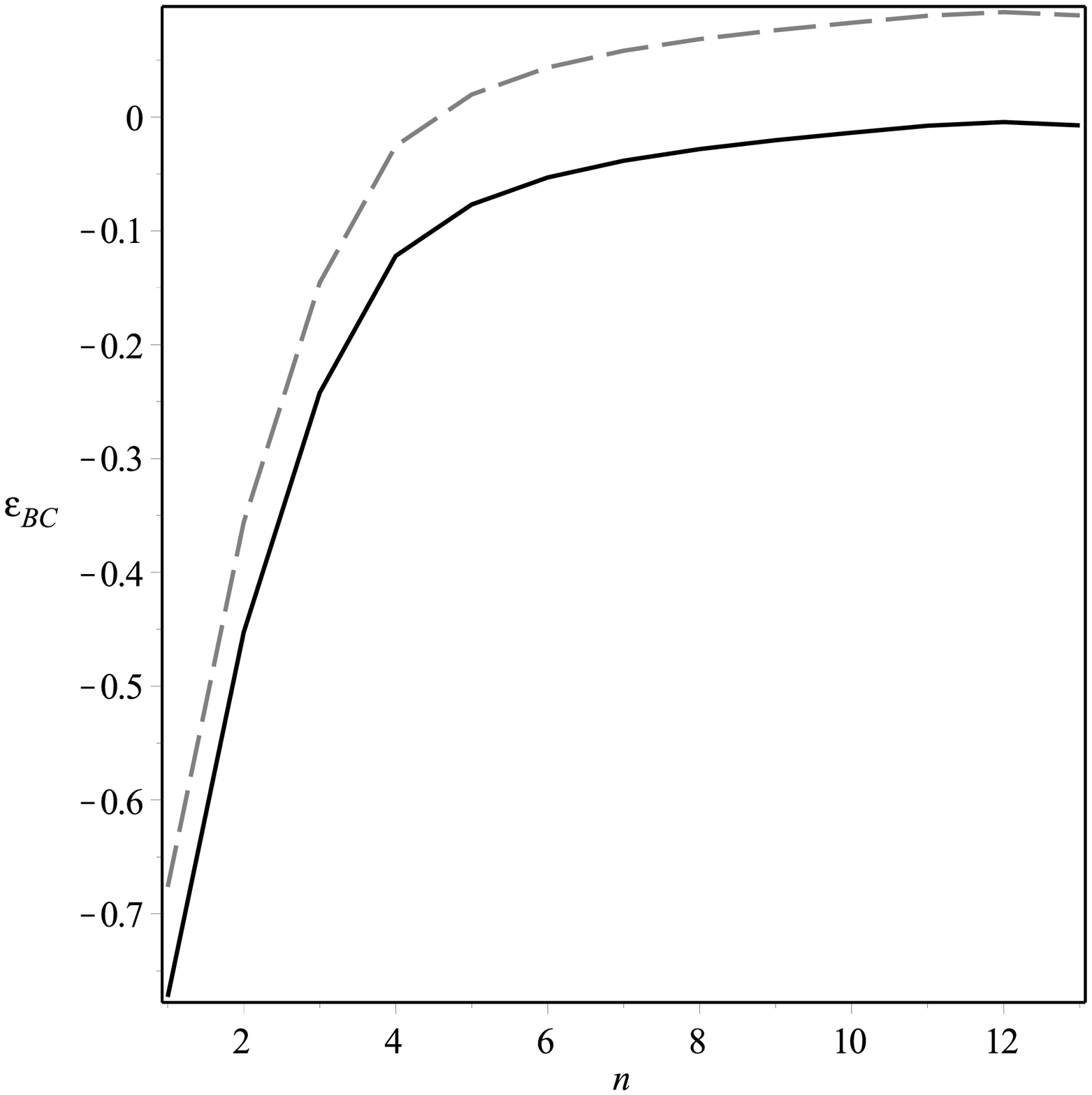}}
\subfigure[]{\includegraphics[width=121px,height=120px]{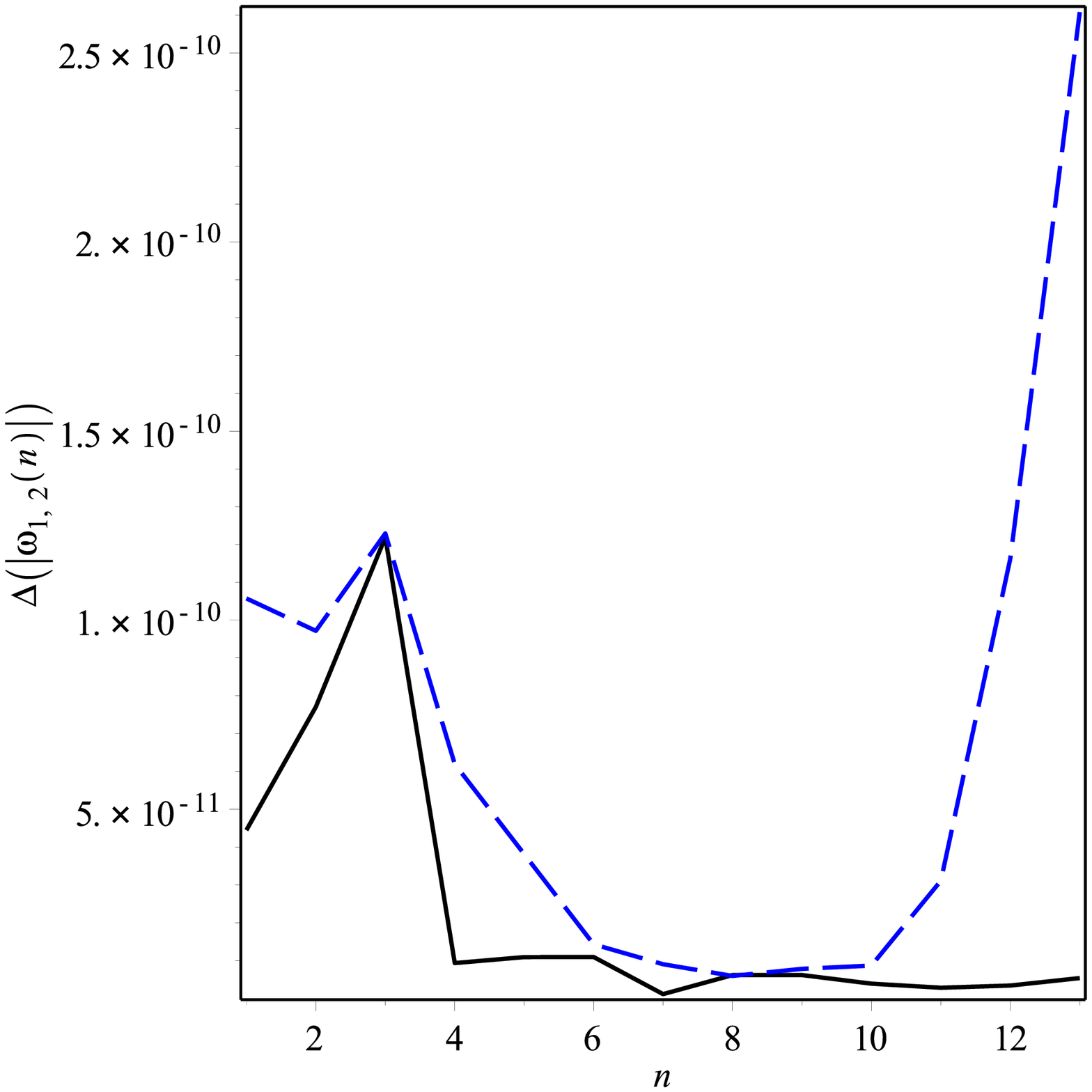}}
\caption{a) The dependence of $\Delta(\omega_{0,9}^{1,2})=\|\omega_{0,9}^{1,2}-\omega_{0,9}^{BC}\|$ on $\epsilon$ for $\epsilon\!=\!-\!0.014..0.55$. With solid and dashed vertical lines are denoted the positions of BC1 and BC2.
b) The position of the two branch cuts for the modes $n=0..12$ in terms of $\epsilon$. With black solid line is BC1, with blue dashed line -- BC2 	\newline
c) As a way to measure the average change in the frequencies due to $\epsilon$ inside the intervals where they are only roots of the radial function,  we plotted the absolute value of the difference between two frequencies of the same type (i.e. $\omega_{0,n}^{1}$ or $\omega_{0,n}^{2}$), evaluated at two points for $\epsilon$ with $\delta \epsilon=0.020$ (for $n=0..12$). With black solid line we denote $\Delta(\omega_{0,n}^{1})$, with blue dashed line $\Delta(\omega_{0,n}^{2})$ Clearly, while this quantity tends to a constant for the frequencies  $\omega_{0,n}^{1}$, it demonstrates different behavior for $\omega_{0,n}^{2}$.  \newline }
\label{m0eps}
\end{figure}

As a case study we present the numerical results for $m=0,n=9$ which demonstrates all the key properties of the dependence of the frequencies on $\epsilon$. For this mode, we have two frequencies with positive real parts and two with negative real ones: 
{\footnotesize
\begin{align*}
\omega_{0,9}^{1,\pm}=\pm0.09906454016+4.59696440777i\\
\omega_{0,9}^{2,\pm}=\pm0.15729928169+4.55601764107i
\end{align*}}
\noindent Since there is symmetry with respect to the imaginary axis and $\epsilon=0$, on Fig. \ref{m0eps} a), one can see the dependence of $\omega_{0,9}^{1,2}$ on $\epsilon$ in the positive half of the interval.  In order to be able to fit the data for both frequencies on the same figure, we have plotted the difference $\Delta(\omega_{0,9})=\omega_{0,9}^{1,2}(\epsilon)-\omega_{0,9}^{1,2}$, along with the two branch cuts affecting these modes: $\Im(r)=0$ and $\Im(r^k)=0$ which we will call BC1 and BC2, denoted with a dashed and a solid vertical line accordingly. If we use the equations of the branch cuts defined in the section {\bf The epsilon-method}, we can introduce the notations $BC1^{\pm}(\omega)=(\pm2(\arg(\pm\omega)-1.4192)/\pi)$ and $BC2^{\pm}=\pm2(\pm\arg(\omega)-3\pi/2)/\pi$
Then, for the intervals of $\epsilon$ into which one can find each appropriate mode, namely:
{\footnotesize
\begin{align*}
&{\omega_{0,9}^{2,-}}\to \epsilon\le-0.0744=BC2^-({\omega_{0,9}^{2,-}})\\
&{\omega_{0,9}^{1,-}}\to \epsilon\le(-0.083,0.013)=(BC2^-({\omega_{0,9}^{1,-}}),BC1^+({\omega_{0,9}^{1,-}}))\\
&\omega_{0,9}^{1}\to \epsilon\in(-0.013,0.083)=(BC1^-({\omega_{0,9}^{1,+}}),BC2^+(\omega_{0,9}^{1,+}))\\
&\omega_{0,9}^{2,+}\to \epsilon\ge0.074=BC2^+(\omega^{2,+})
\end{align*}}
Clearly, there are intervals of $\epsilon$ in which one can find {\em two} roots corresponding to one $n$, both with the same signs of the real parts and with opposite. Note that the limit of applicability of the method in this case is $\epsilon =0.5993$ thus we were not able to check the upper (lower) bound for $\omega_{0,9}^{2,\pm}$.

From the figure it is also clear that while the dependence $\omega_{0,9}^{1,+}(\epsilon)$ appears noise-like implying that the frequency is constant in this interval with more than 10 digits of precision, for $\omega_{0,9}^{2,+}(\epsilon)$ we observe an interval where the frequency remains approximately constant, but also an interval where it demonstrates a different kind of dependence ($\epsilon>0.15$). Very similar behavior occurs for all modes with $n>0$ and gets more pronounced with the increase of $n$ or $a$. 

The case $n=0$ is not included in our discussions, because for it, the numerical routine evaluating the confluent Heun function remains stable only for $\omega_{0,0}^{2,\pm}$ . 

On Fig.\ref{m0eps} b), c) some additional information is presented concerning how the dependence of the modes on $\epsilon$ evolves with the increase of $n$. For example, one observes that the numerical fluctuations in the value of the frequencies decrease with the increase of $n$, even though one can expect an increase of the error of the numerical integration in the complex plane with $n$. 


\section{Appendix: The behavior of the modes for different $\epsilon$ for $a>0$}
\begin{figure}[tb]
\centering
\subfigure{\includegraphics[width=120px,height=120px]{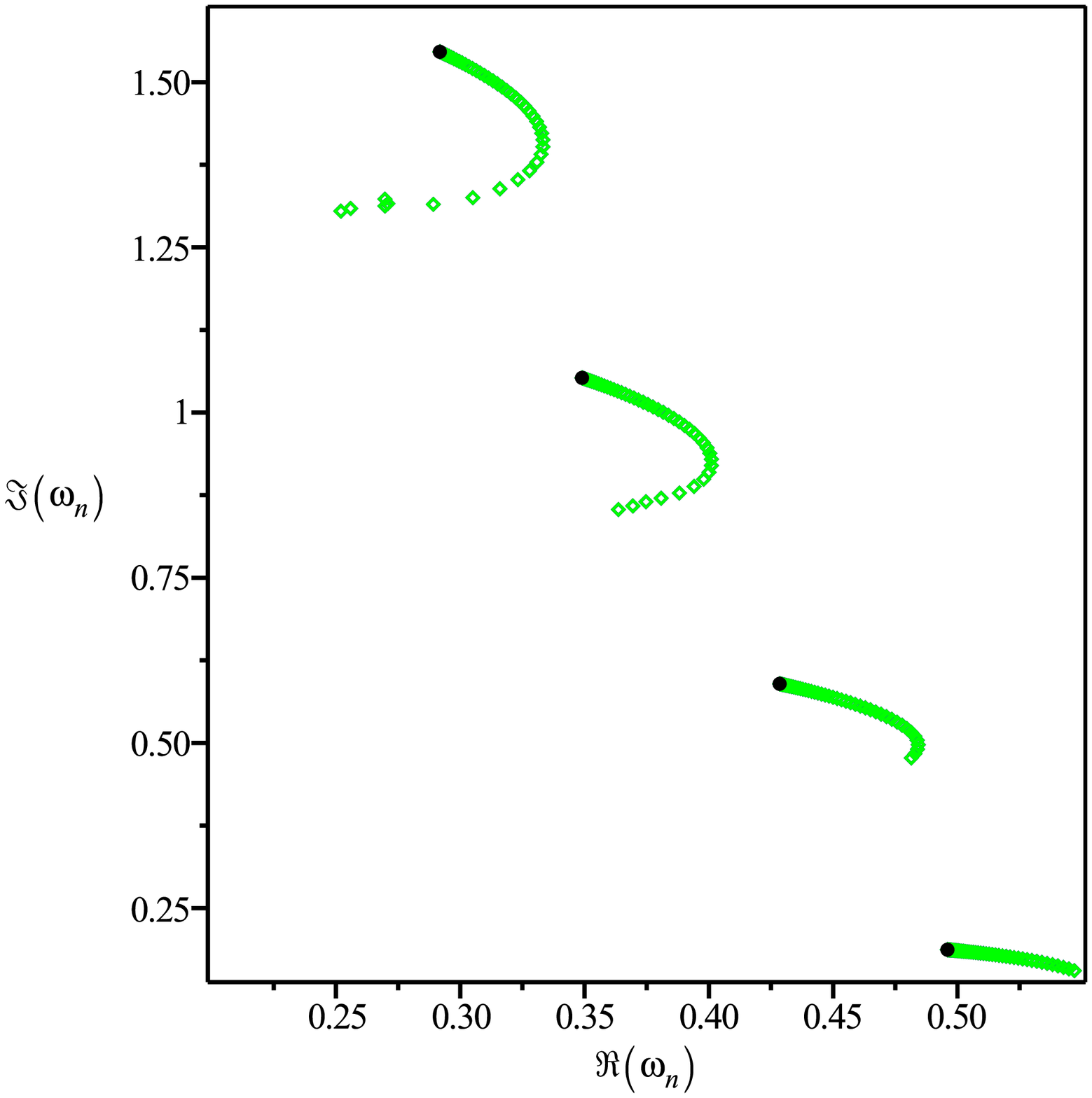}}
\subfigure{\includegraphics[width=120px,height=120px]{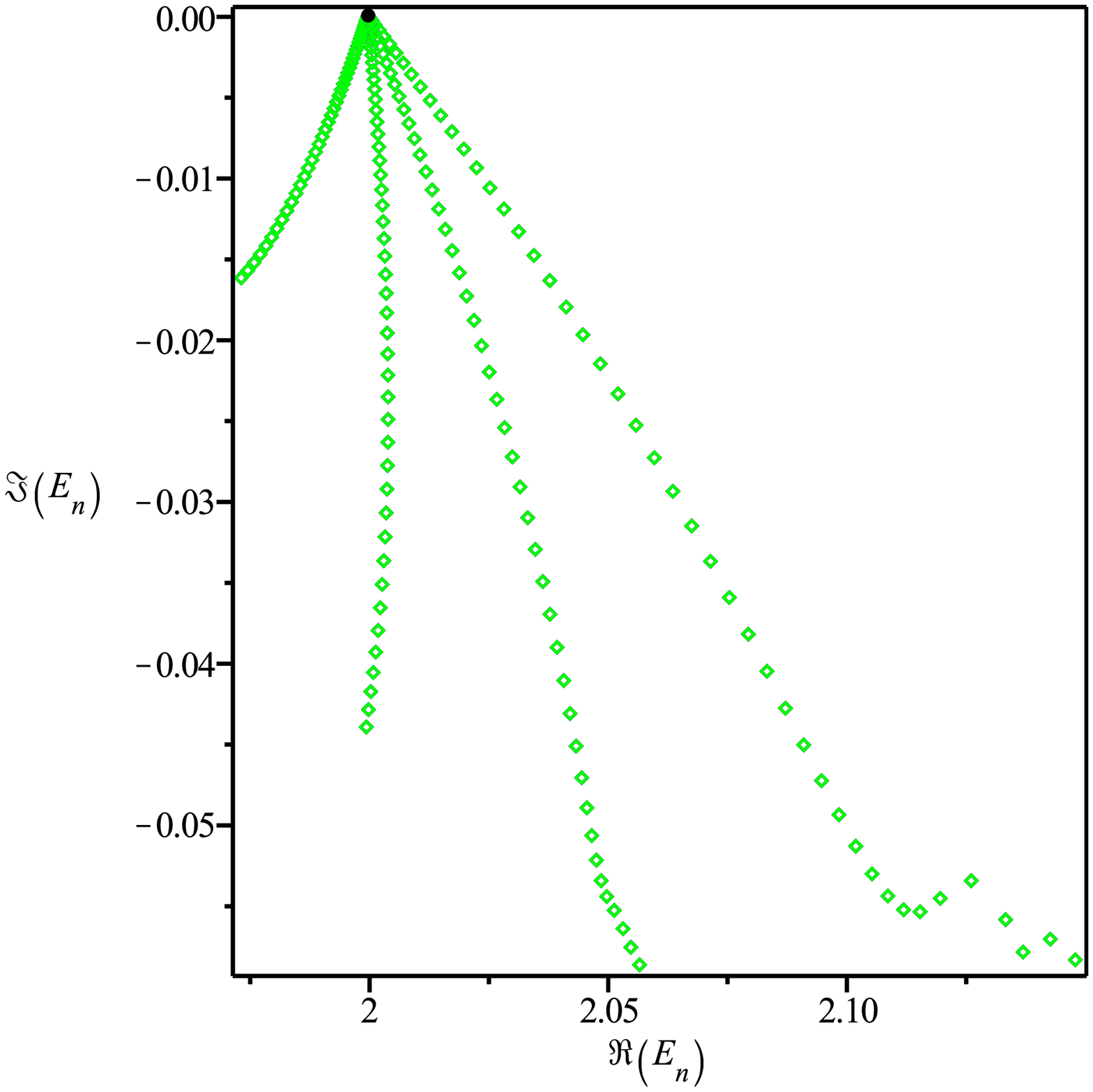}}
\caption{Complex plots of $\omega_{0,n}(a)$ and $E_{0,n}(a)$ for the first 4 modes $n=0..3$, $a=[0,M)$. The points obtained for $\epsilon=0, 0.05, 0.15$ coincide with more than 10 digits thus only one $\epsilon$ is plotted. The black solid circle denotes $a=0$ }
\label{n03_cp}
\end{figure}

\begin{figure}[h]
\vspace{-0cm}
\centering
\subfigure{\includegraphics[width=120px,height=120px]{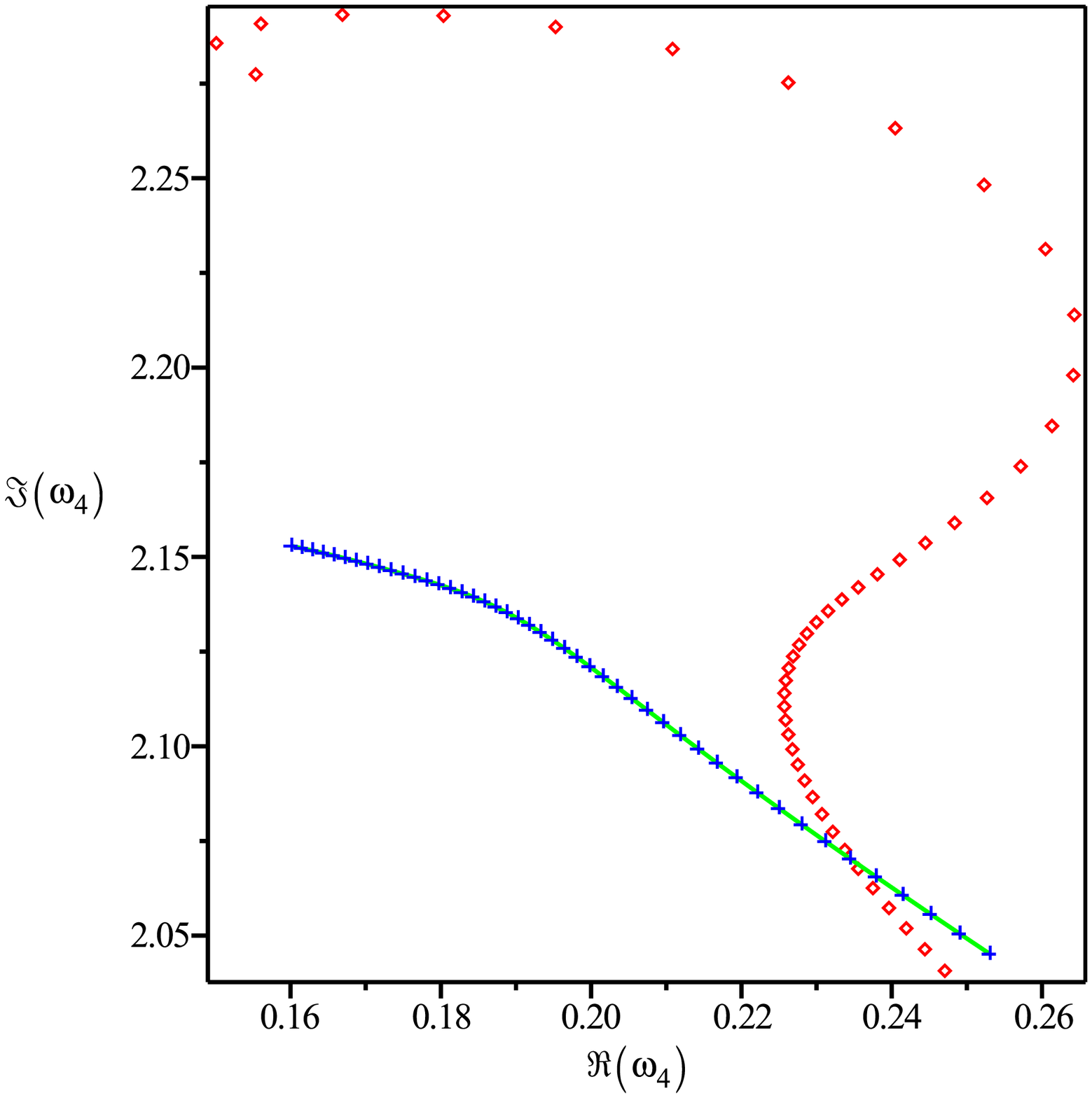}}
\subfigure{\includegraphics[width=120px,height=120px]{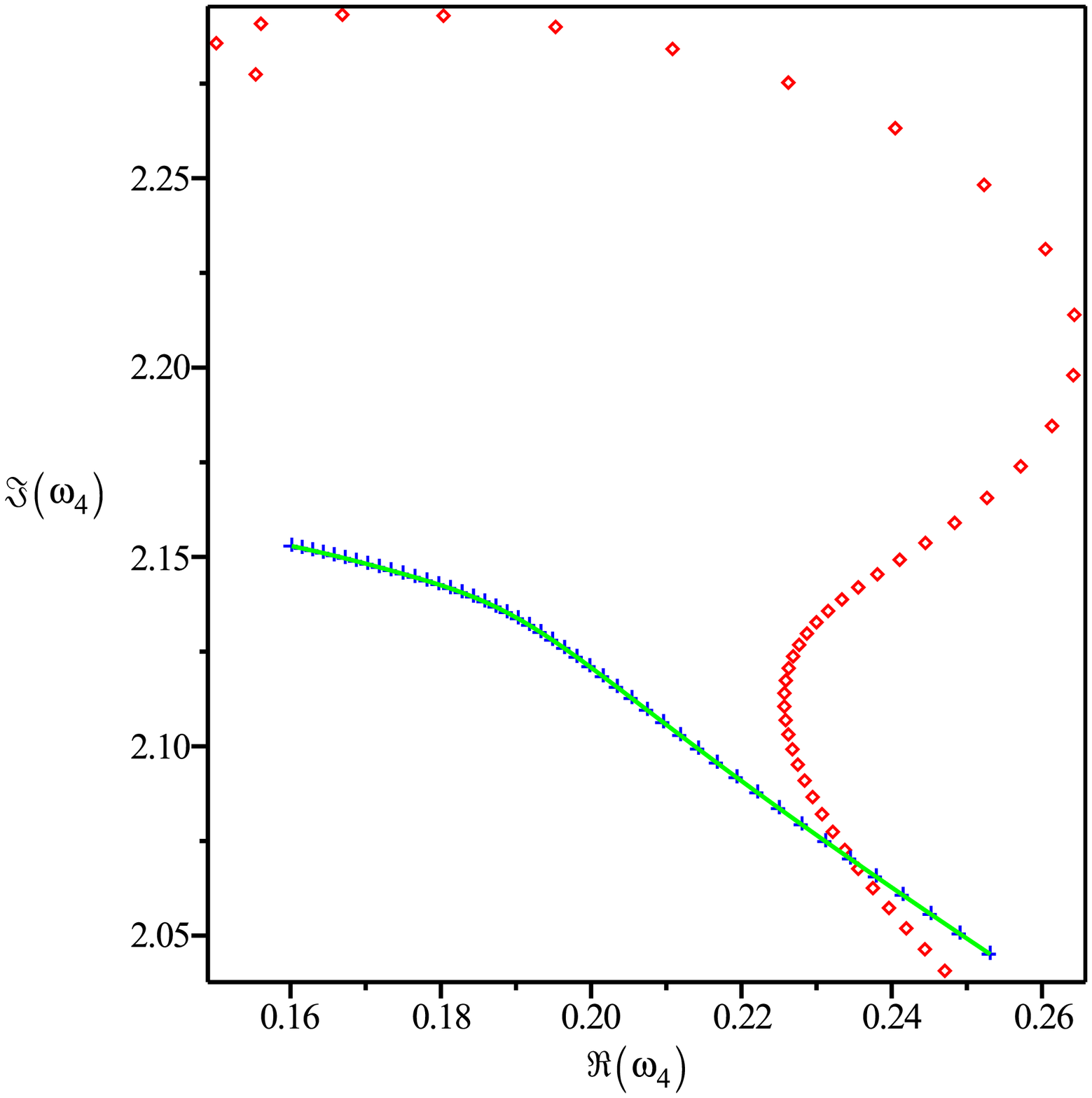}}
\caption{Complex plots of $\omega_{1,4}(a)$ and $E_{1,4}(a)$ for $a=[0,M)$, $\epsilon=0$ (red diamonds), $\epsilon=0.05$(blue crosses) and $\epsilon=0.15$ (green line). There is dramatic deviation of the points obtained for the different values of $\epsilon$}
\label{m1n4}
\end{figure}

When there is rotation, the effect of the two types of branch cuts, BC1 and BC2, continues to dominate the results. 
Studying the modes $n=0..10$ in the case of different $\{l,m,n\}$ for $\epsilon=0,0.05,0.15$ one obtains the following results:
\begin{itemize}
 \item The case $m=0, l=1$.
For modes with $n<4$, $\omega_n^0(a)=\omega_n^{0.05}(a)=\omega_n^{0.15}(a)$ (see Fig. \ref{n03_cp}). The comparison with the control frequencies gives $\|\omega_{n,0}-\omega_{n,0}^B\|<10^{-10},\|E_{n,0}-E_{n,0}^B\|<10^{-10}$ in the whole range for $a$, confirming that in this case the results of the two methods -- the continued fraction and our method-- are the same. 

For $n\ge4$, one observes the two distinct frequencies corresponding to each mode:  $\omega^0(a)=\omega^{0.05}(a)\neq \omega^{0.15}(a)$, with the exception of $n=4,5$ for which $\omega^{0.05}(a)=\omega^{0.15}(a)\neq \omega^0(a)$ (Fig. \ref{n4}). The numerical comparison with the control results $\omega_{n,m}^B,E_{n,m}^B$ show that they coincide with the results for  $\epsilon=0.15$. 

The position of the branch cuts and the corresponding intervals where each mode can be found follows the already described in the nonrotating case. For example, to obtain the modes with negative real parts, one has to use negative $\epsilon$: $\epsilon=0,-0.05,-0.15$.  

The results are symmetrical to those obtained for  $\Re(\omega_{0,n})>0$ with respect to the imaginary axis for $\omega_{0,n}$ and with respect to the real axis for $E_{0,n}$ (for $\Re(\omega_{0,n})>0$, $\Im(E_{0,n})<0$ and vice versa). 
\item The case $m=0, l=2$. 

For $n<7$, we have $\omega^0(a)=\omega^{0.05}(a)= \omega^{0.15}(a)$. For $n=7$,  $\omega^{0.05}(a)=\omega^{0.15}(a)\neq \omega^0(a)$. This behavior continues until $n=12$, where it is $\omega^{0.15}(a)$ that deviates from the other two. The results for $l=2$ can be seen on Fig. \ref{m0l12}.

\item The case $m=1,l=1$.

The modes obtained for the three values of $\epsilon$ coincide up to $n<4$. For $n=4$ (Fig. \ref{m1n4}), $\omega^0(a)$ differs from the other two. For $n\ge6$, it is $\omega^{0.15}(a)$ that differs, while the other two coincide. Note, here the deviation for different $\epsilon$ is much more significant than the case $m=0$ (Fig. \ref{n4}). 
\end{itemize}

\begin{figure}[h]
\vspace{-0cm}
\centering
	\subfigure{\includegraphics[width=150px,height=150px]{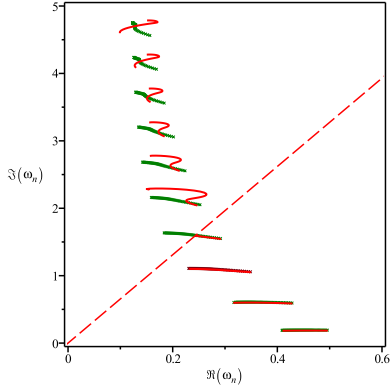}}
\caption{Complex plot of the modes $\omega_{1,n}(a)$ in $a=[0,M)$ for $m=1, l=1, n=0..9$. On the plot are the points obtained for two values of $\epsilon$: $\epsilon=0$ with red lines and $\epsilon=0.15$ with green crosses. The red dashed line corresponds to the branch cut with equation $\Im(\omega)/\Re(\omega)\approx\tan(1.419+\epsilon\pi/2)$ for $\epsilon=0$. One can see that the modes with different $\epsilon$ coincide before reaching the branch cut (i.e. for $n\le3$), and then, the $\epsilon=0$ points begin to differ}
\label{bc}
\end{figure}

\begin{figure}
\vspace{-0cm}
\centering
	\subfigure[]{\includegraphics[width=250px,height=150px]{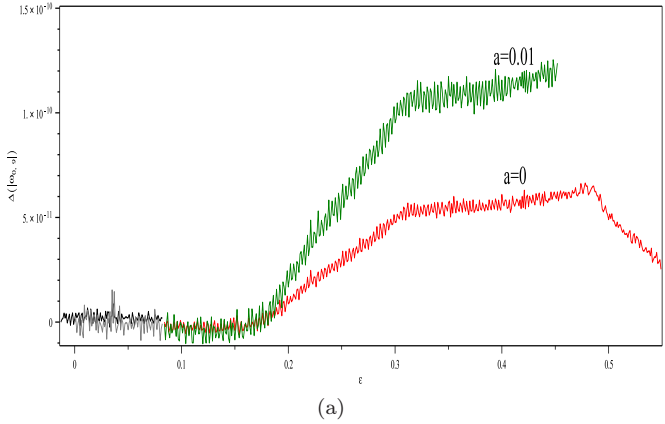}}
\caption{ The dependence of $\Delta(\omega_{1,10})$ on $\epsilon$ for $\epsilon=0.0785..0.088$, $a=0.01$. The solid line denotes the real part, the dotted line -- the imaginary part and $\Delta(\omega_{1,10})(\epsilon)=\omega_{1,10}^{\epsilon}-\omega_{1,10}^{0.0785}$. The dependence $E_{1,10}(\epsilon)$ is similar.}
\label{n10}
\end{figure}

These results show that the peculiarities observed when there is no rotation are inherited by the modes for $a>0$. Studying the dependence of $\omega_{m,n}(\epsilon)$ for each mode in a certain interval of $\epsilon$ is computationally expensive, so we did it only for the cases: $n=9,10$,  $a=0.01$. The results for $n=9$ can be seen on Fig. \ref{n10}. 

Exploration of this dependence to a finer degree in the case $n=10$ in the specific interval $\epsilon=0.0785..0.089$ shows that in it, there are again two pairs of points $[\omega_n,E_n]$ as roots for the {\em same} $\epsilon$ (for $\epsilon\in[0.07862..0.088034]$), namely:

{\footnotesize
\begin{align*}
[0.0680207667+5.1463791539i, 2.0021361645+0.0514150701i],\\
[0.1419210235+5.0686957246i, 2.0028302584+0.0505980908i].
\end{align*}}
Following the formulas for BC1 and BC2 described in the previous section, one finds that the intervals for the first and the second root should be $\epsilon^1\in(-0.0084, 0.0881)$ and $\epsilon^2>0.0786$, thus the interval of coexistence of the two roots should be $\epsilon^*\in (0.0786,0.0881)$, confirmed by our numerical experiments.  

Again, the existence of two sets of QNMs corresponding to one $n$ is unexpected but can be explain with the appearance of BCs in the radial equation, and it repeats for all the modes with $m=0,n>0$. Although it is far from clear whether this behavior represents some new physics in the problem, in any case, it points to a behavior which must be studied more carefully in order to better understand the numerical stability of the EM QNMs. One could argue that the real branch cut coming directly from the TRE is BC1 and BC2 is a numerical artefact, however, for the moment, there is no way to prove or disprove this. What we see from the numerical data is that $\omega^1$ resides in an interval of $\epsilon$ where it remains approximately constant, up to random numerical fluctuations in the 11-th-12th digits.  $\omega^2$, however, demonstrates explicit dependence over $\epsilon$, which the rotation seems to increase. Since $\omega^2$ coincide with our control modes, obtained using different methods,  this result poses once again the question in front of the QNM physics -- which frequencies to use and why. 

\makeatletter
\let\clear@thebibliography@page=\relax
\makeatother


\begin{thebibliography}{00}
\bibitem{RWE} {\sc Regge, ~T., Wheeler ~J. A.}, {\em Stability of a Schwarzschild Singularity}, Phys.Rev {\bf 108}:I.4: 1063-1069 (1957)
%
\bibitem{ZRE} {\sc Zerilli, ~F.~J.}, {\em Effective Potential for Even-Parity Regge-Wheeler Gravitational Perturbation Equations}, Phys.Rev. Lett. {\bf 24}:I.13:  737-738 (1970)
%
\bibitem{vish} {\sc Vishveshwara, ~C. ~V.}, {\em Stability of the Schwarzschild Metric}, Phys.Rev. D {\bf 1}:I.10:  2870-2879 (1970)
%
\bibitem{Teukolsky0} {\sc Bardeen, ~J.M., Press, ~W. H., Teukolsky,~S.~A.}, {\em Rotating Black Holes: Locally Nonrotating Frames, Energy Extraction, and Scalar Synchrotron Radiation}, ApJ {\bf 178}: 347-370 (1972)

%
\bibitem{Teukolsky_0} {\sc Teukolsky,~S.~A.}, {\em Rotating Black Holes: Separable Wave Equations for Gravitational and Electromagnetic Perturbations}, Phys.Rev.Lett. {\bf 29}: 1114-1118 (1972)
%
\bibitem{Teukolsky2} {\sc Teukolsky,~S. A.}, {\em Perturbations of a rotating black hole I Fundamental Equations for Gravitational, Electromagnetic and Neutrino-field Perturbations}, ApJ {\bf 185}: 635-648 (1973)
%
\bibitem{Teukolsky21} {\sc Press ~W.A., Teukolsky ~S. A.}, {\em Perturbations of a Rotating Black Hole. II. Dynamical Stability of the Kerr Metric}, ApJ {\bf 185}: 649-674 (1973)
%
%
\bibitem{teukolsky_} {\sc Teukolsky ~S. ~A., Press ~W. ~H.}, {\em Perturbations of a rotating
black hole. III - Interaction of the hole with gravitational
and electromagnetic radiation}, ApJ {\bf 193}: 443 (1974)
%
\bibitem{chan0} {\sc Chandrasekhar S.},  {\em On the Equations Governing the Perturbations of the Schwarzschild Black Hole},  Proc. Roy. Soc.  London A{\bf 343}: 289-298 (1975)
%
\bibitem{QNM} {\sc Chandrasekhar S., and Detweiler S. L.},  {\em The quasi-normal modes of the Schwarzschild black hole},  Proc. Roy. Soc.  London A{\bf 344}: 441-452 (1975)
%
\bibitem{chan1} {\sc S. Chandrasekhar}, {\em On a transformation of Teukolsky's equation and the
electromagnetic perturbations of Kerr black hole}, Proc. R. Soc. Lond. A {\bf 348}: 39-55 (1976)
%
\bibitem{det1} {\sc S. Detweiler}, {\em On the equations governing the electromagnetic
perturbations of the Kerr black hole}, Proc. R. Soc. London  A {\bf 349}: 217-230 (1976)
%
\bibitem{chan2} {\sc S. Chandrasekhar}, {\em On the equations governing the perturbations of
Reissner - Nordstr\"{o} m black hole}, Proc. R. Soc. London  A {\bf 3365}: 453-465 (1976)
%
\bibitem{QNM2} {\sc Detweiler ~S.}, {\em Black holes and gravitational waves. III - The resonant frequencies of rotating holes}, ApJ:{\bf 239}: 292-295, (1980)
%
\bibitem{QNM0}
    {\sc Chandrasekhar ~S.}, {\em The mathematical theory of black holes},  Clarendon Press/Oxford University Press (International Series of Monographs on Physics. Volume 69), (1983)
%
\bibitem{Leaver} {\sc Leaver ~E. ~W.}, {\em An analytic representation for the quasi-normal modes of Kerr black holes}, Proc. Roy. Soc. London A{\bf 402}: 285-298 (1985)
%
\bibitem{Leaver0} {\sc Leaver ~E. ~W.}, {\em Solutions to a generalized spheroidal wave equation: Teukolsky's equations in general relativity, and the two-center problem in molecular quantum mechanics}, J.Math. Phys.{\bf 27} (5):1238  (1986)
%
\bibitem{Q_N_M} {\sc Andersson~N.}, {\em A numerically accurate investigation of black-hole normal modes},
Proc. Roy. Soc. London A\textbf{439} no.1905: 47-58 (1992)
%
\bibitem{high}{\sc Berti ~E.,Cardoso ~V., Kokkotas ~K.~D., Onozawa ~H.},{\em Highly damped quasinormal modes of Kerr black holes} Phys.Rev. D {\bf 68}: 124018 (2003),   arXiv:hep-th/0307013v2
%
\bibitem{special1} {\sc Berti ~E.}, {\em Black hole quasinormal modes: hints of quantum gravity?}, To be published in the Proceedings of the Workshop on 'Dynamics and Thermodynamics of Black Holes and Naked Singularities' (Milan, May 2004),    arXiv:gr-qc/0411025v1 (2004)
%
\bibitem{extr1}{\sc Hod~S., Keshet ~U.},{\em Intermediate Asymptotics of the Kerr Quasinormal Spectrum},Class.Quant.Grav. {\bf 22}: L71-L76 (2005), arXiv:gr-qc/0505112v1
%
\bibitem{Fiziev1} {\sc Fiziev P.~P.}, {\em Exact Solutions of Regge-Wheeler Equation and Quasi-Normal Modes of Compact Objects}, Class. Quant. Grav. {\bf 23}: 2447-2468 (2006), arXiv:0509123v5 [gr-qc]
%
\bibitem{QNM1}
{\sc Ferrari~V., Gualtieri L.}, {\em Quasi-normal modes and gravitational wave astronomy}, Gen.Rel.Grav.{\bf 40}: 945-970 (2008), arXiv:0709.0657v2 [gr-qc] 
%
\bibitem{extr2}{\sc Hod~S.},{\em Slow relaxation of rapidly rotating black holes},Phys.Rev.D{\bf 78}: 084035, (2008)
arXiv:0811.3806v1 [gr-qc]

%
\bibitem{special3} {\sc Berti~E., Cardoso~V. and Starinets~A.~O.}, {\em Quasinormal modes of black holes and black branes}, Class. Quantum Grav. {\bf 26}: 163001 (108pp) (2009), arXiv:0905.2975v2
%
\bibitem{Fiziev2} {\sc Fiziev~P.~P.},  {\em Teukolsky-Starobinsky identities: A novel derivation and generalizations}, Phys. Rev. D{\bf 80}: 124001 (2009), arXiv:0906.5108 [gr-qc]
%
\bibitem{Fiziev3} {\sc Fiziev~P.~P.},  {\em Classes of exact solutions to the Teukolsky master equation}
                       Class. Quantum Grav. {\bf 27}:  135001 (2010), arXiv:0908.4234v4  [gr-qc]
%
\bibitem{BHB1}{\sc Hod ~S., Hod ~O.}{\em Analytic treatment of the black-hole bomb}, Phys. Rev. D {\bf 81}: 061502 (2010) Rapid communication,    	arXiv:0910.0734v1 [gr-qc]
%
\bibitem{QNM21}
{\sc Konoplya, ~R. ~A., Zhidenko, ~A.}, {\em Quasinormal modes of black holes: from astrophysics to string theory}, Reviews of Modern Physics, {\bf 83}: 793 - 836, issue 3 (2011), arXiv:1102.4014v1 [gr-qc] (2011)
%
\bibitem{GW7}{\sc Hod ~S.} {\em Quasinormal resonances of a massive scalar field in a near-extremal Kerr black hole spacetime}   Physical Review D {\bf 84}: 044046 (2011), arXiv:1109.4080v1 [gr-qc]

\bibitem{GW3}{\sc Schnittman, ~J.~D.} {\em Electromagnetic counterparts to black hole mergers} 
Classical and Quantum Gravity, {\bf 28}, Issue 9, pp. 094021 (2011)
 arXiv:1010.3250v1 [astro-ph.HE]
%
\bibitem{LIGO1}{\sc Abbott ~B.~P. et al., LIGO scientific collaboration }{\em Directional limits on persistent gravitational waves using LIGO S5 science data}, arXiv:1109.1809v2 [astro-ph.CO]
%
\bibitem{LIGO2}{\sc The LIGO Scientific Collaboration, the Virgo Collaboration }{\em Search for gravitational waves from binary black hole inspiral, merger and ringdown} Phys.Rev.D{\bf 83}:122005 (2011), arXiv:1102.3781v1 [gr-qc]	
%
\bibitem{LIGO3}{\sc The LIGO Scientific Collaboration, J. Abadie et al. }{\em Search for Gravitational Wave Bursts from Six Magnetars} Astrophys.J.{\bf 734}:L35 (2011), arXiv:1011.4079v2 [astro-ph.HE]
%
\bibitem{LIGO4}{\sc The LIGO Scientific Collaboration }{\em A search for gravitational waves associated with the August 2006 timing glitch of the Vela pulsar} Phys.Rev.D{\bf 83}:042001 (2011);  Phys.Rev.D{\bf 83}:069902 (2011),
   arXiv:1011.1357v3 [gr-qc]
%
\bibitem{LIGO6}{\sc LIGO Scientific Collaboration}{\em First search for gravitational waves from the youngest known neutron star}, Astrophys.J.{\bf 722}:1504 (2010), arXiv:1006.2535v2 [gr-qc]
%
\bibitem{LIGO7}{\sc the LIGO Scientific Collaboration, the Virgo Collaboration}{\em Search for Gravitational Waves from Compact Binary Coalescence in LIGO and Virgo Data from S5 and VSR1}, Phys.Rev.D{\bf 82}:102001 (2010),
   arXiv:1005.4655v1 [gr-qc]
%
\bibitem{LIGO8}{\sc the LIGO Scientific Collaboration, the Virgo Collaboration}{\em All-sky search for gravitational-wave bursts in the first joint LIGO-GEO-Virgo run}, Phys.Rev.D{\bf 81}:102001 (2010)
arXiv:1002.1036v2 [gr-qc]
%
\bibitem{LIGO5}{\sc Alexander Dietz (LIGO Scientific Collaboration and the Virgo Collaboration) }{\em Searches for inspiral gravitational waves associated with short gamma-ray bursts in LIGO's fifth and Virgo's first science run} 
arXiv:1006.3393v1 [gr-qc]
%
\bibitem{LIGO9}{\sc the LIGO Scientific Collaboration, the Virgo Collaboration}{\em Search for gravitational-wave bursts associated with gamma-ray bursts using data from LIGO Science Run 5 and Virgo Science Run 1}, Astrophysical Journal {\bf 715}:1438-1452 (2010), arXiv:0908.3824v2 [astro-ph.HE]
%
\bibitem{GRB1} {\sc Nysewander M., Fruchter A. S., Pe’er. A.}, {\em A Comparison
of the Afterglows of Short- and Long-Duration Gamma-Ray Bursts}, ApJ.{\bf 701}:824-836 (2009), arXiv:0806.3607
[astro-ph] (2009)
%
\bibitem{GRB2}{\sc Lv H., Liang E., Zhang B. , Zhang B.}, {\em A New Classifica-
tion Method for Gamma-Ray Bursts}, ApJ.{\bf 725}:1965-1970 (2010),arXiv:1001.0598v1
[astro-ph.HE] 
%
\bibitem{GW1}{\sc D.M. Coward, B. Gendre, P.J. Sutton, E.J. Howell, T. Regimbau, M. Laas-Bourez, A. Klotz, M. Boer, M. Branchesi} {\em Toward an optimal search strategy of optical and gravitational wave emissions from binary neutron star coalescence}, MNRAS, {\bf 415}:L26-L30 (2011) arXiv:1104.5552v1 [astro-ph.HE]
%
\bibitem{GW_}{\sc Rezzolla ~L.,  Giacomazzo ~B., Baiotti ~L., Granot J., Kouveliotou C., Aloy M.A.}{\em The missing link: Merging neutron stars naturally produce jet-like structures and can power short Gamma-Ray Bursts},Astrophys. J. Lett. {\bf 732}: L6 (2011), arXiv:1101.4298v2 [astro-ph.HE]
%
\bibitem{GW4}{\sc Bogdanovic ~T. , Bode ~T., Haas ~R., Laguna ~P., Shoemaker ~D.} {\em Properties of Accretion Flows Around Coalescing Supermassive Black Holes}, Classical and Quantum Gravity, {\bf 28}:094020 (2011), arXiv:1010.2496v2 [astro-ph.CO]
%
\bibitem{GW5}{\sc Moesta ~P., Alic ~D., Rezzolla ~L., Zanotti ~O., Palenz ~C.} {\em On the detectability of dual jets from binary black holes},    arXiv:1109.1177v1 [gr-qc]
%
\bibitem{LIGO10}{\sc the LIGO Scientific Collaboration, the Virgo Collaboration}{\em Implementation and testing of the first prompt search for electromagnetic counterparts to gravitational wave transients},    arXiv:1109.3498v1 [astro-ph.IM]
%
\bibitem{GW8}{\sc N.L. Christensen, for the LIGO Scientific Collaboration, the Virgo Collaboration} {\em Multimessenger Astronomy}, For the proceedings for the 46th Rencontres de Moriond and GPhyS Colloquium on Gravitational Waves and Experimental Gravity, arXiv:1105.5843v1 [gr-qc]
%
\bibitem{headon}{\sc Jaramillo ~J.~L., Macedo ~R. ~P., Moesta ~P., Rezzolla ~L.}{\em Black-hole horizons as probes of black-hole dynamics I: post-merger recoil in head-on collisions}, Submitted to PRD,    arXiv:1108.0060v1 [gr-qc]
%
\bibitem{headon0}{\sc Rezzolla ~L., Macedo ~R.~P., Jaramillo ~J.~L.}{\em Understanding the "anti-kick" in the merger of binary black holes},Phys.Rev.Lett.{\bf 104}:221101 (2010) arXiv:1003.0873v2 [gr-qc]
%
\bibitem{bin1}{\sc Anninos ~P., Hobill ~D., Seidel ~E., Smarr ~L., Suen W.-M.}{\em The Collision of Two Black Holes},Phys.Rev.Lett. {\bf 71}:2851-2854 (1993), arXiv:gr-qc/9309016v1
%
\bibitem{bin2}{\sc Buonanno ~A., Cook G.B., Pretorius ~F.}{\em Inspiral, merger and ring-down of equal-mass black-hole binaries},Phys.Rev.D{\bf 75}:124018 (2007), arXiv:gr-qc/0610122v2
%
\bibitem{bin3}{\sc Schnittman ~J.~D., Buonanno ~A. , van Meter ~J.~R., Baker ~J.~G. , Boggs ~W.~D. , Centrella ~J.,  Kelly ~B.J., McWilliams ~S.~T. }{\em Anatomy of the binary black hole recoil: A multipolar analysis},Phys.Rev.D{\bf 77}:044031 (2008), arXiv:0707.0301v2 [gr-qc]
%
\bibitem{bin4}{\sc Shibata ~M., Taniguchi ~K.}{\em Merger of black hole and neutron star in general relativity: Tidal disruption, torus mass, and gravitational waves},Phys.Rev.D{\bf 77}:084015 (2008),
arXiv:0711.1410v1 [gr-qc]
%
\bibitem{bin5}{\sc Lousto ~C.~O., Nakano ~H., Zlochower ~Y., Campanelli ~M.}{\em Intermediate-mass-ratio black hole binaries: intertwining numerical and perturbative techniques} Phys.Rev.D{\bf 82}:104057 (2010), arXiv:1008.4360v2 [gr-qc]
%

\bibitem{special31} {\sc Berti ~E., Cardoso ~V., Will ~C.~M. }, {\em On gravitational-wave spectroscopy of massive black holes with the space interferometer LISA}, Phys.Rev.D {\bf 73}:064030 (2006), arXiv:0512160v2 [gr-qc]
%
\bibitem{NB1} {\sc Schutz ~B.~F., Centrella ~J., Cutler ~C., Hughes ~S.~A.}, {\em Will Einstein Have the Last Word on Gravity?}, astro2010: The Astronomy and Astrophysics Decadal Survey (2010), arXiv:0903.0100v1 [gr-qc]
%
\bibitem{NB2} {\sc Chirenti ~C. ~B. ~M.~H., Rezzolla ~L. }, {\em  How to tell gravastar from black hole}, Class. Quant. Grav. {\bf 24}: 4191-4206 (2007), arXiv:0706.1513v2 [gr-qc]
%
\bibitem{NB3} {\sc Chirenti ~C. ~B. ~M.~H., Rezzolla ~L.  }, {\em Ergoregion instability in rotating gravastars}, Phys.Rev.D {\bf 78}:084011 (2008), arXiv:0808.4080v1 [gr-qc]
%
\bibitem{NB4} {\sc Pani ~P., Berti ~E., Cardoso ~V., Chen ~Y., Norte ~R. }, {\em Gravitational wave signatures of the absence of an event horizon: Nonradial oscillations of a thin-shell gravastar }, Phys.Rev.D {\bf 80}:124047 (2009) , arXiv:0909.0287v2 [gr-qc]
%
\bibitem{spectra} {\sc Staicova ~D., Fiziev ~P.}, {\em The Spectrum of Electromagnetic Jets from Kerr Black Holes and Naked Singularities in the Teukolsky Perturbation Theory},Astrophysics and Space Science, {\bf 332}:385-401 (2010), arXiv:1002.0480v2 [astro-ph.HE]
%
\bibitem{GW6}{\sc Lyutikov ~M., McKinney ~J.~C.} {\em Slowly balding black holes}, Phys. Rev. D {\bf 84}:084019  (2011),    arXiv:1109.0584v1 [astro-ph.HE] 
%
\bibitem{time-scale}{\sc Gao, He and Zhang,Bin-Bin and Zhang,Bing} {\em Evidence Of Superposed Variability Components In GRB Prompt Emission Lightcurves},  arXiv:1103.0074v2 [astro-ph.HE], (2011)
%
\bibitem{GRB}{\sc Zhang ~B.}{\em Open Questions in GRB Physics} Comptes Rendus Physique {\bf 12}: 206-225 (2011), arXiv:1104.0932v1 [astro-ph.HE]
%
\bibitem{rot} {\sc Lei ~W.-H., Zhang ~B.}, {\em Black hole Spin in Sw J1644+57 and Sw J2058+05
}, ApJ {\bf L27}: 740 (2006), 	arXiv:1108.3115v2 [astro-ph.HE]
%
\bibitem{rot1} {\sc McClintock ~J.~E., Narayan ~R., Davis ~S.W., Gou ~L., Kulkarni ~A., Orosz ~J.A., Penna ~R.F.,  Remillard ~R.A., Steiner ~J.F.}, {\em Measuring the Spins of Accreting Black Holes
}, To appear in Classical and Quantum Gravity; Special volume for GR19, eds. D. Marolf and D. Sudarsky, arXiv:1101.0811v2 [astro-ph.HE],(2011) 
%
\bibitem{rot2} {\sc Brenneman ~L.W., Reynolds ~C.S.}, {\em Constraining Black Hole Spin Via X-ray Spectroscopy
}, Astrophys.J.{\bf 652}:1028-1043 (2006), arXiv:astro-ph/0608502v1
%
\bibitem{Fiziev4} {\sc Fiziev~P.~P.},  {\em Novel relations and new properties of confluent Heun's functions and their derivatives of arbitrary order}, J. Phys. A: Math. Theor. {\bf 43}:035203 (2010),    arXiv:0904.0245 [math-ph]
%
\bibitem{heun3_} {\sc Slavyanov ~S.~Y., Lay ~W.}, {\em Special Functions, A Unified Theory Based on Singularities}
(Oxford: Oxford Mathematical Monographs) (2000)
%
\bibitem{heun} {\sc Heun ~K.}, Math. Ann. {\bf 33}: 161  (1889)

\bibitem{heun1_} {\sc Decarreau ~A., Dumont-Lepage ~M. ~Cl., Maroni ~P., Robert ~A. and Roneaux ~A.},  Ann. Soc. Buxelles {\bf 92}: 53 (1978)

\bibitem{heun2_} {\sc Decarreau ~A., Maroni ~P. and Robert ~A.}, 1978 Ann. Soc. Buxelles 92 151. 1995 {\em Heun's Differential Equations ed Roneaux A}, Oxford: Oxford Univ. Press (1995)
%
\bibitem{arxiv1} {\sc Fiziev ~P., Staicova ~D.}. {\em Solving systems of transcendental equations involving the Heun functions.}, American Journal of Computational Mathematics Vol. 02 : 02, pp.95 (2012)
%
\bibitem{BC_new} {\sc Casals ~M., Ottewill ~A.~C.}. {\em Analytic Investigation of the Branch Cut of the Green Function in Schwarzschild Space-time.}, arXiv:1210.0519 [gr-qc]
%
\bibitem{arxiv3} {\sc Fiziev ~P., Staicova ~D.}. {\em Application of the confluent Heun functions for finding the QNMs of nonrotating black hole}, Phys. Rev. D {\bf 84}: 127502 (2011), arXiv:1109.1532 [gr-qc]
%
\bibitem{arxiv}
{\sc Fiziev ~P., Staicova ~D. }, {\em Two-dimensional generalization of the Muller root-finding algorithm and its applications},  arXiv:1005.5375v2 [cs.NA] (2011)
%
\bibitem{special2} {\sc Maassen van den Brink ~A}. {\em Analytic treatment of black-hole gravitational waves at the algebraically special frequency}, Phys. Rev. D {\bf 62}: 064009 (2000), arXiv:gr-qc/0001032v1
 %
\bibitem{AS} {\sc Leung ~P.~T., Maassen van den Brink ~A., Mak ~K.~W., Young ~K.}. {\em Unconventional Gravitational Excitation of a Schwarzschild Black Hole}, Class.Quant.Grav. {\bf 20}: L217 (2003), arXiv:gr-qc/0301018v4
%
\bibitem{AS1} {\sc Onozawa ~H.},{\em A detailed study of quasinormal frequencies of the Kerr black hole}, Phys.Rev. D {\bf 55}: 3593-3602 (1997), arXiv:gr-qc/9610048v1
%
\bibitem{BC_New2} {\sc Yang ~H., Zhang ~F., Zimmerman ~A., Nichols ~D.~A., Berti ~E., Chen ~Y.},{\em Branching of quasinormal modes for nearly extremal Kerr black holes}, arXiv:1212.3271 [gr-qc]
\end{thebibliography}
\end{document}